\documentclass[review]{elsarticle}

\usepackage{hyperref}
\pdfoutput=1	
\usepackage{lineno}	
\usepackage{color}
\usepackage{amsmath}
\usepackage{graphicx}
\usepackage{subfigure}
\usepackage{rotating}
\usepackage{multirow, makecell}
\usepackage{geometry}
\begin{document}

%\graphicspath{{figures/}}

\begin{frontmatter}

\title{Further studies on numerical instabilities of Godunov-type schemes for strong shocks}
			
\author[mymainaddress]{Wenjia Xie\corref{mycorrespondingauthor}}
\cortext[mycorrespondingauthor]{Corresponding author}
\ead{xiewenjia@nudt.edu.cn}
\author[mymainaddress]{Zhengyu Tian}	
\author[mymainaddress]{Ye Zhang}
\author[mymainaddress]{Hang Yu}
\author[mymainaddress]{Fan Yang}
	
\address[mymainaddress]{College of Aerospace Science and Engineering, National University of Defense Technology, Hunan 410073, China}
					
\begin{abstract}
In this paper, continuous research is undertaken to explore the underlying mechanism of numerical shock instabilities of Godunov-type schemes for strong shocks. By conducting dissipation analysis of Godunov-type schemes and a sequence of numerical experiments, we are able to clarify that the instability may be attributed to insufficient entropy production inside the numerical shock structure. As a result, a general entropy-control technique for improving the robustness of various Godunov-type schemes at strong shocks is developed. It plays a part in guaranteeing that enough entropy is produced inside the numerical shock structure. Furthermore, such a modified approach does not introduce any additional numerical dissipation on linear degenerate waves to suppress the shock instability. Numerical results that are obtained for various test cases indicate that the proposed methods have a good performance in terms of accuracy and robustness.
\end{abstract}
			
			
\end{frontmatter}
		
\section{Introduction}
\label{sec1}
Numerical shock prediction is one of the most important issues in computational fluid dynamics. Generally, the simulation of compressible fluid flows containing shocks can be implemented in two alternative approaches: shock-fitting and shock-capturing. The shock-fitting technique is able to provide very accurate solutions and has considerable advantages in efficiency in terms of computational cost. Unfortunately, though considerable progress has been achieved in the shock-fitting technique, it has always been unpopular due to clear limitations in simulating three-dimensional flow fields and complex shock patterns \cite{bonfiglioli2016moretti}. Nowadays, the shock-capturing prevails over the shock-fitting in practical simulations of fluid flows involving shock waves. Based on the mathematical theory of weak solutions, the shock-capturing method is able to naturally recognize shock without special treatment. The Godunov-type schemes with their clear physical interpretations are one of the most famous shock-capturing methods. However, modern Godunov-type schemes that have minimal dissipation on discontinuities are still limited in the cases where strong shocks exist. The solutions of flows involving strong shock waves are often characterized by the appearance of numerical anomalies. The most commonly observed shock anomaly is the carbuncle phenomenon.

Since its first discovery by Peery and Imlay \cite{PEERY1988}, the carbuncle has been extensively studied by many researchers during the past three decades. Readers are referred to references \cite{ismail2006toward,shen2014,Rodionov2017,Xie2017} and references therein for detail literature reviews of this subject. Although the carbuncle commonly refers to a spurious solution of blunt body calculations in which a protuberance grows ahead of the bow shock along the stagnation line \cite{quirk1994contribution}, it has been generally considered to be a concrete manifestation of numerical shock instabilities. A number of studies have shown that these numerical shock instabilities are more prone to occur in cases where numerical methods own minimal dissipation on discontinuities and the shocks are sufficiently strong. Apart from the compressible Euler equations of gas dynamics, recent studies have shown that these numerical shock instabilities also plague numerical approximations to other hyperbolic systems of conservation laws, such as the shallow water equations \cite{bader2014carbuncle,kemm2014note,navas2018improved} and the MHD equations \cite{hanawa2008}. Actually, the stability of numerical approximations to systems of conservation laws has not been well established \cite{Lax2008,Fjordholm2015}. Currently, the entropy stability is the only rigorous theory of the numerical instability for numerical approximations to systems of conservation laws. Godunov scheme \cite{godunov1959finite} and some Godunov-type schemes (e.g. the Roe scheme with Harten's entropy fix \cite{harten1997high}, the HLLE/HLLEM schemes \cite{Einfeldt1988,Einfeldt1991} and the Osher-Solomon scheme \cite{Osher1982}) satisfy a discrete version of the entropy conditions which ensure that the computed entropy will increase across shocks \cite{tadmor2003entropy,tadmor2016entropy}. The entropy conditions rescue Godunov-type schemes from admitting nonphysical expansion shocks. However, in some cases, it may not suffice to ensure the convergence of approximate solutions. It has been theoretically established that Godunov and most Godunov-type methods are still unstable for steady shock waves even for the one-dimensional case \cite{Bultelle1998,Barth1989}. Recent results in studies \cite{de2009euler,chiodaroli2015global,baba2018non} even demonstrate that entropy solutions may not be unique. Consequently, numerical approximations that are designed to satisfy the entropy conditions may not converge to the entropy solutions we desire. That is the reason why authors in \cite{elling2005nonuniqueness,elling2009carbuncle,Robinet2000Shock} suspect that carbuncles may be the entropy solutions of the Euler system and have their own physical background. Considering the failure of entropy stable schemes on solving a steady shock in one dimension, Ismail et al. \cite{ismail2006toward,ismail2009affordable} are the first to investigate the effects of the entropy production on shock stability. They find that proper entropy should be produced inside the shock structure to avoid excessive smearing on discontinuities and spurious oscillations. With suitable amount of diffusion on entropy conservative flux \cite{roe2006affordable}, their entropy-consistent solver successfully eliminates the one-dimensional shock instability even though it does not remove the multidimensional shock instability, particularly the carbuncle phenomenon.

The importance of numerical shock structure to trigger the shock instability has also been emphasized by many other researchers \cite{Wada1997,xu2001dissipative,Dumbser2004,chauvat2005shock,zaide2012numerical}. In our previous work \cite{Xie2017}, we make efforts to clarify the strong relation between the onset of numerical shock instability and the shock structure. It has been clarified that the spatial location of the shock instability originates from the intermediate states inside the shock structure. Furthermore, it has been demonstrated that if the mass flux across the normal shock is correctly preserved during the computation, then the shock wave can be captured stably. These are useful for developing possible strategies for suppressing numerical shock instabilities. For example, the robustness of numerical schemes can be improved by introducing multidimensional dissipation to damp perturbation errors inside the numerical shock structure and then help maintain the correct mass flux across the normal shock. Such a strategy is commonly used to cure shock instabilities of low diffusion approximate Riemann solvers. For example, the hybrid technique \cite{quirk1994contribution,coquel1995hybrid,Kim2009,Kim2010Realization,zhang2017robust} that introduces more numerical dissipation in the vicinity of shocks to suppress instabilities by hybridizing a low dissipative solver and a more dissipative one. In some rotated Riemann solvers \cite{REN20031379,nishikawa2008very,zhang2016evaluation}, artificial dissipation is introduced in the vicinity of shocks to improve the robustness of shock-capturing schemes. Several authors \cite{SIMON2018144,Xie2017} resort to modifying the diffusion term of the complete Riemann solvers to include more numerical dissipation on linear degenerate waves to cure the shock instability. Moreover, some recent studies \cite{powers2015physical,Rodionov2017,rodionov2018artificial} have demonstrated that the carbuncle instability can be cured by imposing some artificial viscosity analogous to physical diffusion on shock-capturing schemes. Although introducing additional numerical dissipation or artificial viscosity has been well-demonstrated to be effective in eliminating the shock instability, such numerical viscosity will probably smear discontinuities and decrease overall accuracy. Actually, it is still a challenge to determine suitable amount of additive diffusion for low diffusion approximate Riemann solvers in the sense that it is large enough to make the resulting solvers carbuncle free and minimum enough to resolve the discontinuities as crisply as possible. Thus, it is an attractive issue on how to cure the carbuncle problem of shock-capturing schemes without resorting to additional numerical dissipation.

Kitamura et al. \cite{kitamura2013towards} develop a new pressure flux for AUSM-family schemes. They find that the dissipation inside the numerical shock structure must be proportional to Mach number. Hence, more numerical diffusion is included to smear the shock profile, but the accuracy on resolving contact and shear waves are maintained. In a recent work of Sangeeth and Mandal \cite{SIMON2019477}, a simple cure for numerical shock instability in the HLLC Riemann solver \cite{Toro1994} is proposed. By a newly developed selective wave modification strategy, they succeed to eliminate the shock instability problem of the HLLC scheme without compromising on its contact and shear preserving ability. Zaide et al. \cite{zaide2011shock,zaide2012numerical} notice that the shock with the unphysical intermediate states are highly sensitive to perturbations. To circumvent shock anomalies, they introduce the interpolated fluxes \cite{zaide2012flux} that do not depend on intermediate states inside the shock. Numerical results demonstrate that such a technique is effective in eliminating several shock anomalies, including the carbuncle problem in one dimension. Apart from circumventing the intermediate state, another attractive approach is to limit the propagation of perturbations that are generated by the intermediate cell. As discussed in the previous study \cite{Xie2017}, the instability is triggered by the erroneous mass flux behind shocks. It is caused by the perturbation errors that are generated inside the numerical shock structure and propagate downstream. Such an observation paves the way for suppressing the shock instability from a new perspective. That is, we may control the transportation of the erroneous perturbations from the intermediate states to the downstream cells, thus suppressing the instability. This technique is more attractive because it does not introduce additional numerical dissipation on linear degenerate waves to enhance the stability. In a recent study \cite{Xie2019}, we have shown that the robustness of the Roe method \cite{roe1997approximate} can be improved by controlling the pressure perturbations that are transported from the intermediate states inside the shock structure to the downstream regions. In the current study, we will continue to explore the mechanism of the numerical shock instability and provide an in-depth explanation for the rationality of this strategy. A focus is on clarifying the connection between the entropy production inside the numerical shock structure and the stability, which is inspired by the pioneer work of Ismail et al. \cite{ismail2009affordable}. Combining the numerical dissipation analysis of Godunov-type schemes and a sequence of numerical experiments, we are able to clarify that the numerical instability at strong shocks can be attributed to inappropriate entropy production inside the shock structure. As a result, a general entropy-control technique is developed to improve the robustness of various Godunov-type schemes at strong shocks. Different from common cures for carbuncles which rely on multidimensional dissipation, the current modified approach does not introduce any additional numerical dissipation to improve the robustness, thus maintaining the accuracy of numerical methods on discontinuities.

The outline of the rest of this paper is as follows. In section 2, governing equations of compressible flows and their related finite volume discretization are presented. Two classical HLL-type schemes are also reviewed in the same section. In section 3, we present a mechanism study of the numerical shock instability. In section 4, the shock instability problem is revisited through a linearized perturbation analysis. A general technique for improving the robustness of Godunov-type schemes at strong shocks is proposed in section 5. The accuracy and robustness of the proposed methods are tested in section 6. Section 7 contains conclusions and an outlook to future developments.

\section{Governing equations and finite volume discretization}
\label{sec2}
The two-dimensional Euler equations may be written in integral form as
\begin{equation}\label{eq2.0.1}
    \frac{\partial}{\partial t} \int_{\Omega} \mathbf{U} d{\Omega} + \oint_{\partial \Omega} \mathbf{F} dS = 0,
\end{equation}
where $\partial \Omega$ denote boundaries of the control volume $\Omega$. The state vector and flux vector are defined as
\begin{equation}\label{eq2.0.2}
\mathbf{U} = \left[ {\begin{array}{{c}}
\rho \\
{\rho u}\\
{\rho v}\\
{\rho e}
\end{array}} \right],
\quad
\mathbf{F} = \left[ {\begin{array}{{c}}
\rho q \\
{\rho u q + p n_x}\\
{\rho v q + p n_y}\\
{\left(\rho e+p\right)q}
\end{array}} \right],
\end{equation}
where $\rho$, $e$, and $p$ represent density, specific total energy and pressure respectively, and ${\bf{u}} = \left( {u,v} \right)$ is the flow velocity. The directed velocity, $q=un_x+vn_y$, is the component of velocity acting in the $\mathbf{n}$ direction, where $\mathbf{n}={\left[ {{n_x},\;{n_y}} \right]^T}$ is the outward unit vector normal to the surface element $dS$. The equation of state is in the form
\begin{equation}\label{eq2.0.3}
p = \left(\gamma-1\right) \rho \left[e-\frac{1}{2}\left(u^2+v^2\right)\right],
\end{equation}
where $\gamma$ is the specific heat ratio.
We consider discretizing the system (\ref{eq2.0.1}) with a cell-centered finite-volume method over a two-dimensional domain subdivided into some structured quadrilateral cells. The semi-discrete finite volume scheme over a particular control volume ${\Omega _i}$  can be written as
\begin{equation}\label{eq2.0.4}
\frac{{{\rm{d}}{{\bf{U}}_i}}}{{{\rm{d}}t}} + \frac{1}{{\left| {{\Omega _i}} \right|}}\sum\limits_{{\Gamma _{ij}} \subset \partial {\Omega _i}} {\left| {{\Gamma _{ij}}} \right|} {{\bf{F}}_{ij}} = 0.
\end{equation}
In the above expression, ${{\bf{U}}_i}$ is the cell average of ${\bf{U}}$ on ${\Omega _i}$ , $\left| {{\Omega _i}} \right|$  denotes the volume of  ${\Omega _i}$. ${\Gamma _{ij}}$ denotes the common edge of two neighboring cells ${\Omega _i}$ and ${\Omega _j}$, and $|{\Gamma _{ij}}|$ is the length of face ${\Gamma _{ij}}$. The flux  ${{\bf{F}}_{ij}}$ is the calculated numerical flux that is supposed to be constant along the individual face  ${\Gamma _{ij}}$. In the following section, we give a brief review of two classical approximate Riemann solvers for determining the numerical flux ${{\bf{F}}_{ij}}$ at each cell interface, i.e., HLLE scheme \cite{Einfeldt1988,Einfeldt1991} and HLLEM scheme \cite{Einfeldt1988}.
	
\subsection{HLL approximate Riemann solver}
\label{sec2.1}
The HLL Riemann solver proposed by Harten, Lax, and van Leer \cite{harten1983upstream} is one of the most reliable approaches to solve the Riemann problem approximately. Different from the exact Riemann solution with a large amount of detail, the HLL solver assumes an average intermediate state $\bf{U}^*$ between the fastest and slowest waves, $S_L$, $S_R$. The resulting approximate Riemann solution can be written as
\begin{equation}\label{eq2.1.1}
{\bf{\omega}}(x/t;{\bf U}_L,{\bf U}_R)=
\begin{cases}
{\bf U}_L       & \text{$x/t < S_L$}     \\
{\bf U}^*       & \text{$S_L <x/t < S_R$}  \\
{\bf U}_R       & \text{$S_R < x/t$}
\end{cases},
\end{equation}
where the approximate intermediate state ${\bf{U}}^*$  is defined to be consistent with the following integral form of the conservation law, i.e.,
\begin{equation}\label{eq2.1.2}
\int_{ - \Delta x /2}^{\Delta x /2} {\bf{\omega}}(x/t;{\bf U}_L,{\bf U}_R) \;dx = \Delta x/2\left( {{{\bf{U}}_L} + {{\bf{U}}_R}} \right) - \Delta t{\bf{F}}\left( {{{\bf{U}}_R}} \right) + \Delta t{\bf{F}}\left( {{{\bf{U}}_L}} \right)
\end{equation}
for $\Delta x/2 > \Delta t \cdot \max \left\{ {\left| S_L \right|,\left| S_R \right|} \right\}$. The cell averages at the next time level are then obtained after computing the approximate Riemann solvers at the cell interfaces, i.e.,
\begin{equation}\label{eq2.1.3}
{\bf{U}}_i^{n + 1} = \frac{1}{{\Delta x}}\int_0^{\Delta x/2} {{\bf{\omega }}_{i - 1/2}^n\left( {x/\Delta t} \right)dx}  + \frac{1}{{\Delta x}}\int_{ - \Delta x/2}^0 {{\bf{\omega }}_{i + 1/2}^n\left( {x/\Delta t} \right)dx}.
\end{equation}
The above equation can be rewritten in the conservative form as
\begin{equation}\label{eq2.1.4}
{\bf{U}}_i^{n + 1} = {\bf{U}}_i^n - \frac{{\Delta t}}{{\Delta x}}\left( {{\bf{F}}_{i + 1/2}^n - {\bf{F}}_{i - 1/2}^n} \right)
\end{equation}
with the numerical flux function defined by
\begin{equation}\label{eq2.1.5}
{\bf{F}}_{\rm{HLL}}=
\begin{cases}
{\bf F}\left( {{{\bf{U}}_L}} \right)       & \text{$S_L>0$}               \\
\frac{{{S_R}{\bf{F}}\left( {{{\bf{U}}_L}} \right) - {S_L}{\bf{F}}\left( {{{\bf{U}}_R}} \right)}}{{{S_R} - {S_L}}} + \frac{{{S_L}{S_R}}}{{{S_R} - {S_L}}}\left( {{{\bf{U}}_R} - {{\bf{U}}_L}} \right)       & \text{$S_L \le 0 \le S_R$}    \\
{\bf F}\left( {{{\bf{U}}_R}} \right)       & \text{$S_R<0$}
\end{cases}.
\end{equation}
To completely determine the interface flux, the wave speeds need to be estimated. Concerning the stability and positivity preserving property, Einfeldt et al. \cite{Einfeldt1991} proposed a way of computing the wave speeds,
\begin{equation}\label{eq2.1.6}
S_L=\min(q_L-a_L,\widehat{q}-\widehat{a}),\quad S_R=\max(q_R+a_R,\widehat{q}+\widehat{a})
\end{equation}
where $\widehat{(\cdot)}$ is Roe's averaged variable with respect to ${\bf U}_L$ and ${\bf U}_R$, $a_L$ and $a_R$ are the sound speeds of the left and the right states. The estimate (\ref{eq2.1.6}) can return the exact shock velocity in the case of an isolated shock \cite{Batten19972}. Davis \cite{Davis1988} also suggests another simpler and more diffusive estimate,
\begin{equation}\label{eq2.1.7}
S_L=\min(q_L-a_L,q_R-a_R),\quad S_R=\max(q_L+a_L,q_R+a_R).
\end{equation}
Practically, the flux function (\ref{eq2.1.5}) can be further rewritten into a unified form
\begin{equation}\label{eq2.1.8}
{\bf{F}}_{\rm{HLL}}=\frac{{{S_R^+}{\bf{F}}\left( {{{\bf{U}}_L}} \right) - {S_L^-}{\bf{F}}\left( {{{\bf{U}}_R}} \right)}}{{{S_R^+} - {S_L^-}}} + \frac{{{S_L^-}{S_R^+}}}{{{S_R^+} - {S_L^-}}}\left( {{{\bf{U}}_R} - {{\bf{U}}_L}} \right)
\end{equation}
with the wave speeds $S_L^- = \min(S_L, 0)$ and $S_R^+ = \max(S_R, 0)$.

\subsection{HLLEM approximate Riemann solver}
\label{sec2.2}
Due to the assumption of a two-wave configuration, the HLL scheme introduces excess dissipation on linear waves. Therefore, it provides poor resolution of physical features such as contact surfaces, shear waves and material interfaces. The difficulty with this method can be resolved by modifying the intermediate state in (\ref{eq2.1.1}) through a linear distribution approach. For the HLLEM scheme, the solution of the Riemann problem is approximated as
\begin{equation}\label{eq2.2.1}
{\bf{\omega}}(x/t;{\bf U}_L,{\bf U}_R)=
\begin{cases}
{\bf U}_L               & \text{$x/t < S_L$}  \\
{\bf U}^*+(x-\overline{q} t) ({\hat{\delta}}^*_2 \hat{\alpha}_{2} \widehat{{\bf R}}_{2} + {\hat{\delta}}^*_3 \hat{\alpha}_{3} \widehat{{\bf R}}_{3} )     & \text{$S_L \le x/t \le S_R$} \\

{\bf U}_R               & \text{$S_R < x/t$}
\end{cases}
\end{equation}
where $\overline{q}$ denotes the numerical approximation of the velocity at the contact discontinuity, it is defined as a simple arithmetic average of wavespeeds $S_L$ and $S_R$,
\begin{equation}\label{eq2.2.2_2}
	\overline{q}=\frac{S_L+S_R}{2}.
\end{equation}
$\widehat {\bf{R}}_2$  and $\widehat {\bf{R}}_3$  are the second and third right eigenvectors of the flux Jacobian evaluated at a Roe's averaged state $\widehat {\bf{U}}$. $\widehat \alpha _2$ and $\widehat \alpha _3$  are the coefficients of the projection of ${{\bf{U}}_R} - {{\bf{U}}_L}$  onto these eigenvectors, i.e.,
\begin{equation}\label{eq2.2.2}
{{\bf{U}}_R} - {{\bf{U}}_L} = \sum\limits_{k = 1}^4 {\widehat \alpha _k\widehat {\bf{R}}_k}
\end{equation}
with
\begin{equation}\label{eq2.2.3}
\widehat \alpha _2 = \Delta \rho  - \frac{{\Delta p}}{{{{\widehat a}^2}}},\qquad
\widehat \alpha _3 = \widehat \rho \left( {-{n_y}\Delta u + {n_x}\Delta v} \right),
\end{equation}
and
\begin{equation}\label{eq2.2.4}
\widehat {\bf{R}}_2 = {\left[ {\begin{array}{*{20}{c}}
1&{\widehat u}&{\widehat v}&{\left( {{{\widehat u}^2} + {{\widehat v}^2}} \right)/2}
\end{array}} \right]^{\rm{T}}},\begin{array}{*{20}{c}}
{}&{}
\end{array}\widehat {\bf{R}}_3 = {\left[ {\begin{array}{*{20}{c}}
0&{ - {n_y}}&{{n_x}}&{ - \widehat u{n_y} + \widehat v{n_x}}
\end{array}} \right]^{\rm{T}}},
\end{equation}
The parameters ${\hat{\delta}}^* _2$ and ${\hat{\delta}}^* _3$ denote anti-diffusion coefficients which play a role in controlling the amount of anti-diffusion in the linear degenerate fields,
\begin{equation}\label{eq2.2.2_5}
	{\hat{\delta}}^*_k=\frac{2}{T(S_R-S_L)} {\hat{\delta}}_k  \quad \text{for} \quad k=2,3
\end{equation}
with
\begin{equation}\label{eq2.2.2_6}
	{\hat{\delta}}_2 = {\hat{\delta}}_3 = \frac{{\widehat a}}{{\widehat a + \left| {\widehat {{q}} } \right|}}.
\end{equation}
One should notice that the Roe's averaged velocity $\widehat q$ instead of $\overline{q}$ is used to calculate ${\hat{\delta}} _2$ and ${\hat{\delta}} _3$, which resolves the stationary contact discontinuity exactly \cite{park2003dissipation}.
The numerical scheme corresponding to the Riemann solver (\ref{eq2.2.1}) can be written again in conservation form as that in (\ref{eq2.1.4}) with the following flux function,
\begin{equation}\label{eq2.2.6}
{\bf{F}}_{{\rm{HLLEM}}} = \frac{{{S_R^+}{\bf{F}}\left( {{{\bf{U}}_L}} \right) - {S_L^-}{\bf{F}}\left( {{{\bf{U}}_R}} \right)}}{{{S_R^+} - {S_L^-}}} + \frac{{{S_L^-}{S_R^+}}}{{{S_R^+} - {S_L^-}}}\left( {{{\bf{U}}_R} - {{\bf{U}}_L} - {\hat{\delta}} _2\widehat \alpha _2\widehat {\bf{R}}_2 - {\hat{\delta}} _3\widehat \alpha _3\widehat {\bf{R}}_3} \right).
\end{equation}
One can observe that the HLLEM scheme will reduce to the HLLE scheme if the anti-diffusion coefficients ${\hat{\delta}} _2$ and ${\hat{\delta}} _3$ are eliminated.

\section{Exploring the mechanism of numerical shock instability}
\label{sec3}
The carbuncle phenomenon is conventionally referred to the distorted shock ahead of the blunt body in a supersonic or hypersonic flow. In order to analyze the carbuncle, it is wise to resort to a simplified test case which shares essential characteristics of the blunt body carbuncle and can be analyzed in a simplified manner. The problem considered here is the steady normal shock problems in one and two dimensions \cite{Kitamura2009}. It has been well demonstrated that if a scheme encounters the instability for the steady shock, then it will also suffer from the carbuncle \cite{Dumbser2004,ismail2006toward,Kitamura2009}. In the current section, we start by analyzing the instability characteristics of the one-dimensional steady shock problem, i.e., the one-dimensional carbuncle. It is a one-dimensional manifestation of the carbuncle which is often deemed as a multidimensional phenomenon. Attempts are made to explore the connection between the entropy production inside the numerical shock structure and the instability. To this end, we conduct a mechanism study through two interlinked approaches: an analysis of the numerical entropy production and numerical experiments. Then we generalize the approach to the multidimensional case using the dimensional splitting method at the end of this section.

\subsection{One-dimensional carbuncle problem}
\label{sec3.1}
\subsubsection{ Numerical experiment setup}
\label{sec3.1.1}
The one-dimensional carbuncle is a steady shock problem which is a dimensionally-reduced case of its two-dimensional counterpart: the 1.5D steady normal shock test \cite{Kitamura2009}. Here, we repeat the setup of this numerical test for self-containedness. Readers are suggested to refer to references \cite{Kitamura2009,Xie2017} for more detailed descriptions.

\begin{figure}[htbp]
	\centering
	\includegraphics[width=3.8in]{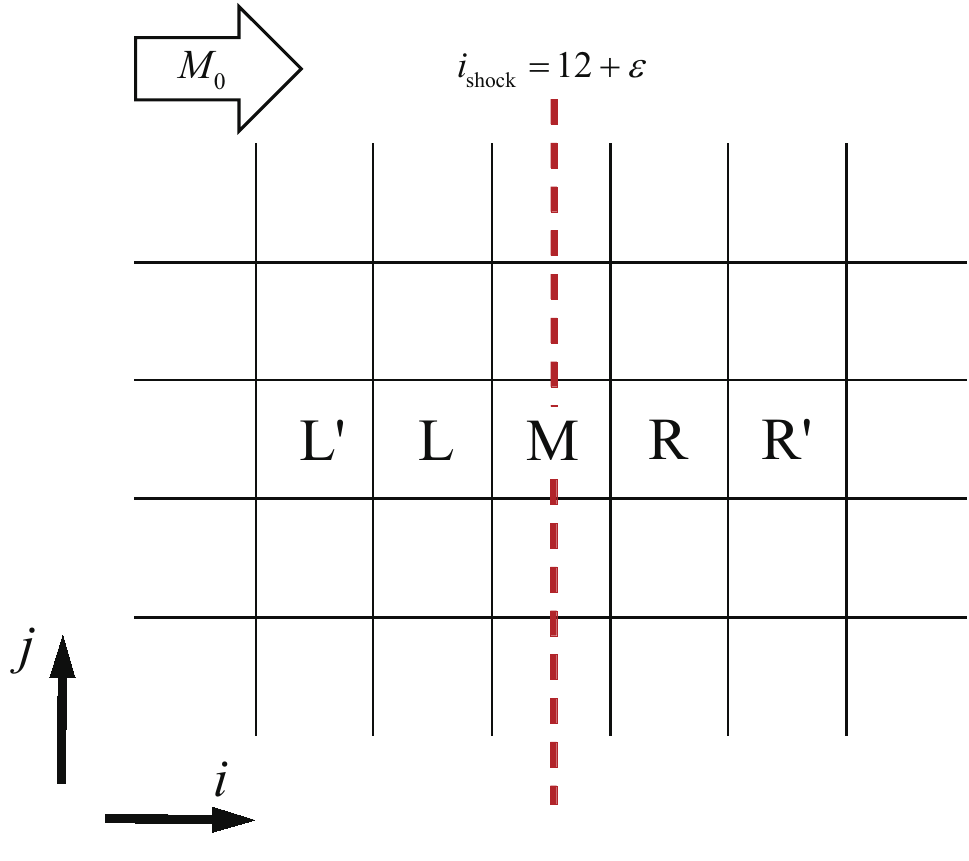}
	\caption{Schematic illustration of the computational grid and conditions for steady normal shock problem.}
	\label{fig3-1-1}
\end{figure}

In Fig. \ref{fig3-1-1}, a schematic diagram is presented to show the computational grid and conditions for the steady shock problem. For the one-dimensional case, only one cell is used in the $j$ direction and $50$ equally spaced cells are used in the $i$ direction. The initial conditions are prescribed for left $(L:i\leq 12)$ and right $(R:i\geq 14)$ following the Rankine-Hugoniot conditions across the normal shock as
\begin{equation}\label{eq3.1}
{\bf{U}}_{L}=
\begin{pmatrix}
1\\
1\\
0\\
\frac{1}{\gamma(\gamma-1)M_0^2}+\frac{1}{2}
\end{pmatrix}, \quad {\bf{U}}_{R}=
\begin{pmatrix}
f(M_0)\\
1\\
0\\
\frac{g(M_0)}{\gamma(\gamma-1)M_0^2}+\frac{1}{2f(M_0)}
\end{pmatrix},
\end{equation}
with
\begin{equation}\label{eq3.2}
f(M_0)=\left(\frac{2}{(\gamma+1)M_0^2}+\frac{\gamma-1}{\gamma+1}\right)^{-1},\quad g(M_0)=\frac{2\gamma M_0^2}{\gamma+1}-\frac{\gamma-1}{\gamma+1},
\end{equation}
where ${M_0}$  and $\gamma $ represent the upstream Mach number and the gas specific heat ratio. The inflow boundary conditions are set to freestream values. The mass flux at the ghost cell is prescribed as
\begin{equation}\label{eq3.3}
(\rho u)_{imax+1,j}=(\rho u)_0=1
\end{equation}
in order for the mass in the whole computational domain to remain constant and to fix the shock at the same position \cite{ismail2009affordable}. Meanwhile, other values are simply extrapolated. The intermediate states within the shock are assumed to lie on the Hugoniot curve \cite{chauvat2005shock}. The internal shock conditions $(M:i=13)$ are as follows:
\begin{equation}\label{eq3.4}
\begin{aligned}
\rho_M & =(1-\alpha_{\rho})\rho_L+\alpha_{\rho}\rho_R \\
   u_M & =(1-\alpha_{u})u_L+\alpha_u u_R \\
   p_M & =(1-\alpha_{p})p_L+\alpha_p p_R
\end{aligned}
\end{equation}
with
\begin{equation}\label{eq3.5}
\begin{aligned}
{\alpha _\rho } & = \varepsilon \\
{\alpha _u} & = 1 - \left( {1 - \varepsilon } \right){\left( {1 + \varepsilon \frac{{M_0^2 - 1}}{{1 + \left( {\gamma  - 1} \right)M_0^2/2}}} \right)^{ - 1/2}}{\left( {1 + \varepsilon \frac{{M_0^2 - 1}}{{1 - 2\gamma M_0^2/\left( {\gamma  - 1} \right)}}} \right)^{ - 1/2}}\\
{\alpha _p} & = \varepsilon {\left( {1 + \left( {1 - \varepsilon } \right)\frac{{\gamma  + 1}}{{\gamma  - 1}}\frac{{M_0^2 - 1}}{{M_0^2}}} \right)^{ - 1/2}}
\end{aligned}
\end{equation}
where $\varepsilon  = 0.0,\;...\;,\;0.9$  is a discrete weighting average. It describes the initial state of the internal cell and is called shock position here. We apply the first-order finite volume method with HLLE/HLLEM schemes defined in section {\ref{sec2}} on this problem for various combinations of Mach numbers and shock positions $\varepsilon $.

\subsubsection{The profile of a one-dimensional carbuncle}
\label{sec3.1.2}
It was demonstrated both analytically \cite{Bultelle1998,Barth1989} and numerically \cite{Kitamura2009} that Riemann solvers which are able to produce stationary discrete shocks with a single interior point are not always stable. As with the Godunov scheme \cite{godunov1959finite} or the Roe scheme \cite{roe1997approximate}, the HLLE and HLLEM solvers with wave speeds (\ref{eq2.1.6}) are able to resolve shocks with only one intermediate state inside the numerical shock structure. They suffer from the one-dimensional carbuncle which happens for strong enough shock, over a range of shock positions. We repeat the one-dimensional steady shock test in \cite{Kitamura2009} to evaluate the behaviors of the HLLE/HLLEM schemes for a wide Mach number spectrum and different shock positions. The results are also presented again for the convenience of further discussion. In Table \ref{table1}, stability results of numerical schemes for 1D steady shock problem are summarized. As expected, both schemes produce unstable results for certain shock positions $\varepsilon=0.1 \sim 0.3$. Compared with the HLLEM scheme, the HLLE flux omits the contact wave in the solution of the Riemann problem, thus it contains excess numerical dissipation on contact discontinuities. However, it is demonstrated in Table \ref{table1}  that such numerical dissipation is of little use against the one-dimensional carbuncle.

\begin{table}
\centering
	\caption{1D test results of Godunov-type flux functions (S: stable and converged, U: unstable).}
	\label{table1}
		\begin{tabular}{cccccccccccc}
			\hline
			Riemann solver & Test problem & $\varepsilon=0.0$ & 0.1 & 0.2 & 0.3 & 0.4 & 0.5 & 0.6 & 0.7 & 0.8 & 0.9 \\
			\hline
			HLLEM & 1D & U & U & U & U & S & S & S & S & S & S \\
			HLLE  & 1D & U & U & U & U & S & S & S & S & S & S \\
			\hline
		\end{tabular}	
\end{table}

\begin{figure}[htbp]
	\centering
	\subfigure[]{
		\begin{minipage}[b]{0.48\textwidth}
			\includegraphics[width=1.0\textwidth]{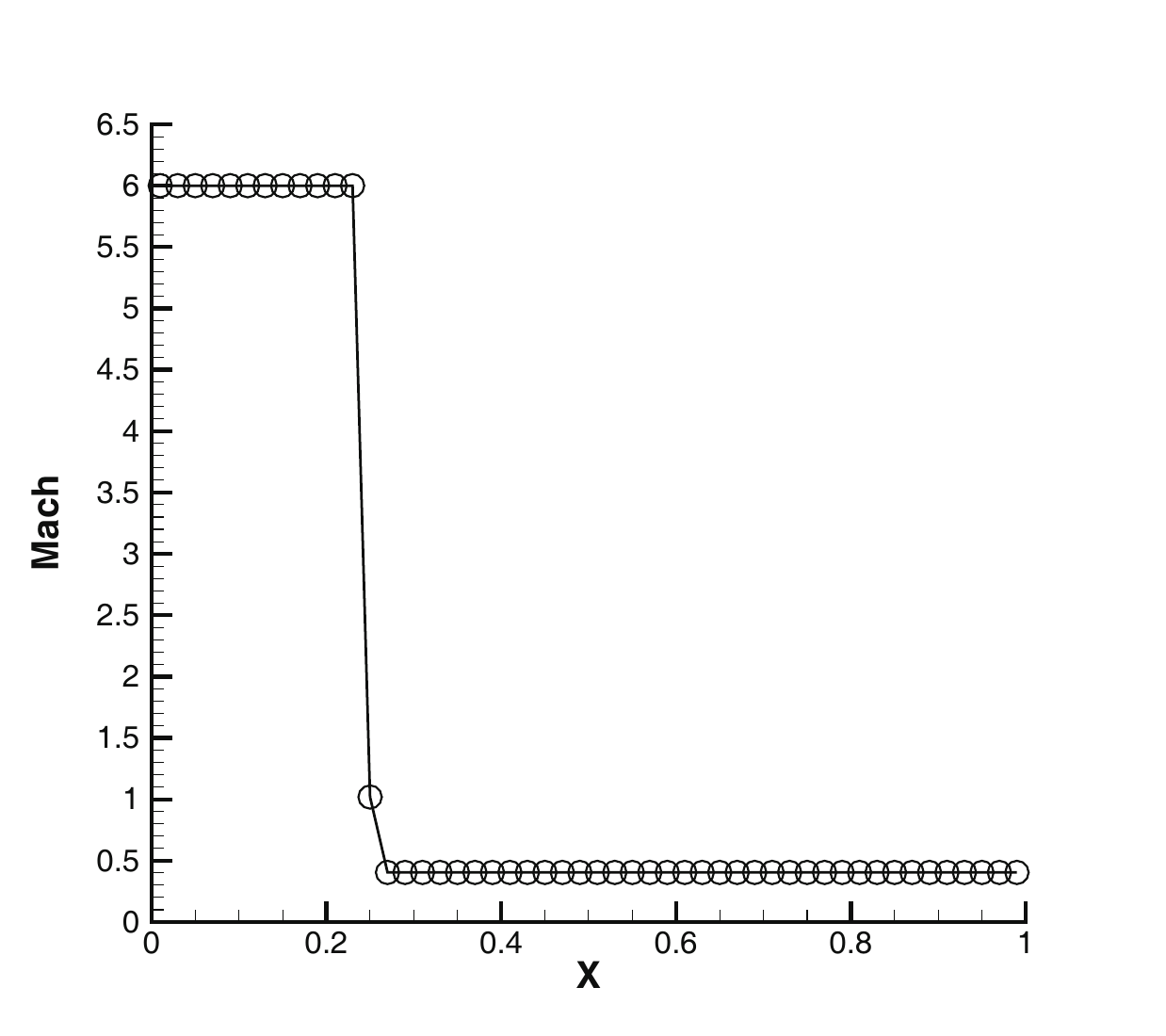}
		\end{minipage}
	}
	\subfigure[]{
		\begin{minipage}[b]{0.48\textwidth}
			\includegraphics[width=1.0\textwidth]{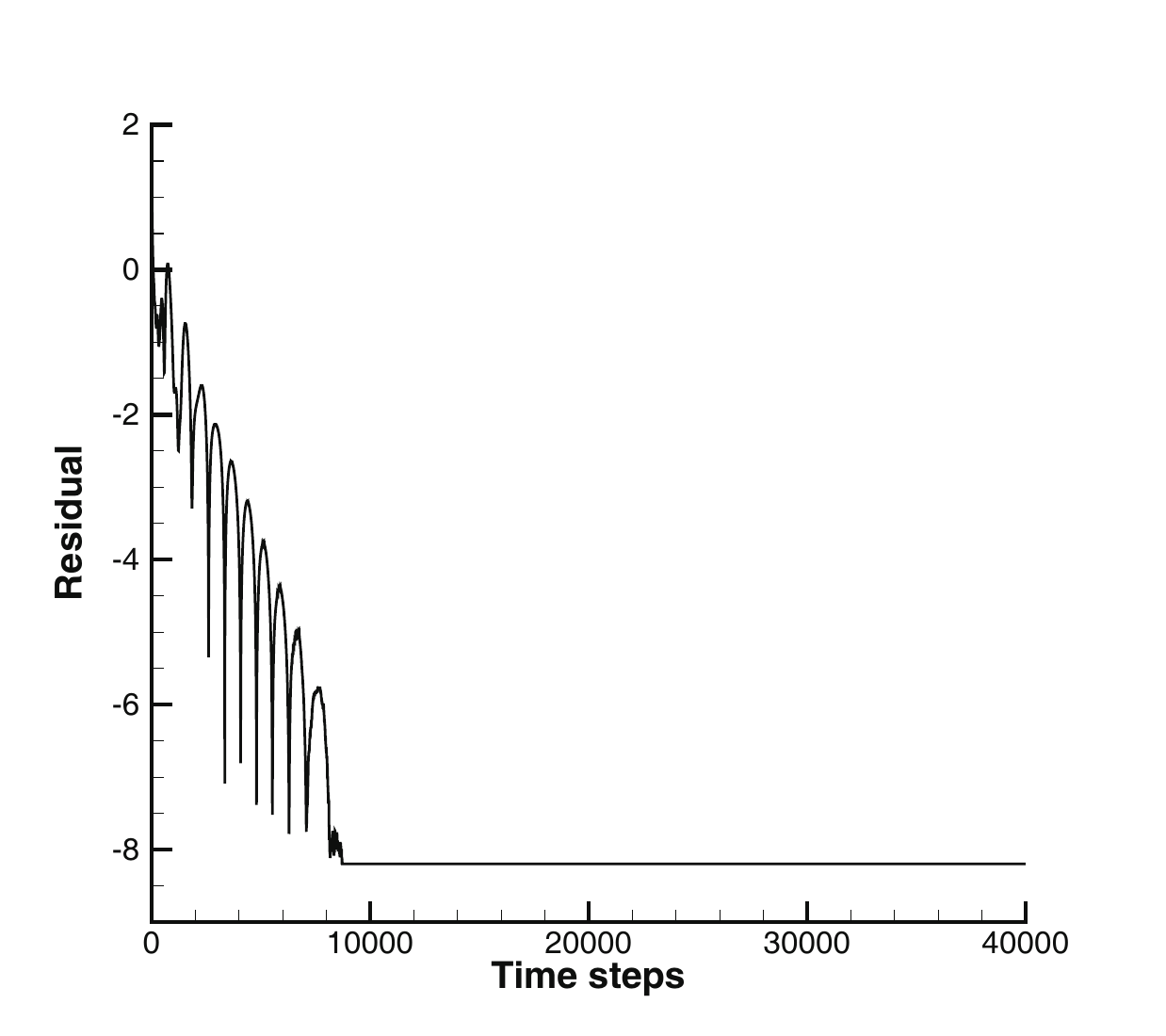}
		\end{minipage}
	}
	\caption{One-dimensional shock profile predicted by the HLLEM scheme ( ${M_0} = 6.0$, $\varepsilon  = 0.5$.), (a) Mach number, (b) residual histories.}
	\label{fig3-1-2}
\end{figure}

\begin{figure}[htbp]
	\centering
	\subfigure[]{
		\begin{minipage}[b]{0.48\textwidth}
			\includegraphics[width=1.0\textwidth]{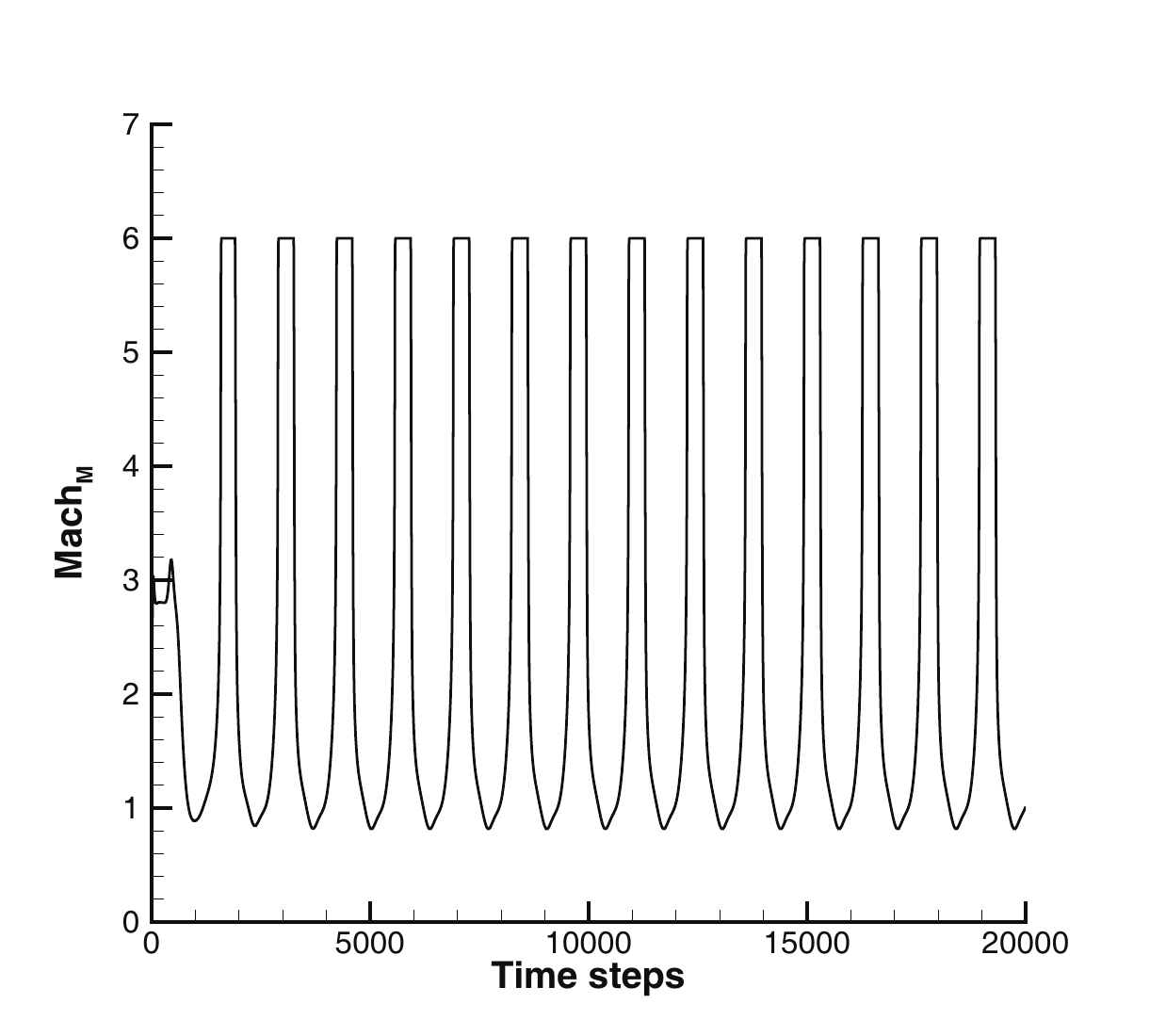}
		\end{minipage}
	}
	\subfigure[]{
		\begin{minipage}[b]{0.48\textwidth}
			\includegraphics[width=1.0\textwidth]{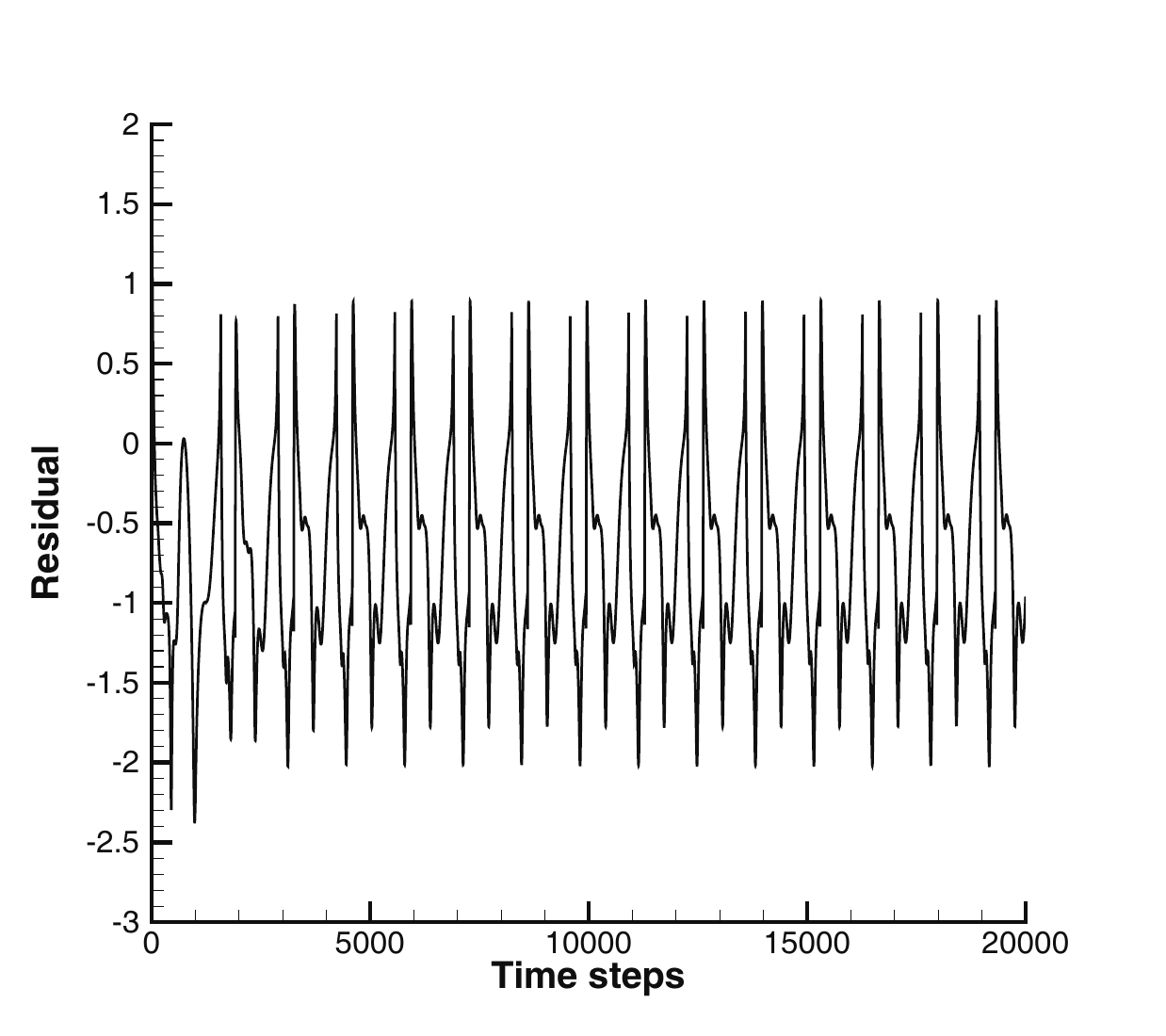}
		\end{minipage}
	}
	\caption{One-dimensional carbuncle predicted by the HLLEM scheme (${M_0} = 6.0$, $\varepsilon = 0.3$. Only the solutions before 20000 time steps are shown for clarity), (a) Mach number, (b) residual histories.}
	\label{fig3-1-3}
\end{figure}

In Fig. \ref{fig3-1-2}, solutions of the HLLEM scheme for a stable shock position are presented. As shown, the computation converges to a weak solution of the Euler equations. An intermediate state exists inside the shock structure which is bounded by the left and right primitive variables. However, the solution fails to converge to a steady state for certain unstable shock positions. Fig. \ref{fig3-1-3} demonstrates essential behaviors of the one-dimensional carbuncle. For the unstable case with setup in section \ref{sec3.1.1}, the intermediate state oscillates inside the shock and enters into an approximate limit cycle involving the shedding of spurious waves and their reflection from the downstream boundary. The HLLE scheme demonstrates similar results that are not shown for clarity. One should note that these unstable behaviors of the one-dimensional carbuncle have also been discussed in previous work by Ismail \cite{ismail2006toward,ismail2009affordable,wahi2013numerical} and Zaide \cite{zaide2011shock,zaide2012numerical} respectively in the context of the Roe scheme \cite{roe1997approximate}. It is a common anomalous phenomenon for Godunov-type schemes, particularly the one that has minimal dissipation on shocks.

\subsection{A heuristic discussion of numerical entropy production and shock instability}
\label{sec3.2}
As we have demonstrated above, for Godunov-type schemes that capture steady shocks with a single interior point, the instability is inevitable. One may argue that adding the dissipation corresponding to nonlinear waves can spread the shock profile, thus suppressing the shock instability. This is the case. In our own experience, shock-capturing schemes that yield a steady shock structure with more than two internal cells are generally stable for the one-dimensional carbuncle. The list includes the Osher scheme \cite{Osher1982}, van Leer scheme \cite{van1982flux}, Roe scheme with Harten's entropy fix \cite{harten1997high}, to name a few. Whereas, the finite discretization of the Euler equations is expected to capture the shock with minimal smearing. Thus, it is still attractive to improve the stability of Godunov-type schemes that resolve the shock with only one intermediate point at most.

In terms of continuum mechanics on which the Euler equations are established, the shock wave is regarded as a jump discontinuity. Theoretical studies have established that inviscid shock waves for the ideal gas are uniformly stable in one and multiple space dimensions \cite{erpenbeck1962stability,majda2012compressible}. However, due to the nature of finite discretization, it is impossible to capture the shock with zero thickness. The captured shock occupies several cells that is often greater than the physical width. Hence, the shock instability of numerical approximations to the Euler systems may come from a loss of information between the continuous and discrete levels. Without explicit knowledge of these discretization errors, it is hard to theoretically investigate the instability of corresponding numerical methods.

However, the state of the intermediate point inside the numerical shock structure is defined and controlled by artificial dissipation introduced by the numerical scheme. Hence, the intermediate state is a direct representation of the characteristics of the numerical dissipation. Thus, it paves the way for exploring the mechanism of numerical shock instability through analyzing the characteristics of the shock structure. In our previous study \cite{Xie2017}, it has been shown that the shock will be stabilized if the intermediate state is fixed to a proper value. Such a value is obtained by the solution of the steady normal shock with an artificial condition that enforces the mass flux consistency (i.e., $(\rho u)_L=(\rho u)_R$) across the normal shock. However, such a technique cannot be implemented in a practical simulation. There is still an important question that needs to be answered. What the intermediate state should be ? Certain useful criterion, which can be used to guarantee that the intermediate state lies inside the stable spectrum, is needed.

The entropy condition plays a decisive role in the numerical prediction of the system of conservation laws. For the Euler systems, the entropy condition requires that the entropy should increase across the shock front. However, such a condition does not guarantee that the entropy is produced in right amounts. The numerical shock profile can be considered to be an approximation of a viscous shock profile governed by the Euler equations with an additional viscosity introduced by the numerical discretization. Analogous to the physical viscosity, the numerical viscosity will also provide a dissipative mechanism that transforms the energy into heat. The whole system should obey the second thermodynamics law and tends to a stable state with the maximum entropy. Without a guarantee that enough entropy is produced inside the shock, the numerical shock structure will be unstable.

\begin{figure}[htbp]
	\centering
	\includegraphics[width=3.8in]{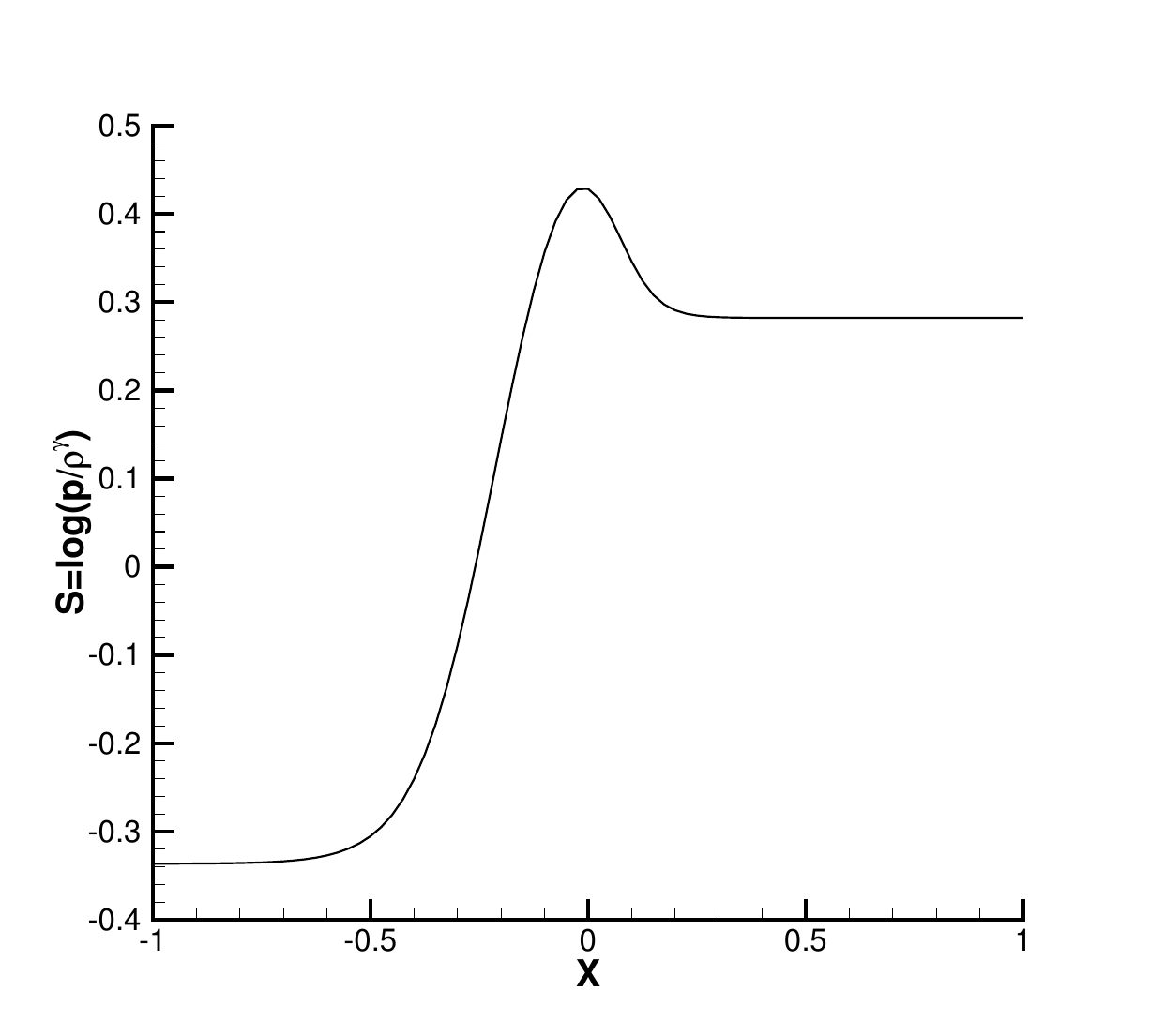}
	\caption{Exact entropy profile of viscous Navier-Stokes shock: $Pr=3/4, Re_{\infty}=25, M_{\infty}=3.5, \gamma=1.4, T_{\infty}=400K$.}
	\label{fig3-2-1}
\end{figure}

Actually, the viscous shock structure of the Navier-Stokes equations may shed some light on this problem. In Fig. \ref{fig3-2-1}, the entropy profile for the NS shock has been plotted, it is generated by an open code provided by Nishikawa \cite{Nishikawa}. It is demonstrated that an entropy overshoot is observed inside the shock. One should note that theoretical studies \cite{humpherys2009spectral,humpherys2010,Humpherys2017} have established that planar viscous Navier-Stokes shock waves for the ideal gas are also uniformly stable in one and multiple space dimensions. Thus, the key factor to the shock instability lies on the characteristics of the dissipation introduced in the numerical procedure and the entropy production can be used to determine what the proper dissipation for stabilizing the shock is. In the following sections, we will demonstrate that if enough entropy production is guaranteed inside the shock, then the instability can be successfully eliminated. More entropy production will make the solution converge to a stable state more quickly.

\subsection{Dissipative mechanism in Godunov-type schemes}
\label{sec3.3}
As discussed in the above section, the numerical shock instability may be attributed to insufficient entropy production inside the numerical shock structure. To increase the entropy production around shocks, there are two alternative approaches. One is to impose more numerical dissipation on nonlinear waves to smear the shock profile, thus more entropy production can be obtained. The other is to reduce the numerical dissipation on entropy waves in the vicinity of shocks, then a larger entropy state will be produced inside the numerical shock structure. In the current study, we prefer the latter one and try to avoid adding additional dissipation corresponding to the nonlinear wave to stabilize shocks. There are at least two reasons to do that. Firstly, more numerical dissipation is harmful for simulation that needs high accuracy. Secondly, the dissipation corresponding to nonlinear waves helps little in eliminating the multidimensional shock instability. Actually, the ultimate goal of this study is to stabilize shocks without introducing any additional numerical dissipation corresponding to whether nonlinear waves or linear waves to the target Riemann solver. To facilitate a further investigation of the relation between entropy production and shock stability, we first conduct an analysis of numerical dissipation in the HLL-family schemes. And then, two modified schemes are constructed by modifying the dissipative terms in the target HLLEM scheme to control the entropy production. An essential role is played by the modified equation associated with these discrete schemes, which is found to be relevant even for solutions containing shock waves.

\subsubsection{Numerical dissipation analysis of HLL-family schemes}
\label{sec3.3.1}
In this section, the dissipative mechanisms of Godunov-type schemes particularly the HLL-family schemes are carefully analyzed. First, we consider the approximate solution of the Riemann problem for the HLLEM flux defined in Eq. (\ref{eq2.2.1}). It can be written as
\begin{equation}\label{eq3.3.1.1}
  {\bf{\omega}}(x/t;{\bf U}_L,{\bf U}_R)=
  \begin{cases}
  {\bf U}_L               & \text{$x/t < S_L$}  \\
  {\bf U}^*+(x-{\widehat{u}} t) {\hat{{\delta}}}^*_2 \widehat{\alpha}_{2} \widehat{{\bf R}}_{2}     & \text{$S_L <x/t < S_R$} \\
  {\bf U}_R               & \text{$S_R < x/t$}
  \end{cases}
\end{equation}
where the dissipative term corresponding to shear waves has been omitted in the one-dimensional case. The resulting flux function in the corresponding conservation form (\ref{eq2.1.4}) can be obtained,
\begin{equation}\label{eq3.3.1.2}
{\bf{F}}_{\rm{HLLEM}}=
\begin{cases}
{\bf F}\left( {{{\bf{U}}_L}} \right)       & \text{$S_L>0$}               \\
\frac{{{S_R}{\bf{F}}\left( {{{\bf{U}}_L}} \right) - {S_L}{\bf{F}}\left( {{{\bf{U}}_R}} \right)}}{{{S_R} - {S_L}}} + \frac{{{S_L}{S_R}}}{{{S_R} - {S_L}}}\left( {{{\bf{U}}_R} - {{\bf{U}}_L} - {\hat{{\delta}}}_2\widehat \alpha _2\widehat {\bf{R}}_2 } \right)       & \text{$S_L \le 0 \le S_R$}    \\
{\bf F}\left( {{{\bf{U}}_R}} \right)       & \text{$S_R<0$}
\end{cases}
\end{equation}
where the following simplified wave speed estimates are used for facilitating further analysis,
\begin{equation}\label{eq3.3.1.3}
  S_L=\widehat{u}-\widehat{a},\qquad S_R=\widehat{u}+\widehat{a}.
\end{equation}
It is justified for the one-dimensional steady shock problem considered here. One should note that the HLLEM scheme ({\ref{eq3.3.1.2}}) with the wave speed estimates (\ref{eq3.3.1.3}) is identical to the Roe scheme \cite{roe1997approximate}, so results of the current analysis can also be applied to the Roe scheme. The above flux function for the subsonic case may also be written in the form
\begin{equation}\label{eq3.3.1.4}
  {\bf{F}}_{\rm{HLLEM}} = \frac{1}{2}\left( {{{\bf{F}}_L} + {{\bf{F}}_R}} \right) - \frac{1}{2}{ {\bf{Q}}_{i + 1/2}}\left( {{{\bf{U}}_R} - {{\bf{U}}_L}} \right)
\end{equation}
According to Einfeldt\cite{Einfeldt1988}, the numerical viscosity matrix ${\bf{Q}}_{i + 1/2}$ can be rewritten by
\begin{equation}\label{eq3.3.1.5}
  {\bf{Q}}_{i + 1/2}=\frac{S_L+S_R}{S_R-S_L}\widehat{{\bf{A}}}-\frac{2{S_L}{S_R}}{S_R-S_L}{\bf{I}}+\frac{2{S_L}{S_R}}{S_R-S_L}{\hat{\delta}}_2\widehat{\bf{R}}{\bf{T}}\widehat{\bf{R}}^{-1}
\end{equation}
with
\begin{equation}\label{eq3.3.1.6}
{\bf{T}}={\left({\begin{array}{*{20}{c}}
  0&0&0\\
  0&1&0\\
  0&0&0
\end{array}}\right)}.
\end{equation}
Here, $\widehat{\bf{A}}$ is Roe Jacobian matrix and $\widehat{\bf{R}}=(\widehat{\bf{R}}_1,\widehat{\bf{R}}_2,\widehat{\bf{R}}_3)$. The last term on the right of Eq. $(\ref{eq3.3.1.5})$ is negative and plays a role in reducing numerical diffusion on entropy waves. One possible way to increase the entropy production is to increase the coefficient $\hat{{\delta}}_2$, thus less diffusion will be imposed on entropy waves. To demonstrate this, the modified equation associated with the HLLEM scheme is considered.

In reference \cite{wang2014numerical}, the modified equation for the HLL scheme is obtained through a simple approach. Here, we follow closely the work of Wang et al. \cite{wang2014numerical} and extend the strategies therein to the HLLEM method. Furthermore, the characteristic form of the modified equation is also derived. It serves as an essential role in analyzing the artificial viscosity on the individual characteristic waves, particularly the entropy waves. For the steady shock problem in the current study, we focus only on the numerical dissipation introduced by the flux function. Therefore, only the semi-discrete equations are studied. The HLLEM scheme applied to the one-dimensional Euler system can be written in semi-discrete conservative form as
\begin{equation}\label{eq3.3.1.7}
  \frac{{\rm{d}}}{{\rm{d}}t}{{\bf{U}}_i} = - \frac{1}{{\Delta x}} \left[ {{{\bf{F}}_{i + 1/2}} - {{\bf{F}}_{i - 1/2}}} \right]
\end{equation}
where ${{\bf{F}}_{i + 1/2}}$ is the numerical flux which is the approximation of the flux through the interface $x_{i+1/2}$. $\Delta x$ is the spatial cell size and $x_i=i\Delta x$, $i=1,...,N$. For the Godunov-type schemes in (\ref{eq3.3.1.4}), (\ref{eq3.3.1.5}) and (\ref{eq3.3.1.6}), the system (\ref{eq3.3.1.7}) can be recast into the equivalent viscosity form
\begin{equation}\label{eq3.3.1.8}
  \frac{{\rm{d}}}{{{\rm{d}}t}}{{\bf{U}}_i} =  - \frac{1}{{2\Delta x}}\left[ {{{\bf{F}}_{i + 1}} - {{\bf{F}}_{i - 1}}} \right] + \frac{1}{{2\Delta x}}\left[ {{ {\bf{Q}}_{i + 1/2}}\left( {{{\bf{U}}_{i + 1}} - {{\bf{U}}_i}} \right) - {{\bf{Q}}_{i - 1/2}}\left( {{{\bf{U}}_i} - {{\bf{U}}_{i - 1}}} \right)} \right]
\end{equation}
where ${\bf{Q}}_{i+1/2}$ is the numerical viscosity coefficient matrix. The modified equation of (\ref{eq3.3.1.8}) has the form
\begin{equation}\label{eq3.3.1.9}
  \widetilde{\bf{U}}_t + {\bf{F}\left({\widetilde{\bf{U}}}\right)}_x = \frac{1}{2} { \bf{Q}\left({\widetilde{\bf{U}}}\right) }  {\widetilde{\bf{U}}_{xx}} \Delta x
\end{equation}
where $\widetilde{\bf{U}}$ is assumed to be a polynomial interpolated through the mesh values of $\bf{U}$ in the neighborhood of the mesh point $x_i$. The second order and even higher order error terms are suppressed in Eq. (\ref{eq3.3.1.9}) in the context of modified equations.

To derive the modified equation associated with the HLLEM scheme, the numerical viscosity coefficient matrix needs to be determined, it can be written by sending ${\bf{U}}_j^n$ and ${\bf{U}}_{j+1}^n$ to ${\bf{U}}$,
\begin{equation}\label{eq3.3.1.10}
{\bf{Q}}_{\rm{HLLEM}} \left( {\bf{U}} \right) = \frac{{{S_R} + {S_L}}}{{{S_R} - {S_L}}} {\bf{A}} \left( {\bf{U}} \right) - \frac{{2{S_L}{S_R}}}{{{S_R} - {S_L}}}{\bf{I}} + \frac{{2{S_R}{S_L}}}{{{S_R} - {S_L}}} {\delta _2} {\bf{R}}\left( {\bf{U}} \right){\bf{T}}{\bf{R}}^{-1}\left( {\bf{U}} \right)
\end{equation}
where $\bf{A}\left(\bf{U}\right)$ is the Jacobian of the flux function $\bf{F}\left(\bf{U}\right)$ and ${\bf{A}}\left( {\bf{U}} \right) = \frac{{\partial {\bf{F}}\left( {\bf{U}} \right)}}{{\partial {\bf{U}}}}$, $\bf{I}$ is the identity matrix, $S_L:=S_L\left({\bf{U}},{\bf{U}}\right)$ and $S_R:=S_R\left({\bf{U}},{\bf{U}}\right)$.

We want to analyze the artificial viscosity on the individual characteristic waves, especially the entropy wave. To this end, we derive the characteristic form of the modified equation. The right eigenvectors of the Jacobian ${\bf{A}}\left( {\bf{U}} \right)$ are $\bf{R}_1$, $\bf{R}_2$, $\bf{R}_3$  and their corresponding eigenvalues are
\begin{equation}\label{eq3.3.1.11}
  \lambda_1=u-a, \qquad \lambda_2=u, \qquad \lambda_3=u+a.
\end{equation}
The transformation matrices $\bf{R}$ and its inverse $\bf{R}^{-1}$ are applied to the modified Eq. (\ref{eq3.3.1.9}), by sending ${\widetilde {\bf{U}}}$ to ${\bf{U}}$,
\begin{equation}\label{eq3.3.1.12}
{{\bf{R}}^{ - 1}}{ {\bf{U}}_t} + \left( {{{\bf{R}}^{ - 1}}{\bf{AR}}} \right){{\bf{R}}^{ - 1}}{ {\bf{U}}_x} = \frac{1}{2}\left( {{{\bf{R}}^{ - 1}}{\bf{Q}}\left( { {\bf{U}}} \right){\bf{R}}} \right){{\bf{R}}^{ - 1}}{ {\bf{U}}_{xx}}\Delta x .
\end{equation}
The Eq. (\ref{eq3.3.1.12}) can be further simplified by using the relations ${\bf{A}} = {\bf{R\Lambda }}{{\bf{R}}^{ - 1}}$ and ${\rm{d}} {\bf{W}} = {{\bf{R}}^{ - 1}} {\rm{d}} {\bf{U}}$, the modified equation in characteristic variables can be obtained as
\begin{equation}\label{eq3.3.1.13}
  {{\bf{W}}_t} + \Lambda {{\bf{W}}_x} = \frac{1}{2}{\bf{V}}{{\bf{W}}_{xx}}\Delta x
\end{equation}
where ${\bf{\Lambda }} = {\rm{diag}}\left( {u - a,u,u + a} \right)$ is the matrix of eigenvalues, the characteristic variables are defined by
\begin{equation}\label{eq3.3.1.14}
{\rm{d}}{\bf{W}} = \left( {\begin{array}{*{20}{c}}
{{\rm{d}}{w_1}}\\
{{\rm{d}}{w_2}}\\
{{\rm{d}}{w_3}}
\end{array}} \right) = \left( {\begin{array}{*{20}{c}}
{\frac{1}{{2{a^2}}}{\rm{d}}p - \frac{\rho }{{2a}}{\rm{d}}u}\\
{{\rm{d}}\rho  - \frac{{{\rm{d}}p}}{{{a^2}}}}\\
{\frac{1}{{2{a^2}}}{\rm{d}}p + \frac{\rho }{{2a}}{\rm{d}}u}
\end{array}} \right),
\end{equation}
$\bf{V}$ is the viscosity matrix of the characteristic form that is expressed as
\begin{equation}\label{eq3.3.1.15}
  {{\bf{V}}} = \frac{{{S_L} + {S_R}}}{{{S_R} - {S_L}}}{\bf{\Lambda }}\left( {\bf{U}} \right) - \frac{{2{S_L}{S_R}}}{{{S_R} - {S_L}}}{\bf{I}} + \frac{{2{S_L}{S_R}}}{{{S_R} - {S_L}}}{\delta _2}{\bf{T}}.
\end{equation}
The Eq. ({\ref{eq3.3.1.13}}) can be written in a form where each individual characteristic wave is explicit presented
\begin{equation}\label{eq3.3.1.16}
  \frac{\partial w_i }{\partial t} + {\lambda}_i \frac{\partial w_i }{\partial x} = \frac{1}{2} \mu_i \frac{\partial^2}{\partial x^2} w_i \Delta x,
\end{equation}
where $\mu_i$ can be interpreted as numerical viscosity coefficients for the individual characteristic waves. To assess the numerical entropy change, we should only consider the coefficient of artificial viscosity corresponding to the entropy wave ${\rm{d}}w_2$,
\begin{equation}\label{eq3.3.1.17}
  \mu_2 = \frac{{{S_L} + {S_R}}}{{{S_R} - {S_L}}}{{\lambda}}_2 - \frac{{2{S_L}{S_R}}}{{{S_R} - {S_L}}} + \frac{{2{S_L}{S_R}}}{{{S_R} - {S_L}}}{\delta _2}.
\end{equation}
It can be observed that the last term on the right of Eq. (\ref{eq3.3.1.17}) is nonpositive and plays a role in reducing numerical diffusion on the entropy wave. For the HLL flux, the coefficient $\delta_2$ equals zero. That is the reason why it contains excessive dissipation on the entropy wave. One possible way to increase the entropy production is to increase the coefficient ${\delta}_2$, thus less diffusion will be applied on the entropy wave. However, this will not work, because ${\hat{{\delta}}}_2$ used in Eq. ({\ref{eq3.3.1.5}}) should be chosen such that a TVD-type condition for the viscosity matrix ${\bf{Q}}_{i+1/2}$ is valid \cite{Einfeldt1988},
\begin{equation}\label{eq3.3.1.18}
  {\lambda}^{*}_k \geq \left| {\hat{\lambda} _k} \right|, \quad \text{for} \quad k=2,
\end{equation}
where ${\lambda}^{*}_k$ represents the eigenvalues of ${\bf{Q}}_{i + 1/2}$ and ${\hat{\lambda} _k}$ denotes the eigenvalues of $\widehat{\bf{A}}$. It follows from ($\ref{eq3.3.1.18}$) that the anti-diffusion coefficient ${\hat{\delta}}_2$ should satisfy the following condition
\begin{equation}\label{eq3.3.1.19}
  {\hat{{\delta}}}_2 \leq \frac{{\widehat a}}{{\widehat a + \left| {\widehat {u}} \right|}}.
\end{equation}
Considering the definition of ${\widehat{{\delta}}}_2$ presented in Eq. (\ref{eq2.2.2_6}), it is not desirable to increase the magnitude of the coefficient ${\hat{{\delta}}}_2$. Hence, we resort to modifying the dissipative terms in the wave strength ${\hat{\alpha}}_2$ to control the numerical dissipation on the entropy wave.

\subsubsection{Two alternative modified schemes}
\label{sec3.3.2}
In this section, two alternative modified schemes which contain less numerical dissipation on entropy waves are constructed. To avoid smearing the shock profile, the solutions ${\bf U}_L$ and ${\bf U}_R$ of the Riemann problem (\ref{eq3.3.1.1}) for the supersonic case should not be altered. Here, two alternative approaches are considered. First, the average state ${\bf{U}}_{i+1/2}$ in (\ref{eq3.3.1.1}) can be modified as
\begin{equation}\label{eq3.3.2.1}
    \omega_{\rm{HLLEM}-\rho}(x/t;{\bf U}_L,{\bf U}_R)=
    \begin{cases}
    {\bf U}_L               & \text{$x/t < S_L$} \\
    {\bf U}^* + (x-{\hat{u}} t){\hat{\delta}}^*_2 \widehat{\alpha}^{\ast}_{2} \widehat{{\bf R}}_{2}   & \text{$S_L <x/t < S_R$} \\
    {\bf U}_R               & \text{$S_R < x/t$}
    \end{cases}
\end{equation}
with the following modified wave strength as
\begin{equation}\label{eq3.3.2.2}
  \widehat{\alpha}^{\ast}_2=f_{\rho}\cdot\Delta\rho-\frac{\Delta p}{\widehat{a}^2}.
\end{equation}
To indicate this modification, the modified scheme is denoted as the HLLEM-${\rho}$ scheme. The modification of the average state ${\bf{U}}_{i+1/2}$ does not change the integral ({\ref{eq2.1.2}}), therefore the Riemann solver (\ref{eq3.3.2.1}) can still be written in the following conservation form
\begin{equation}\label{eq3.3.2.3}
  {\bf{F}}_{i+1/2}^{{\rm{HLLEM}-{\rho}}} = \frac{{{S_R}{\bf{F}}\left( {{{\bf{U}}_L}} \right) - {S_L}{\bf{F}}\left( {{{\bf{U}}_R}} \right)}}{{{S_R} - {S_L}}} + \frac{{{S_L}{S_R}}}{{{S_R} - {S_L}}}\left( {{{\bf{U}}_R} - {{\bf{U}}_L} - \hat{\delta}_2\widehat \alpha^{\ast} _2\widehat {\bf{R}}_2 } \right)
\end{equation}
with the modified wave strength $\widehat{\alpha}^{\ast}_2$ defined in $(\ref{eq3.3.2.2})$.

Alternatively, the pressure difference term in $\widehat{\alpha}_2$ can also be modified. The new scheme denoted as the HLLEM-$p$ can be expressed by
\begin{equation}\label{eq3.3.2.4}
    {\bf{F}}_{i+1/2}^{{\rm{HLLEM}}-p} = \frac{{{S_R}{\bf{F}}\left( {{{\bf{U}}_L}} \right) - {S_L}{\bf{F}}\left( {{{\bf{U}}_R}} \right)}}{{{S_R} - {S_L}}} + \frac{{{S_L}{S_R}}}{{{S_R} - {S_L}}}\left( {{{\bf{U}}_R} - {{\bf{U}}_L} - \delta _2\widehat \alpha^{\ast} _2\widehat {\bf{R}}_2 } \right)
\end{equation}
with a modified wave strength defined as
\begin{equation}\label{eq3.3.2.5}
  \widehat{\alpha}^{\ast}_2=\Delta\rho-f_p\cdot\frac{\Delta p}{\widehat{a}^2}.
\end{equation}
The coefficients $f_{\rho}$ and $f_p$ are used to determine numerical diffusion on entropy waves. In order to demonstrate how they will work, the dissipation mechanisms of these Godunov-type schemes are analyzed. To this end, we apply the strategies presented in section {\ref{sec3.3.1}} to derive modified equations of these two new schemes. Here, the HLLEM-$p$ scheme is used to demonstrate the main procedure, the derivation of the modified equation of the HLLEM-$\rho$ scheme is described in Appendix A.

As demonstrated in the above section, the modified equation of the HLLEM-$p$ scheme can be derived if its numerical viscosity coefficient
matrix is determined. The flux function defined in Eq. (\ref{eq3.3.2.4}) and (\ref{eq3.3.2.5}) can be rewritten by
\begin{equation}\label{eq3.3.2.6}
  {\bf{F}}_{{\rm{HLLEM}}-p} = \frac{1}{2}\left( {{{\bf{F}}_L} + {{\bf{F}}_R}} \right) - \frac{1}{2}{ {\bf{Q}}_{i + 1/2}}\left( {{{\bf{U}}_R} - {{\bf{U}}_L}} \right)
\end{equation}
where the numerical viscosity matrix ${\bf{Q}}_{i + 1/2}$ is defined by
\begin{equation}\label{eq3.3.2.7}
  {\bf{Q}}_{i + 1/2}=\frac{S_L+S_R}{S_R-S_L}\widehat{{\bf{A}}}-\frac{2{S_L}{S_R}}{S_R-S_L}{\bf{I}}+\frac{2{S_L}{S_R}}{S_R-S_L}\hat{{\delta}}_2\widehat{\bf{R}}{\bf{T}}\widehat{\bf{R}}^{-1}-\frac{2{S_L}{S_R}}{S_R-S_L}\hat{{\delta}}_2\widehat{\bf{R}}{\bf{T}}\widehat{\bf{R}}^{*,-1}
\end{equation}
with the following relation
\begin{equation}\label{eq3.3.2.8}
  \widehat{\alpha} _2^*{{\widehat{\bf{R}}}_2} = \widehat{\bf{R}}{\bf{T}}\widehat{\bf{R}}^{*,-1} \cdot \left( {{{\bf{U}}_{R}} - {{\bf{U}}_{L}}} \right).
\end{equation}
The modified equation corresponding to the HLLEM-$p$ scheme can be written as
\begin{equation}\label{eq3.3.2.9}
  {\bf{U}}_t + {\bf{F}\left({{\bf{U}}}\right)}_x = \frac{1}{2} {{\bf{Q}}_{{\rm{HLLEM}}-p}\left({{\bf{U}}}\right) } {{\bf{U}}_{xx}} \Delta x
\end{equation}
where the viscosity matrix can be obtained,
\begin{equation}\label{eq3.3.2.10}
  {\bf{Q}}_{{\rm{HLLEM}}-p} \left( {\bf{U}} \right) = \frac{{{S_R} + {S_L}}}{{{S_R} - {S_L}}} {\bf{A}} \left( {\bf{U}} \right) - \frac{{2{S_L}{S_R}}}{{{S_R} - {S_L}}}{\bf{I}} + \frac{{2{S_R}{S_L}}}{{{S_R} - {S_L}}} {\delta _2} {\bf{R}}\left( {\bf{U}} \right){\bf{T}}{\bf{R}}^{-1}\left( {\bf{U}} \right)-\frac{2{S_L}{S_R}}{S_R-S_L}{\delta}_2{\bf{R}} \left( {\bf{U}} \right) {\bf{T}}{\bf{R}}^{*,-1}\left( {\bf{U}} \right) .
\end{equation}
The modified equation in characteristic variables can be obtained following the procedure presented in Eq. (\ref{eq3.3.1.12}), it can be written as
\begin{equation}\label{eq3.3.2.11}
  {{\bf{W}}_t} + \Lambda {{\bf{W}}_x} = \frac{1}{2}{\bf{V}}{{\bf{W}}_{xx}}\Delta x + (f_p-1) \frac{{{S_{\rm{L}}}{S_R}}}{{{S_{\rm{R}}} - {S_{\rm{L}}}}}{\delta _2}{\bf{T}} {\bf{W}}_{xx}^2 \Delta x
\end{equation}
where the following relation has been used,
\begin{equation}\label{eq3.3.2.11a}
  \rm{d}\bf{W}=\rm{d}{\bf{W}}^1-\rm{d}{\bf{W}}^2
\end{equation}
with
\begin{equation}\label{eq3.3.2.11b}
{\rm{d}}{{\bf{W}}^1} = \left( {\begin{array}{*{20}{c}}
  {{\rm{d}}{w_1}}\\
  {{\rm{d}}w_2^1}\\
  {{\rm{d}}{w_3}}
  \end{array}} \right) = \left( {\begin{array}{*{20}{c}}
  {\frac{1}{{2{a^2}}}{\rm{d}}p - \frac{\rho }{{2a}}{\rm{d}}u}\\
  {{\rm{d}}\rho }\\
  {\frac{1}{{2{a^2}}}{\rm{d}}p + \frac{\rho }{{2a}}{\rm{d}}u}
  \end{array}} \right), \quad
{\rm{d}}{{\bf{W}}^2} = \left( {\begin{array}{*{20}{c}}
  0\\
  {{\rm{d}}w_2^2}\\
  0
  \end{array}} \right) = \left( {\begin{array}{*{20}{c}}
  0\\
  {\frac{{{\rm{d}}p}}{{{a^2}}}}\\
  0
\end{array}} \right) .
\end{equation}
To access the numerical viscosity corresponding to the entropy wave, we need to consider the following modified equation for the entropy wave ${\rm{d}}w_2$ that is extracted from Eq. (\ref{eq3.3.2.11}),
\begin{equation}\label{eq3.3.2.12}
  \frac{\partial w_2 }{\partial t} + {\lambda}_2 \frac{\partial w_2 }{\partial x} = \frac{1}{2} \mu_2 \frac{\partial^2}{\partial x^2} w_2 \Delta x+(f_p-1) \frac{{{S_L}{S_R}}}{{{S_R} - {S_L}}}{\delta _2} \frac{\partial^2}{\partial x^2} w_2^2 \Delta x,
\end{equation}
where the numerical viscosity coefficient  $\mu_2$ is defined in Eq. (\ref{eq3.3.1.17}).
Considering (\ref{eq3.3.2.11a}) and (\ref{eq3.3.2.11b}), Eq. (\ref{eq3.3.2.12}) can be splitted into the following subsystems,
\begin{align}
  \frac{{\partial w_2^1}}{{\partial t}} + {\lambda _2}\frac{{\partial w_2^1}}{{\partial x}} & = \frac{1}{2}{\mu _2}\frac{{{\partial ^2}}}{{\partial {x^2}}}w_2^1\Delta x, \label{eq3.3.2.15} \\
  \frac{{\partial w_2^2}}{{\partial t}} + {\lambda _2}\frac{{\partial w_2^2}}{{\partial x}} &= \frac{1}{2}{\mu _2}\frac{{{\partial ^2}}}{{\partial {x^2}}}w_2^2\Delta x + \left( {{f_p} - 1} \right)\frac{{{S_L}{S_R}}}{{{S_R} - {S_L}}}{\delta _2}\frac{{{\partial ^2}}}{{\partial {x^2}}}w_2^2\Delta x. \label{eq3.3.2.16}
\end{align}
Due to the nonpositivity of the last term at the right side of Eq. (\ref{eq3.3.2.16}), the dissipative term plays a role in smearing the wave ${\rm{d}}w_2^2$, thus a reduced $f_p$ will lead to a smaller ${\rm{d}}w_2^2$. Combining with the relations (\ref{eq3.3.2.11a}) and (\ref{eq3.3.2.11b}), an increased entropy will be obtained. Similarly, an increased $f_{\rho}$ can also lead to an increased entropy, which is indicated by the dissipation analysis of HLLEM-$\rho$ flux in Appendix A. However, one should be noted that the stability condition (\ref{eq3.3.1.18}) for the contact discontinuity is no longer valid when $f_{\rho}$ is greater than one. Thus, the HLLEM-${\rho}$ flux can not be used for the multidimensional case where contact discontinuities may exist. Whereas, it works well for the mentioned steady shock problem in one dimension.

\subsection{Experimental study on entropy production and shock instability}
\label{sec3.4}
In the previous section, we have demonstrated that the numerical dissipation of HLLEM scheme on entropy waves can be controlled by modifying the wave strength in the dissipative term. Furthermore, it is indicated that the dissipation on entropy waves can be reduced by increasing the magnitude of density difference $\Delta {\rho}$ or reducing the magnitude of the pressure difference $\Delta p$ in the wave strength ${\widehat{\alpha}}_2$. If these modified schemes are applied at shocks, then a larger entropy production can be obtained inside the numerical shock structure. The coefficients $f_{\rho}$ and $f_p$ play a decisive role in controlling the entropy production. Since it is hard to analytically determine the amount of $f_{\rho}$ and $f_p$, we resort to conducting numerical experiments to determine their proper values.

\begin{figure}[htbp]
	\centering
	\subfigure[]{
		\begin{minipage}[b]{0.48\textwidth}
			\includegraphics[width=1.0\textwidth]{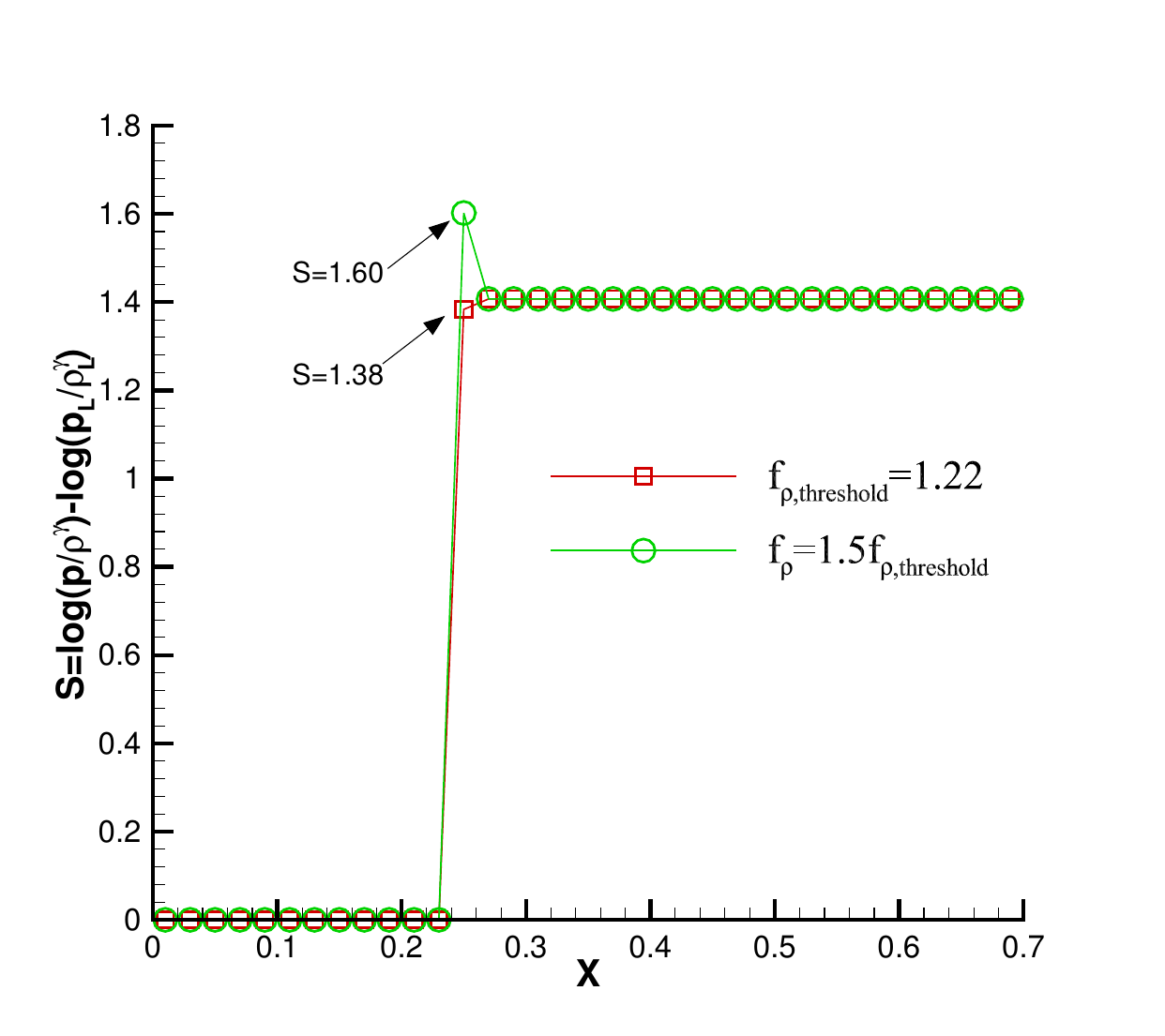}
		\end{minipage}
	}
	\subfigure[]{
		\begin{minipage}[b]{0.48\textwidth}
			\includegraphics[width=1.0\textwidth]{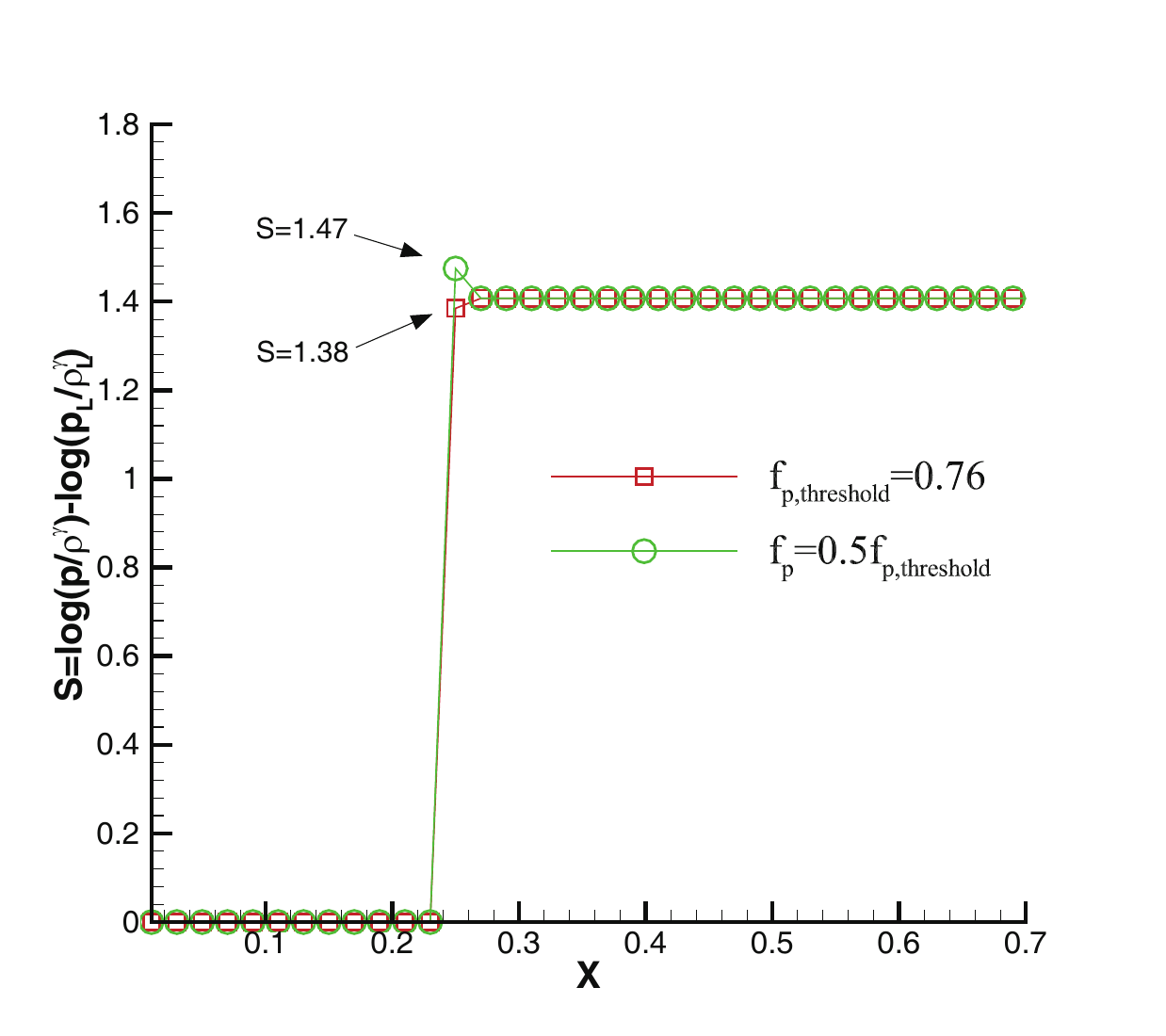}
		\end{minipage}
	}
	\caption{Entropy profiles of 1D steady shock problem predicted by the HLLEM scheme at ${M_0} = 6.0$ and $\varepsilon = 0.3$ with (a) HLLEM-${\rho}$ flux applied at interfaces of cell M (b) HLLEM-${p}$ flux applied at interfaces of cell M.}
	\label{fig3-4-1}
\end{figure}

\begin{figure}[htbp]
	\centering
	\subfigure[]{
		\begin{minipage}[b]{0.48\textwidth}
			\includegraphics[width=1.0\textwidth]{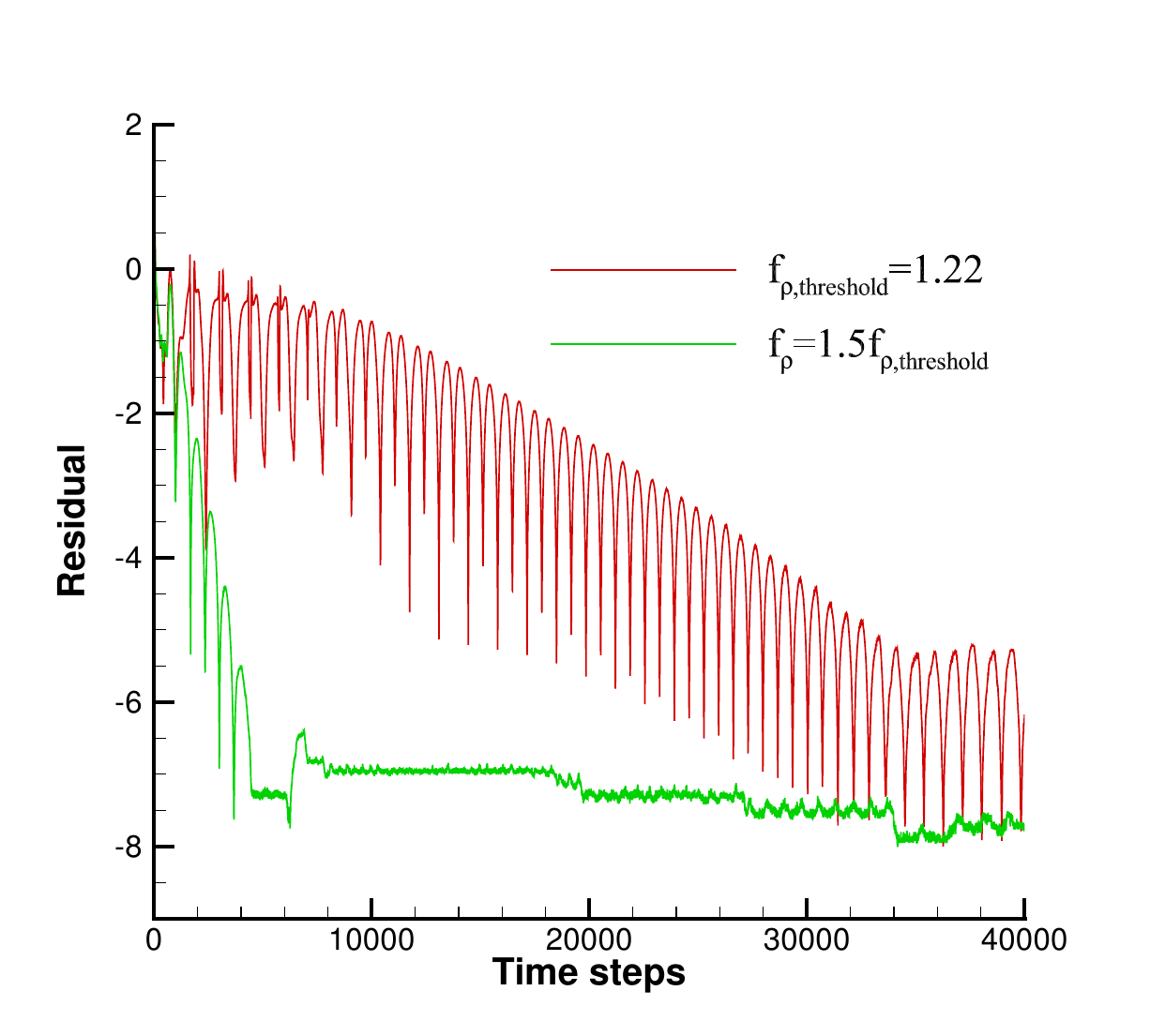}
		\end{minipage}
	}
	\subfigure[]{
		\begin{minipage}[b]{0.48\textwidth}
			\includegraphics[width=1.0\textwidth]{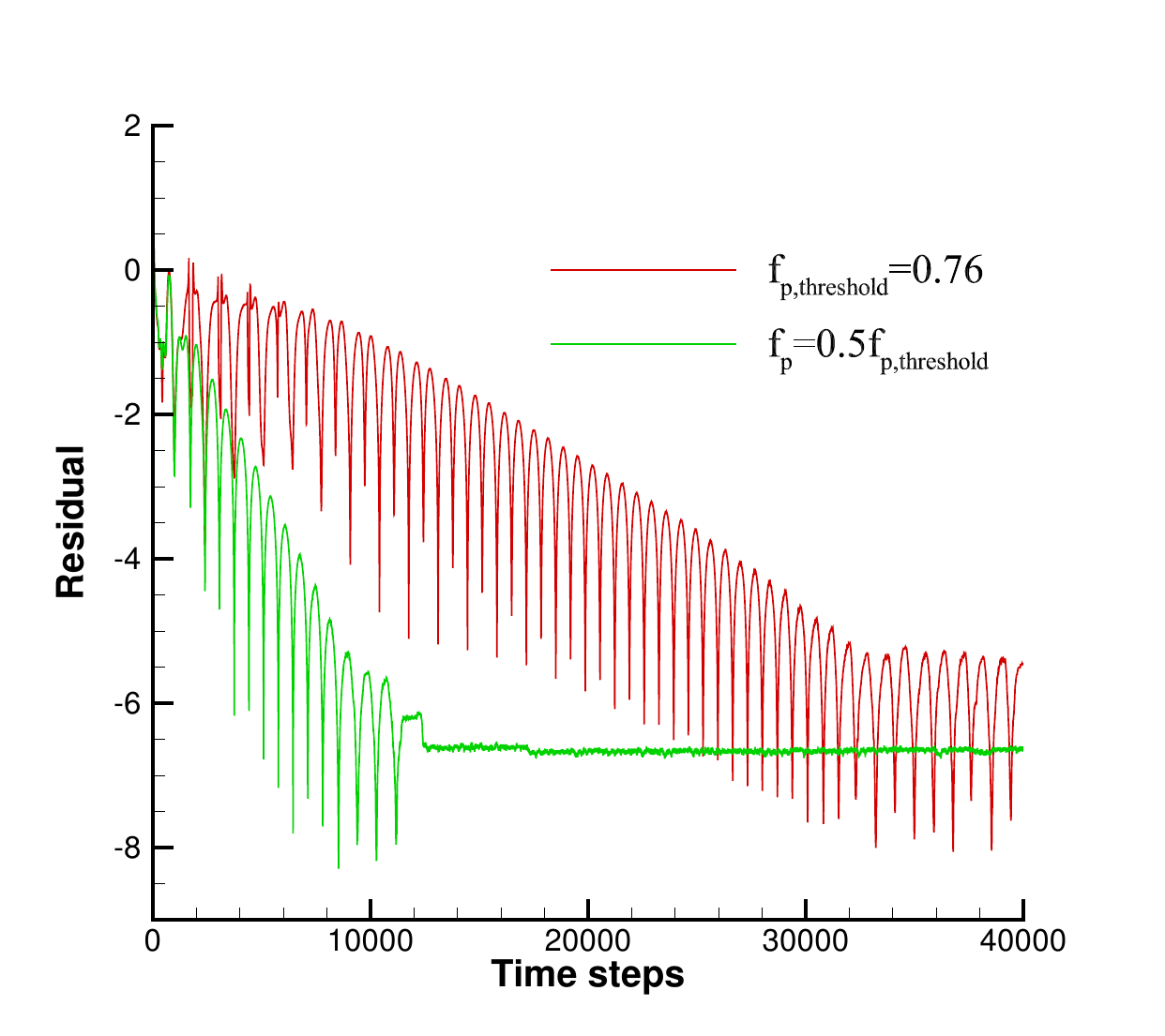}
		\end{minipage}
	}
	\caption{Residual histories for 1D steady shock problem computed by the HLLEM scheme at ${M_0} = 6.0$ and $\varepsilon = 0.3$ with (a) HLLEM-${\rho}$ flux applied at interfaces of cell M, (b) HLLEM-${p}$ flux applied at interfaces of cell M.}
	\label{fig3-4-2}
\end{figure}

\begin{figure}[htbp]
	\centering
	\includegraphics[width=3.8in]{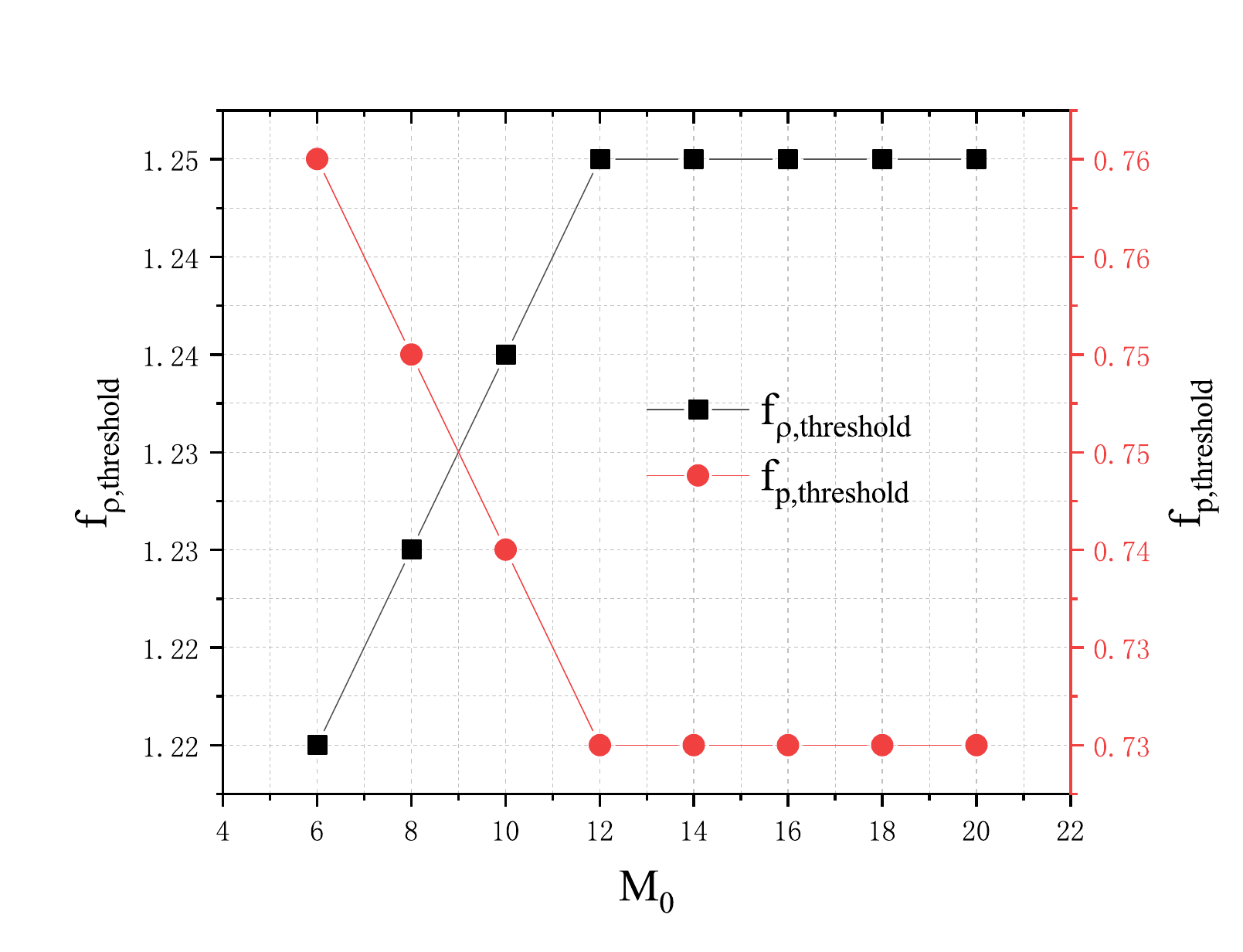}
	\caption{Stable thresholds of coefficients $f_{\rho}$ and $f_p$ vs. Mach number for 1D steady shock problem.}
	\label{fig3-4-3}
\end{figure}

\begin{figure}[htbp]
	\centering
	\includegraphics[width=3.8in]{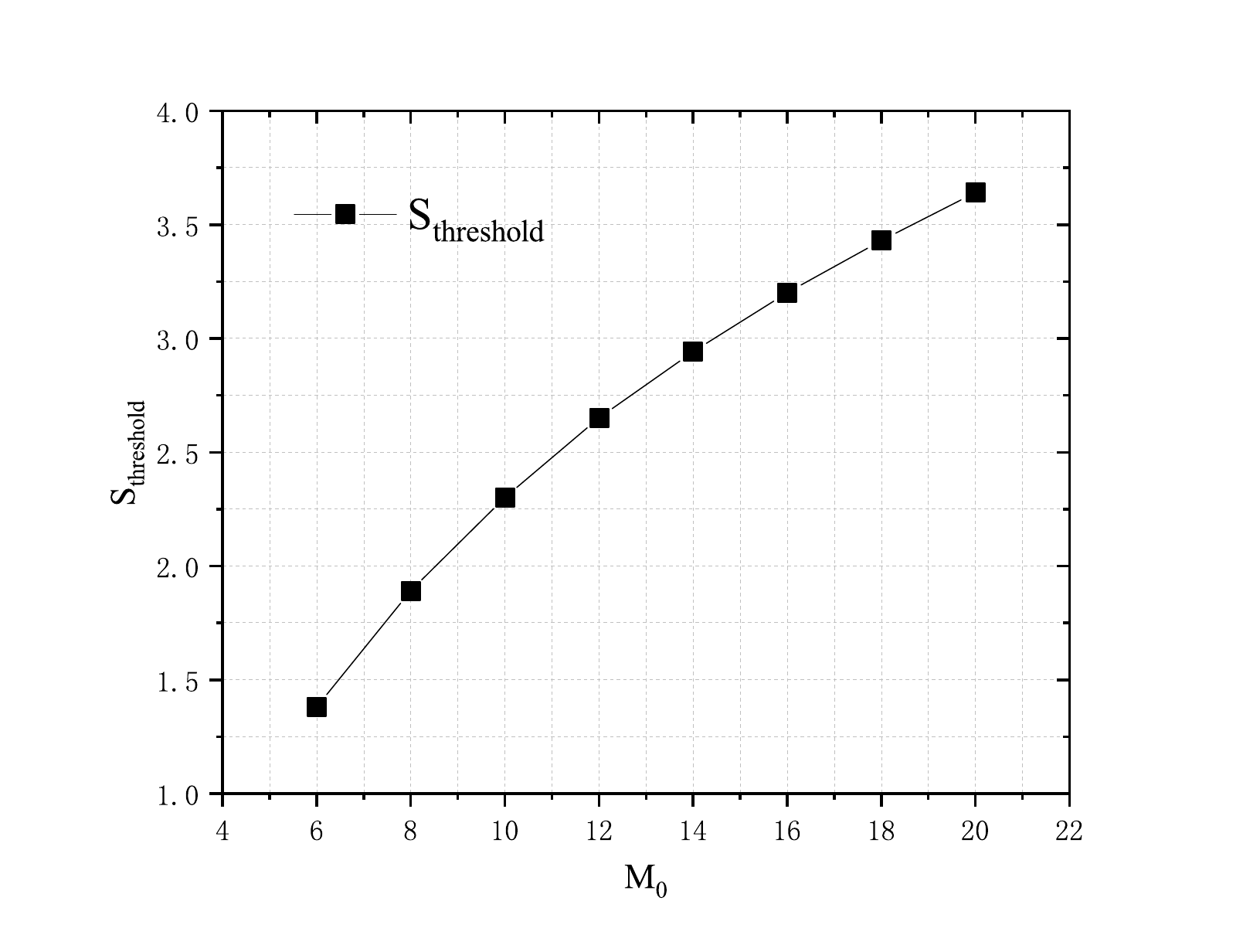}
	\caption{Stable thresholds of entropy $S=\log \frac{p}{{\rho}^{\gamma}}-\log \frac{p_L}{{\rho}_L^{\gamma}}$ vs. Mach number for 1D steady shock problem.}
	\label{fig3-4-4}
\end{figure}

Considering the one-dimensional steady shock problem in section \ref{sec3.1}, the computations are conducted with the first-order HLLEM scheme. A three-stage first-order Runge-Kutta method \cite{van1992design} is used with CFL=0.5 until the time step reaches 40,000. The Mach number is set as ${M_0} = 6.0$ and the shock position is $\varepsilon = 0.3$. The modified schemes HLLEM-${\rho}$ and HLLEM-$p$ are applied at the interfaces of cell M to control the entropy production in it. As demonstrated in section {\ref{sec3.1.2}}, the HLLEM scheme will produce the one-dimensional carbuncle under this condition. Thus, to obtain a stable and converged solution, we gradually increase the magnitude of the coefficient $f_{\rho}$ or reduce the magnitude the coefficient $f_p$ to increase the entropy production inside the intermediate cell M. In Fig. \ref{fig3-4-1}, entropy profiles of stable solutions computed with different coefficients $f_{\rho}$ and $f_p$ are illustrated. The corresponding residual histories are demonstrated in Fig. \ref{fig3-4-2}. As shown in Fig. \ref{fig3-4-1}, a larger entropy $S=\log \frac{p}{{\rho}^{\gamma}}-\log \frac{p_L}{{\rho}_L^{\gamma}}$ will be produced inside the numerical shock structure (i.e., the cell M) in the case that a larger $f_{\rho}$ or a smaller $f_{p}$ is used. Such results are in accordance with the dissipation analysis presented in section \ref{sec3.3.2} and they show that if enough entropy production is guaranteed inside the shock, then the instability can be successfully eliminated. Residual histories presented in Fig. \ref{fig3-4-2} indicate that more entropy production will make the solution converge to a stable state more quickly. It should be noted that similar results are also observed for various combinations of Mach numbers $M_0$ and the shock positions $\varepsilon$. In Fig. \ref{fig3-4-3}, we plot the stable thresholds of $f_{{\rho},threshold}$ and $f_{p,threshold}$ for different Mach numbers $M_0$ and their corresponding entropy thresholds are illustrated in Fig. \ref{fig3-4-4}.

\subsection{Extension to the multidimensional case}
\label{sec3.5}
So far, we have discussed the shock instability for the one-dimensional steady shock problem. In this part, the entropy-control technique developed in the above sections is extended to the two-dimensional case. As shown in Fig. {\ref{fig3-1-1}}, the numerical setup for the two-dimensional steady shock problem is the same as its one-dimensional counterpart except that 25 cells are used in the $j$ direction. This simple numerical test is commonly used to analyze numerical shock instability of Godunov-type schemes. It is believed that numerical methods that produce shock instabilities for this problem also produce carbuncle solutions for the blunt body problem \cite{Dumbser2004,ismail2006toward,Kitamura2009}.

\begin{figure}[htbp]
	\centering
	\includegraphics[width=3.2in]{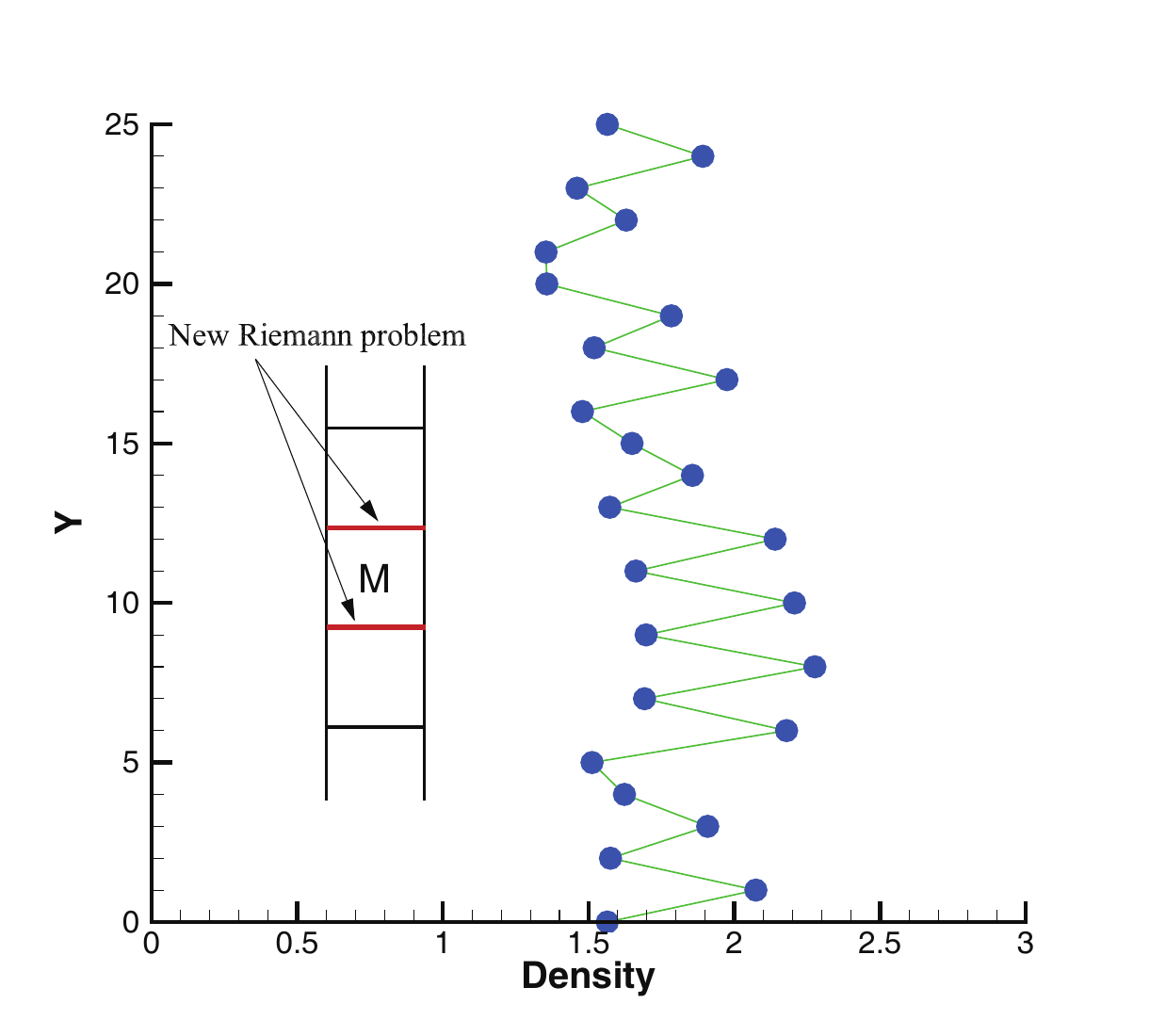}
	\caption{Density distribution along the $j$ direction in the intermediate cells of the shock at the initial stage of instability (predicted by HLLEM scheme, with $M_0=6.0$ and $\varepsilon=0.3$ at $200$ time steps).}
	\label{fig3-5-1}
\end{figure}

As shown in Fig. {\ref{fig3-5-1}}, the conservative quantities inside the numerical shock structure may suffer errors when fluids pass through shock with perturbations. Due to discrepancies of physical quantities in the transverse of shock, new Riemann problems may exist at interfaces (see the red solid lines in Fig. {\ref{fig3-5-1}}). In the context of Godunov-type methods, one-dimensional planar waves will be generated at these interfaces. One should be noted that, physically, the conservative quantities in the transverse direction should be uniform and there should be no mass flux appearing in the $j$ direction due to the nature of quasi-one-dimensional flows. Thus, these waves may be not actually present. During the computation of this two-dimensional problem, dissipation would be added to stabilize these waves which may be much more than is actually required. Inside the numerical shock structure, it leads to insufficient entropy production that triggers the instability. To further validate this heuristic consideration, we also resort to numerical experiments.

\begin{figure}[htbp]
	\centering
	\subfigure[]{
		\begin{minipage}[b]{0.46\textwidth}
			\centering
			\includegraphics[width=1.0\textwidth]{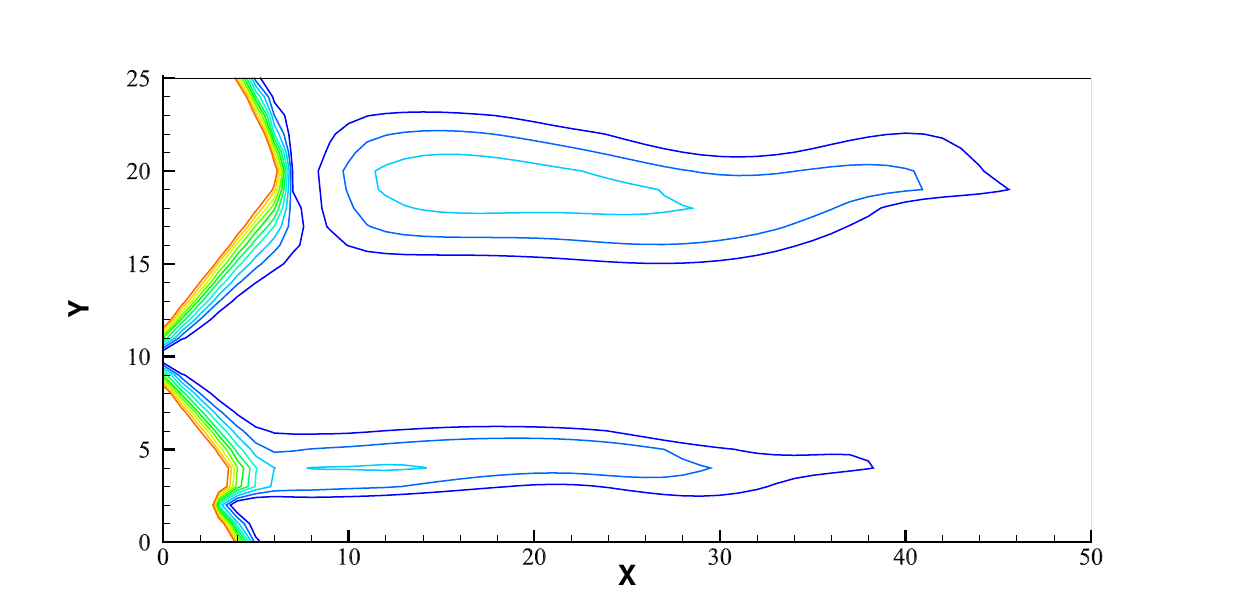}
		\end{minipage}
	}
	\subfigure[]{
		\begin{minipage}[b]{0.46\textwidth}
			\centering
			\includegraphics[width=1.0\textwidth]{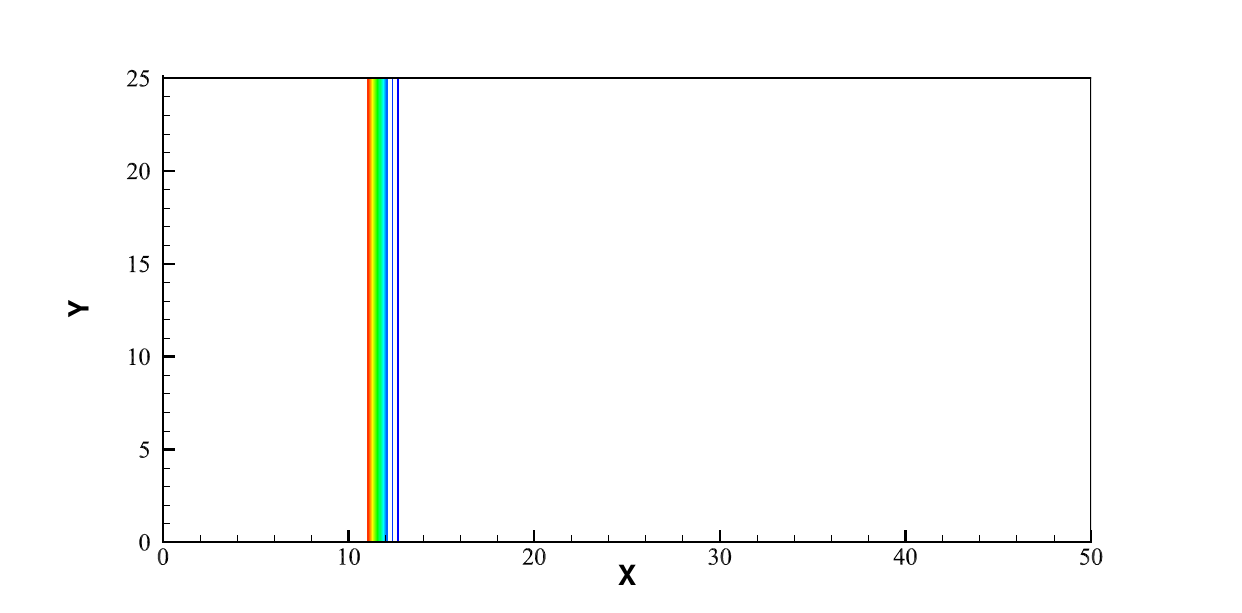}
		\end{minipage}
  }
  \subfigure[]{
		\begin{minipage}[b]{0.46\textwidth}
			\centering
			\includegraphics[width=1.0\textwidth]{fig3_5_2b-eps-converted-to.pdf}
		\end{minipage}
	}
	\caption{Mach number contours for 1.5D steady normal shock problem (grid with $50\times25$ cells, upstream Mach number $M_0=6.0$, shock position $\varepsilon=0.3$), (a) HLLEM scheme, (b) HLLEM scheme with HLLEM-$p$ flux ($f_{p,threshold}=0.33$) applied at all the interfaces of intermediate cells inside the shock, (c) HLLEM scheme with HLLEM-$p$ flux ($f_p=0.0f_{p,threshold}$) applied at all the interfaces of intermediate cells inside the shock.}
	\label{fig3-5-2}
\end{figure}

\begin{figure}[htbp]
	\centering
	\includegraphics[width=3.5in]{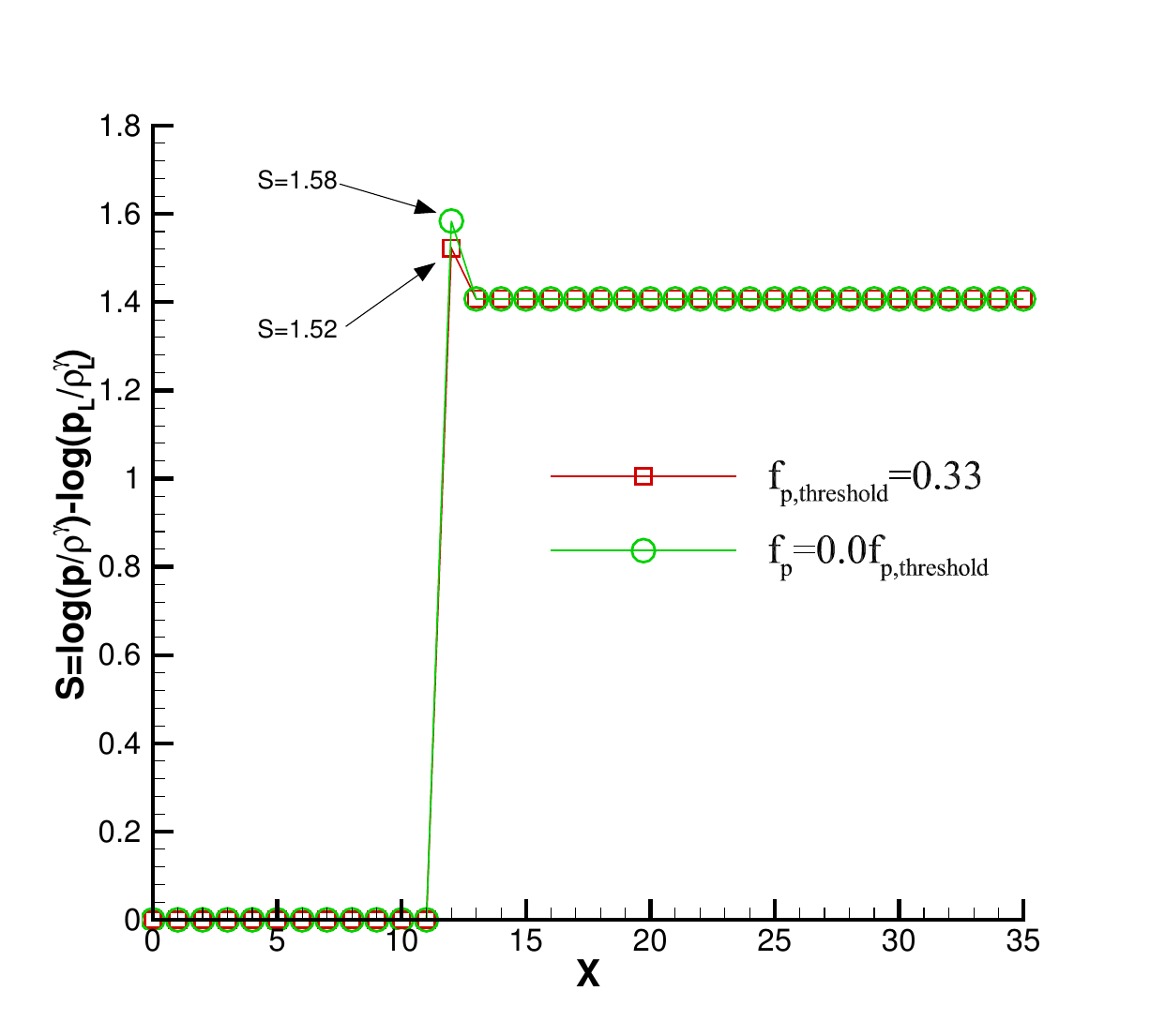}
	\caption{Entropy profiles along the $i$ direction extracted from solutions of 1.5D steady shock problem (predicted by the HLLEM scheme with HLLEM-${p}$ fluxes applied at all the interfaces of intermediate cells inside the shock at ${M_0} = 6.0$ and $\varepsilon = 0.3$).}
	\label{fig3-5-3}
\end{figure}

\begin{figure}[htbp]
	\centering
	\includegraphics[width=3.5in]{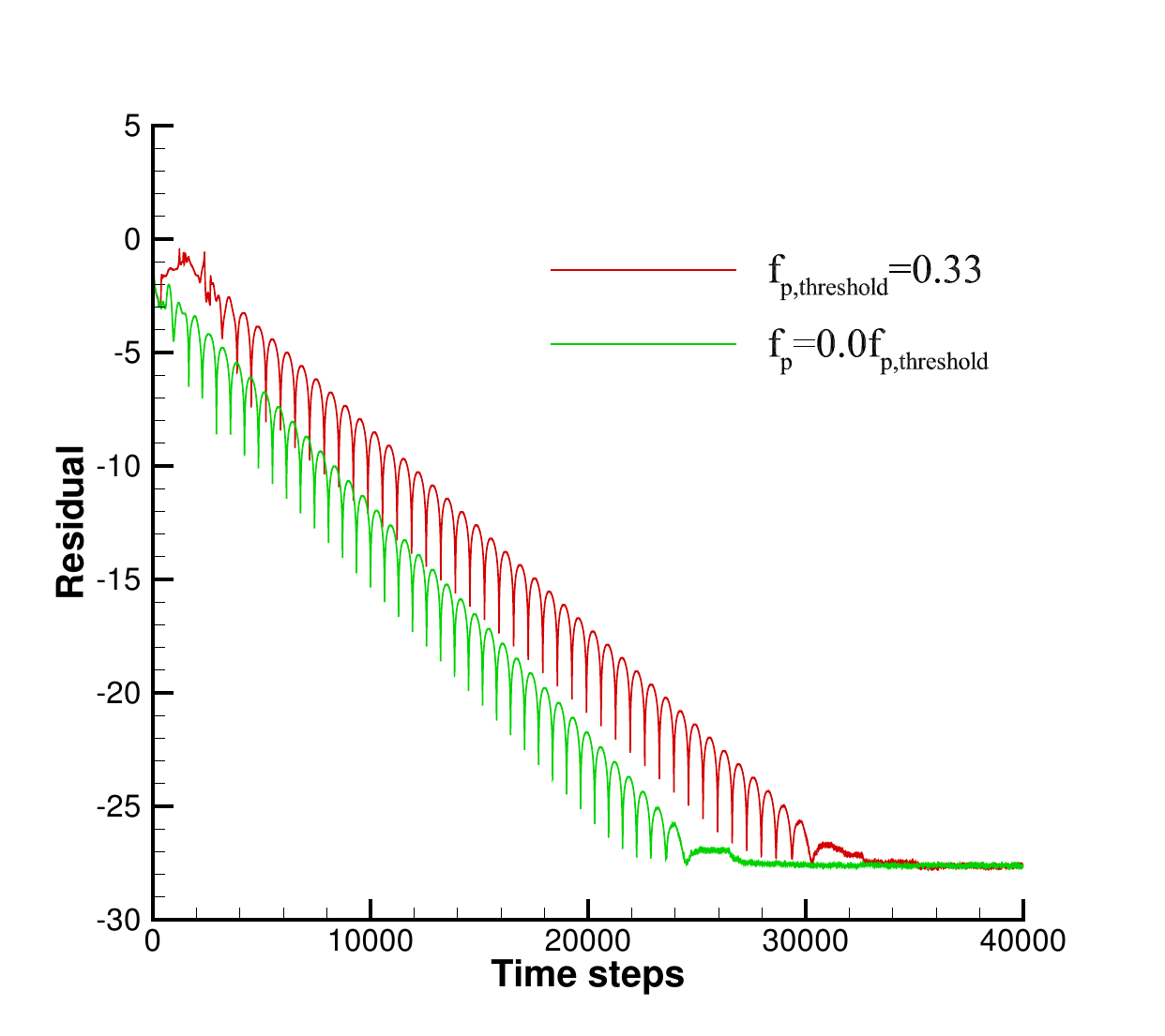}
	\caption{Residual histories for 1.5D steady shock problem computed by the HLLEM scheme with HLLEM-${p}$ flux applied at all the interfaces of intermediate cells inside the shock (${M_0} = 6.0$ and $\varepsilon = 0.3$) .}
	\label{fig3-5-4}
\end{figure}

\begin{figure}[htbp]
	\centering
	\includegraphics[width=3.5in]{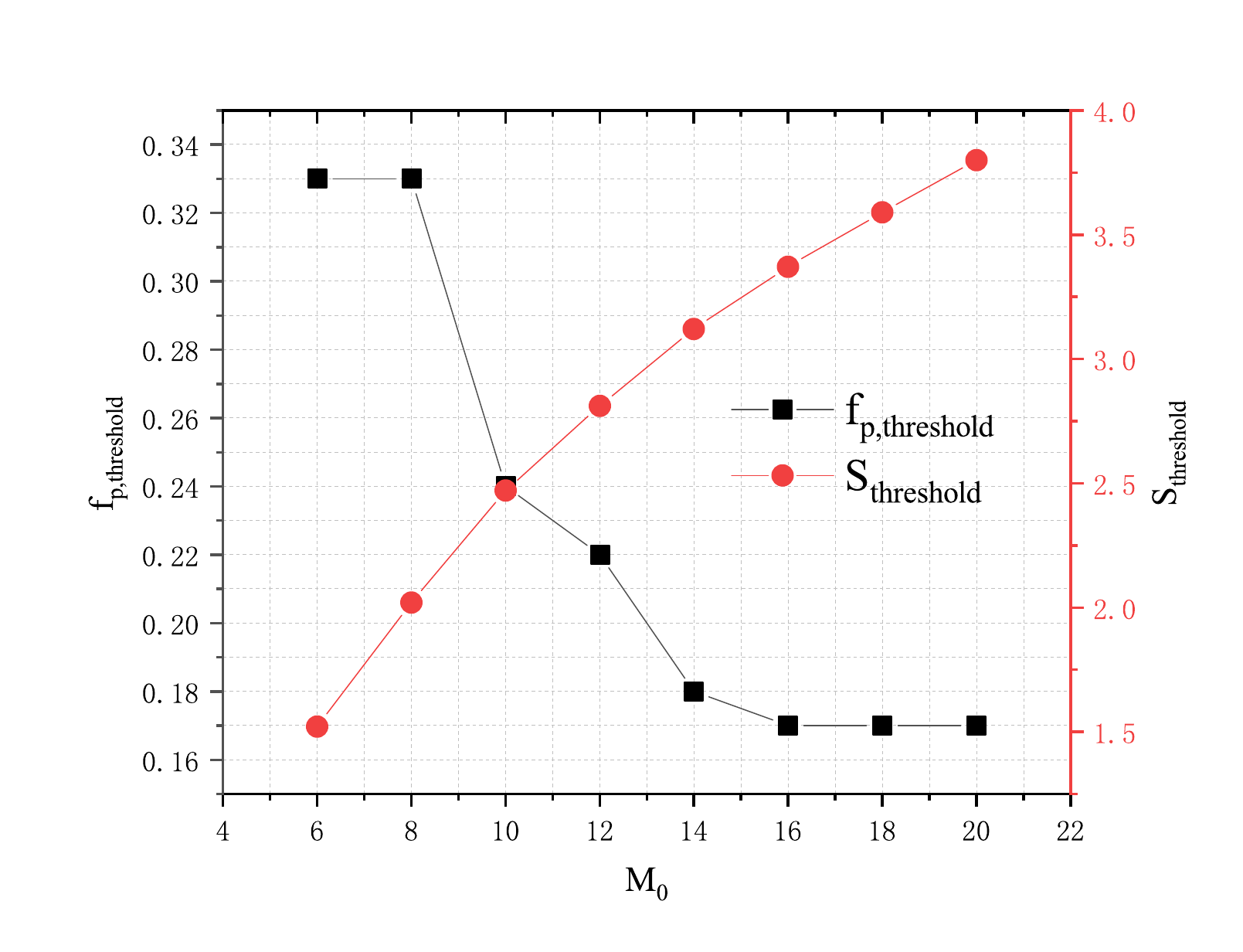}
	\caption{Stable thresholds of coefficient $f_p$ and their corresponding entropy $S=\log \frac{p}{{\rho}^{\gamma}}-\log \frac{p_L}{{\rho}_L^{\gamma}}$ vs. Mach number for 1.5D steady shock problem.}
	\label{fig3-5-5}
\end{figure}

The computations are conducted with the first-order HLLEM scheme for the two-dimensional steady shock problem. A three-stage first-order Runge-Kutta method \cite{van1992design} is used with CFL=0.5 until the time step reaches 40,000. To increase the entropy production, the HLLEM-$p$ fluxes are applied at all the interfaces of the intermediate cells inside the numerical shock structure. Stable and converged solutions can be obtained by reducing the magnitude of the coefficient $f_p$ properly. In Fig. \ref{fig3-5-2}, computed solutions by the original HLLEM scheme and the HLLEM scheme with the HLLEM-$p$ fluxes applied at interfaces of intermediate cells are demonstrated. It can be observed that the computed solution of the HLLEM scheme exhibits obvious shock instabilities, the shock front is severely distorted. With the HLLEM-$p$ flux applied at interfaces of intermediate cells inside the shock, the instability is barely observed and the shock is captured stably without any anomalies. The entropy profile extracted from the stable solutions are plotted in Fig. \ref{fig3-5-3}. It indicates that a smaller $f_p$ leads to larger entropy production inside the numerical shock structure, which is again well accordant with the dissipation analysis presented in section \ref{sec3.3.2}. Residual histories presented in Fig. \ref{fig3-5-4} indicate that more entropy production will make the solution converge to a stable state more quickly. It should be noted that similar results are also observed for various combinations of Mach numbers $M_0$ and the shock positions $\varepsilon$. In Fig. \ref{fig3-5-5}, we also provide the stable thresholds of $f_{p,threshold}$ for different Mach numbers $M_0$ and their corresponding entropy thresholds are also included.

\section{Numerical shock instability revisited: A linearized perturbation analysis}
\label{sec4}
In the previous sections, we focus on analyzing the stability of approximate Riemann solvers to solve steady shocks. It has been demonstrated that the root for such numerical instabilities is due to the improper entropy production inside the shock structure. In this section, we revisit the instability of approximate Riemann solvers for the steady shock problem through a linearized perturbation analysis. This analysis method is first used by Quirk \cite{quirk1994contribution} and then followed by many other researchers \cite{Kim2003,shen2014,Xie2017,SIMON2018144} to explore the mechanism of numerical shock instabilities. One advantage of the linearized perturbation analysis is that it is able to provide us an intuitive way to understand the mechanism of shock instability in the view of perturbations. However, one should always be caution that the linear analysis is supposed to only apply to cases where the perturbation errors are small enough. Thus, it can only provide a qualitative description of the instability.

\subsection{One-dimensional case}
\label{sec4.1}
In section {\ref{sec3}}, the improper entropy production inside the numerical shock structure is found to be the root cause of the one-dimensional carbuncle. By modifying the wave strength of dissipative term in Godunov-type schemes, enough entropy production can be guaranteed inside shocks. In order to further understand the relation between the entropy production and the instability, the instability problem is revisited in the context of the linearized perturbation analysis. In our previous work \cite{Xie2017}, it has been found that if the mass flux across the normal shock is accurately preserved (i.e., $(\rho u)^n_L=(\rho u)^n_R$), then the shock could be stabilized. The erroneous mass flux is originated from the intermediate states inside the shock structure. Thus, we need to clarify how the entropy-control technique could influence the mass flux perturbation behind the shock. Here, the linearized perturbation analysis is conducted again for the numerical flux functions HLLEM-$\rho$ and HLLEM-$p$. Readers are referred to reference \cite{Xie2017} for more details.

At time $t^n$, the instability happens. It is assumed that there are some perturbation errors being generated in the cell M, they are expressed as
\begin{equation}\label{eq4.1.1}
  \rho_M^n=\rho_M^{*,n}+\delta \rho_M^n, \quad (\rho u)_M^n=(\rho u)_M^{*,n}+\delta (\rho u)_M^n, \quad p_M^n=p_M^{*,n}+\delta p_M^n,
\end{equation}
where $\delta (\cdot)$ denote the perturbation errors which represent small discrepancies from the stable steady states. ${\left(  \cdot  \right)^ * }$ denote the stable steady solutions that are assumed to locate at a Hugoniot curve. They have the following relationship with states in the cell L and the cell R,
\begin{equation}\label{eq4.1.2}
  \rho_M^{*,n}=(1-\alpha_{\rho})\rho_L^{*,n}+\alpha_{\rho}\rho_R^{*,n},\quad u_M^{*,n}=(1-\alpha_u)u_L^{*,n}+\alpha_u u_R^{*,n},\quad
p_M^{*,n}=(1-\alpha_p)p_L^{*,n}+\alpha_p p_R^{*,n}.
\end{equation}
where the coefficients $\alpha_{\rho}$, $\alpha_u$ and $\alpha_p$ are defined in Eq. (\ref{eq3.5}).

To explore how the perturbation errors generated in cell M influence the mass flux perturbation in cell R, we need to consider the following conservative scheme
\begin{equation}\label{eq4.1.3}
  \left( {\rho u} \right)_R^{n + 1} = \left( {\rho u} \right)_R^n - \frac{{\Delta t}}{{\Delta x}}\left[ {\left( {\rho {u^2} + p} \right)_{R,R'}^n - \left( {\rho {u^2} + p} \right)_{M,R}^n} \right].
\end{equation}
The subscript $R,R'$ denotes the interface between the cell $R$  and the cell $R'$ , the subscript $M,R$  denotes the interface between the cell $M$  and the cell $R$. For the HLLEM-$p$ scheme, the numerical fluxes at the interfaces in Eq. (\ref{eq4.1.3}) can be written as
\begin{flalign}\label{eq4.1.4}
\begin{split}
\left( {\rho {u^2} + p} \right)_{R,R'}^n &= \left( {\rho {u^2} + p} \right)_R^n, \\
\left( {\rho {u^2} + p} \right)_{M,R}^n  &= \frac{S_R}{S_R - S_M}\left( {\rho {u^2} + p} \right)_M^n - \frac{S_M}{S_R - S_M}\left( {\rho {u^2} + p} \right)_R^n \\
&\quad + \frac{S_M S_R}{S_R - S_M}\left[ {\left( {\rho u} \right)_R^n - \left( {\rho u} \right)_M^n - {\widehat{\delta}}_2 \left( {\rho _R^n - \rho _M^n - {f_p} \frac{ p_R^n - p_M^n }{ {\widehat a}^2} } \right) \widehat u} \right].
\end{split}
\end{flalign}
Here, the subscript M denotes the cell M, the subscript R denotes the cell R  that is at the right side of M. $\widehat{(\cdot)}$ are Roe's averaged variables between states in cells M and R. Inserting (\ref{eq4.1.4}) into (\ref{eq4.1.3}), after some calculations we will obtain that
\begin{equation}\label{eq4.1.5}
\delta \left( {\rho u} \right)_R^{n + 1} - \delta \left( {\rho u} \right)_R^n = {\theta}_{\rho} \cdot \delta {\rho}_M^n + {\theta}_{u} \cdot \delta u_M^n + {\theta}_p \cdot \delta p_M^n
\end{equation}
with
\begin{equation}\label{eq4.1.6}
\begin{aligned}
{\theta}_{\rho} & = \frac{S_R \nu M_0}{\left(S_R - S_M \right)\left( 1+M_0 \right)} \left( {\frac{{{\omega ^2}}}{{{f^2}}} - \frac{{{S_M}\omega }}{f} + \frac{{{S_M}\hat{\delta} _2}}{f}\frac{{1 + \omega \sqrt \sigma  }}{{1 + \sqrt \sigma  }}} \right)               \\
{\theta}_u & = \frac{S_R \nu M_0}{\left(S_R - S_M \right)\left( 1+M_0 \right)} \left( {2\sigma \omega  - {S_M}\sigma f} \right)        \\
{\theta}_p & = \frac{S_R \nu M_0}{\left(S_R - S_M \right)\left( 1+M_0 \right)} \left( {1 - {f_p}\frac{{{S_M}\hat{\delta} _2}}{{{{\widehat a}^2}f}}\frac{{1 + \omega \sqrt \sigma }}{{1 + \sqrt \sigma  }}} \right)
\end{aligned}
\end{equation}
where the coefficients $\omega$ and $\sigma$ are defined as
\begin{equation}\label{eq4.1.7}
\begin{split}
\omega & = f - f \cdot {\alpha _u} + {\alpha _u} , \\
\sigma & = \frac{{1 - {\alpha _\rho } + f \cdot {\alpha _\rho }}}{f} .
\end{split}
\end{equation}
In Eq. (\ref{eq4.1.6}), $\nu$ denotes the Courant number and the coefficients ${\theta}_{\rho}$, ${\theta}_u$ and ${\theta}_p$ are functions of the freestream Mach number $M_0$ and the shock position parameter $\varepsilon$. A similar result can also be obtained for the HLLEM-${\rho}$ flux. The recursive formulas of perturbed quantities can still be written as Eq. (\ref{eq4.1.5}), but different coefficients are obtained as
\begin{equation}\label{eq4.1.8}
  \begin{aligned}
  {\theta}_{\rho} & = \frac{S_R \nu M_0}{\left(S_R - S_M \right)\left( 1+M_0 \right)} \left( {\frac{{{\omega ^2}}}{{{f^2}}} - \frac{{{S_M}\omega }}{f} + {f_{\rho}} \frac{{{S_M}\hat{\delta} _2}}{f}\frac{{1 + \omega \sqrt \sigma  }}{{1 + \sqrt \sigma  }}} \right)               \\
  {\theta}_u & = \frac{S_R \nu M_0}{\left(S_R - S_M \right)\left( 1+M_0 \right)} \left( {2\sigma \omega  - {S_M}\sigma f} \right)        \\
  {\theta}_p & = \frac{S_R \nu M_0}{\left(S_R - S_M \right)\left( 1+M_0 \right)} \left( {1 - \frac{{{S_M}\hat{\delta} _2}}{{{{\widehat a}^2}f}}\frac{{1 + \omega \sqrt \sigma  }}{{1 + \sqrt \sigma  }}} \right)
  \end{aligned}
\end{equation}
where the coefficients $\omega$ and $\sigma$ are still defined in Eqs. (\ref{eq4.1.7}).

Due to the nonpositivity of the wave speed $S_M$, it can be observed from (\ref{eq4.1.6}) that a smaller $f_p$ can lead to a reduced ${\theta}_p$, then a reduced mass flux $\delta (\rho u)_R^{n+1}$ will be obtained. Thus, the instability can be suppressed. Similarly, a larger $f_{\rho}$ leads to a reduced ${\theta}_{\rho}$, then a reduced mass flux $\delta (\rho u)_R^{n+1}$ will also be obtained, which plays a role in eliminating the instability. In conclusion, the entropy-control technique is equal to reducing the perturbation errors of mass flux which contribute to the instability in the view of perturbations analysis.

\subsection{Two-dimensional case}
\label{sec4.2}
The magnitude of the velocity in the transverse direction of the shock wave has been well recognized to be a proper parameter to use to show the magnitude of the multidimensional carbuncle phenomenon \cite{Dumbser2004,henderson2007grid}. For the steady normal shock problem considered in section \ref{sec3}, physically, there should be no mass flux appearing in the transverse direction and the transverse velocity should be zero. Hence, any erroneous mass flux developed in this direction results from the instability.

The dimensional splitting method \cite{leveque2002finite} is used to update the solutions of the two-dimensional normal shock problem. For a particular cell $(i,j)$ inside the shock structure, the updated cell average ${\bf{U}}_{i,j}^ {n+1/2} $  is obtained by solving the one-dimensional conservative scheme in the $x$-direction,
\begin{equation}\label{eq4.2.1}
  {\bf{U}}_{i,j}^ {n+1/2}  = {\bf{U}}_{i,j}^n - \frac{{\Delta t}}{{\Delta x}}\left( {{\bf{F}}_{i + 1/2,j}^n - {\bf{F}}_{i - 1/2,j}^n} \right),
\end{equation}
where ${\bf{F}}_{i + 1/2,j}^n$  is the numerical flux function for the one-dimensional Riemann problem between cells $(i,j)$ and $(i+1,j)$. Then the state ${\bf{U}}_{i,j}^ {n+1/2} $  is used as data to update the solution in the $y$-direction,
\begin{equation}\label{eq4.2.2}
  {\bf{U}}_{i,j}^{n + 1} = {\bf{U}}_{i,j}^ {n+1/2}  - \frac{{\Delta t}}{{\Delta y}}\left( {{\bf{G}}_{i,j + 1/2}^{n+1/2} - {\bf{G}}_{i,j - 1/2}^{n+1/2}} \right)
\end{equation}
where ${\bf{G}}_{i,j + 1/2}^{n+1/2}$ is the numerical flux function between cells $(i,j)$ and $(i,j+1)$. At the beginning of the instability, perturbations are generated inside the shock structure. As shown in Fig. \ref{fig3-5-1}, the perturbed states are distributed along the $j$ direction in a saw-toothed manner. To facilitate further analysis, it is assumed that the states along the $y$ direction inside the shock structure are initialized as follows,
\begin{equation}\label{eq4.2.3}
\rho_{j}^n=\rho_{i,j}^{*,n}-\delta {\rho}^n, \quad \left( {\rho u} \right)_{i,j}^n = {\left( {\rho u} \right)_{i,j}^{*,n}}-\delta ({\rho u})^n, \quad \left( {\rho v} \right)_{i,j}^n = {\left( {\rho v} \right)_{i,j}^{*,n}} - \delta ({\rho v})^n, \quad p_{i,j}^n = {p_{i,j}^{*,n}} - \delta p^n,
\end{equation}
and
\begin{equation}\label{eq4.2.4}
  \rho _{i,j \pm 1}^n = \rho_{i,j \pm 1}^{*,n} + \delta {\rho}^n, \quad \left( {\rho u} \right)_{i,j \pm 1}^n = {\left( {\rho u} \right)_{i,j \pm 1}^{*,n}}+\delta ({\rho u})^n, \quad \left( {\rho v} \right)_{i,j \pm 1}^n = {\left( {\rho v} \right)_{i,j \pm 1}^{*,n}} + \delta ({\rho v})^n, \quad p_{i,j \pm 1}^n = {p_{i,j \pm 1}^{*,n}} + \delta {p^n},
\end{equation}
where ${()}^{*}$ represent the stable steady solutions that are assumed to be uniform along the transverse direction. In what follows, we omit the subscript $i$ for clarity. In the two-dimensional case, we need to clarify how the perturbations will promote the perturbed mass flux in the transverse direction. Hence, the following conservative scheme is considered,
\begin{equation}\label{eq4.2.5}
  \left( {\rho v} \right)_j^{n + 1} = \left( {\rho v} \right)_j^n - \frac{{\Delta t}}{{\Delta y}}\left[ {\left( {\rho {v^2} + p} \right)_{j + 1/2}^n - \left( {\rho {v^2} + p} \right)_{j - 1/2}^n} \right].
\end{equation}
For the flux function HLLEM-$p$, the numerical fluxes at the interfaces in (\ref{eq4.2.5}) can be written as
\begin{flalign}\label{eq4.2.6}
\begin{split}
\left( {\rho {v^2} + p} \right)_{j + 1/2}^n &= \frac{S_{R,j+1/2}}{{S_{R,j+1/2}}-{S_{L,j+1/2}}} \left( {\rho {v^2} + p} \right)_j^n - \frac{S_{L,j+1/2}}{{S_{R,j+1/2}}-{S_{L,j+1/2}}}  \left( {\rho {v^2} + p} \right)_{j + 1}^n\\
&\quad + \frac{{S_{L,j+1/2}}{S_{R,j+1/2}}}{{S_{R,j+1/2}} - {S_{L,j+1/2}}}\left[ {\left( {\rho v} \right)_{j + 1}^n - \left( {\rho v} \right)_j^n - {{\widehat \delta }_{2,j + 1/2}}\left( {{\rho^n _{j + 1}} - {\rho^n _j} - {f_p}\frac{{{p_{j + 1}} - {p_j}}}{{\widehat a_{j + 1/2}^2}}} \right){{\widehat v}_{j + 1/2}}} \right],\\
\left( {\rho {v^2} + p} \right)_{j - 1/2}^n &= \frac{S_{R,j-1/2}}{{S_{R,j-1/2}}-{S_{L,j-1/2}}} \left( {\rho {v^2} + p} \right)_{j - 1}^n - \frac{S_{L,j-1/2}}{{S_{R,j-1/2}}-{S_{L,j-1/2}}} \left( {\rho {v^2} + p} \right)_j^n\\
&\quad + \frac{{S_{L,j-1/2}}{S_{R,j-1/2}}}{{S_{R,j-1/2}} - {S_{L,j-1/2}}}  \left[ {\left( {\rho v} \right)_j^n - \left( {\rho v} \right)_{j - 1}^n - {{\widehat \delta }_{2,j - 1/2}}\left( {\rho _j^n - \rho _{j - 1}^n - {f_p}\frac{{p_j^n - p_{j - 1}^n}}{{{{\widehat a}^2}}}} \right){{\widehat v}_{j - 1/2}}} \right].
\end{split}
\end{flalign}
where ${\widehat {\left(  \cdot  \right)}_{j + 1/2}}$ denote Roe's averaged variables between states in cells $(i,j)$ and $(i,j+1)$. Inserting (\ref{eq4.2.6}) into (\ref{eq4.2.5}), we can obtain the following relationship (see Appendix B for the detailed derivation)
\begin{equation}\label{eq4.2.7}
    \delta \left( {\rho v} \right)_j^{n + 1} - \delta \left( {\rho v} \right)_j^{n} = {\theta _\rho } \cdot \delta \rho ^n{\rm{ + }}{\theta _v} \cdot \delta v^n + {\theta _p} \cdot \delta p^n
\end{equation}
with
\begin{equation}\label{eq4.2.8}
\begin{aligned}
{\theta}_{\rho} & = 2\frac{\nu }{{{v^{*,n}} + {a^{*,n}}}} {{\left( {{v^{*,n}}} \right)}^2}  \\
{\theta}_v & = 2\frac{\nu }{{{v^{*,n}} + {a^{*,n}}}} {\rho ^{*,n}}\frac{{{{\left( {{v^{*,n}}} \right)}^2} + {{\left( {{a^{*,n}}} \right)}^2}}}{{{a^{*,n}}}}       \\
{\theta}_p & = 2\frac{\nu }{{{v^{*,n}} + {a^{*,n}}}} \left[ {\frac{{{v^{*,n}}}}{{{a^{*,n}}}} + {f_p}\frac{{{v^{*,n}}\left( {{a^{*,n}} - {v^{*,n}}} \right)}}{{{a^{*,n}}}}} \right].
\end{aligned}
\end{equation}
It can be observed from (\ref{eq4.2.8}) that a smaller $f_p$ can lead to a reduced ${\theta}_p$, then a reduced mass flux $\delta (\rho v)_j^{n+1}$ will be obtained. Thus, the instability can be suppressed. Similar to its one-dimensional case, the entropy-control technique is equal to reducing the mass flux perturbation errors which contribute to the instability in the view of perturbations analysis.

\section{A general technique for suppressing the shock instability by an entropy control}
\label{sec5}
In above sections, efforts have been devoted to investigating the underlying mechanism of numerical shock instability for the HLL-type schemes. It has been demonstrated that the shock instability can be attributed to improper entropy production inside the numerical shock structure. In this section, we extend the entropy-control technique discussed in section {\ref{sec3.3.2}} to other Godunov-type schemes to demonstrate its generality.

The proposed numerical flux function with an {\bf{E}}ntropy {\bf{C}}ontrol can be expressed as
\begin{equation}\label{eq5.1}
  {\bf{F}}_{\rm{X-EC}}={\bf{F}}_{\rm{X}}+{\bf{F}}_{\rm{EC}}
\end{equation}
where $``\rm{X}"$ denotes certain Godunov-type schemes. The entropy-control term ${\bf{F}}_{\rm{EC}}$ can be derived directly from the mentioned HLLEM-$p$ flux . It is defined by
\begin{equation}\label{eq5.2}
{\bf{F}}_{\rm{EC}} = (f_p-1) \frac{S^-_L S^+_R}{S^+_R - S^-_L} {\widehat{\delta}}_2 \frac{\Delta p}{{\widehat{a}}^2} {\widehat{{\bf{R}}}}_2
\end{equation}
where ${\widehat{\delta}}_2$ represents the anti-diffusion coefficient defined in Eq. (\ref{eq2.2.2_6}), ${\widehat{{\bf{R}}}}_2$ denotes the second right eigenvector defined in Eq. (\ref{eq2.2.4}), $\widehat{a}$ is the sound speed evaluated by Roe's averaged variables, $\Delta p$ is the pressure difference $\Delta p = p_R - p_L$. The wave speeds $S^-_L:=\min(0,S_L)$ and $S^+_R:=\max(0,S_R)$ are evaluated by (\ref{eq2.1.6}), which provides minimal diffusion on isolated shocks. Alternatively, the estimate (\ref{eq2.1.7}) can also be used to compute wave speeds $S^-_L$ and $S^+_R$, they are more diffusive and useful for suppressing possible post-shock oscillations. To complete the evaluation of the resulting flux function ${\bf{F}}_{\rm{X-EC}}$, $f_p$ needs to be determined, it is written by
\begin{equation}\label{eq5.3}
  f_p=\min\left( p_{i,j+1/2},p_{i-1/2,j},p_{i+1/2,j},p_{i-1/2,j+1},p_{i+1/2,j+1}\right)^3,
\end{equation}
with
\begin{equation}\label{eq5.4}
    p_{i,j+1/2}=\min \left( p_{i,j}/p_{i,j+1},p_{i,j+1}/p_{i,j}\right).
\end{equation}
In the vicinity of shock waves, the pressure difference is increased and the function $f_p$ becomes small, the entropy control term ${\bf{F}}_{\rm{EC}}$ is activated to increase the entropy production inside the numerical shock structure, thus enhancing the stability. In the smooth region, the function $f_p$ approaches unit, the modified flux turns into its original form. It should be noted that the entropy-control term ${\bf{F}}_{\rm{EC}}$ does not involve any numerical dissipation on linear degenerate waves, thus it avoids the risk of introducing unnecessary numerical dissipation to contaminate flow structures. In the current study, the entropy-control technique is applied to four classical Godunov-type schemes, i.e., ``X'' denote Godunov's exact Riemann solver \cite{godunov1959finite}, Roe \cite{roe1997approximate}, HLLEM \cite{Einfeldt1988} and HLLC \cite{Toro1994} schemes, which are known to resolve discontinuities with minimal diffusion but are plagued by numerical shock instabilities.

\section{Numerical experiments}
\label{sec6}
In what follows, we apply our approach to a series of numerical experiments already used for the validation of several improved Godunov-type schemes\cite{Xie2017,Xie2019}. The former four problems are selective test cases to access the robustness of the proposed method against strong shock waves. A focus of other problems is on accuracy for resolving shear layers.

\subsection{Quirk's odd-even grid perturbation problem}
\label{sec6.1}
\begin{figure}[htbp]
  \centering  %图片全局居中
  \subfigure[Godunov]{
  \label{fig6.1.7}
  \includegraphics[width=0.48\textwidth]{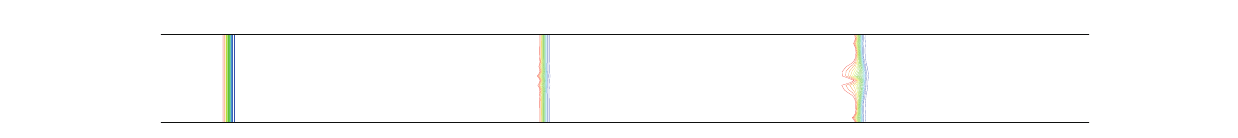}}
  \subfigure[Godunov-EC]{
  \label{fig6.1.8}
  \includegraphics[width=0.48\textwidth]{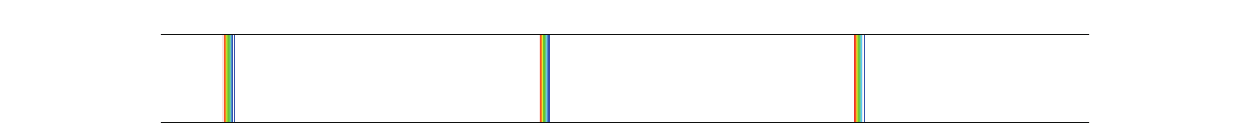}}
  \subfigure[Roe]{
  \label{fig6.1.1}
  \includegraphics[width=0.48\textwidth]{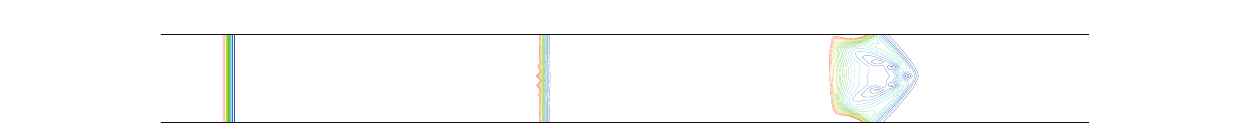}}
  \subfigure[Roe-EC]{
  \label{fig6.1.2}
  \includegraphics[width=0.48\textwidth]{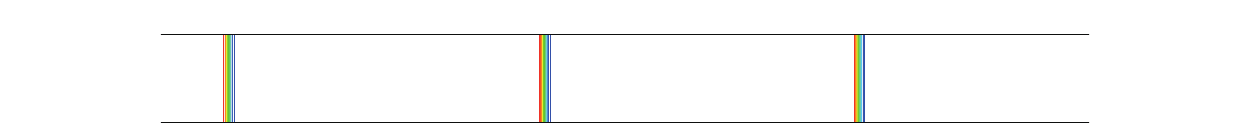}}
  \subfigure[HLLEM]{
  \label{fig6.1.3}
  \includegraphics[width=0.48\textwidth]{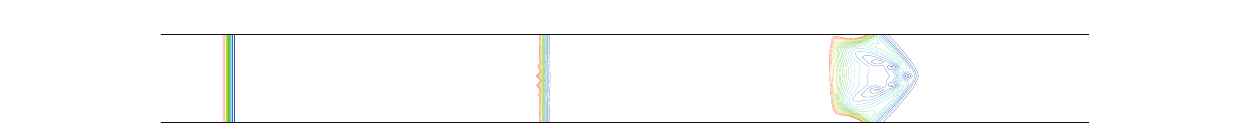}}
  \subfigure[HLLEM-EC]{
  \label{fig6.1.4}
  \includegraphics[width=0.48\textwidth]{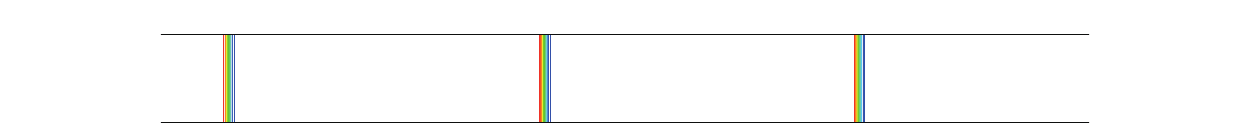}}
  \subfigure[HLLC]{
  \label{fig6.1.5}
  \includegraphics[width=0.48\textwidth]{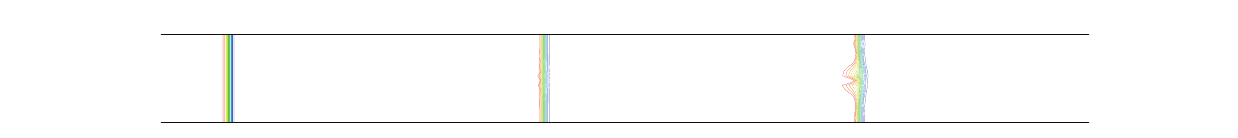}}
  \subfigure[HLLC-EC]{
  \label{fig6.1.6}
  \includegraphics[width=0.48\textwidth]{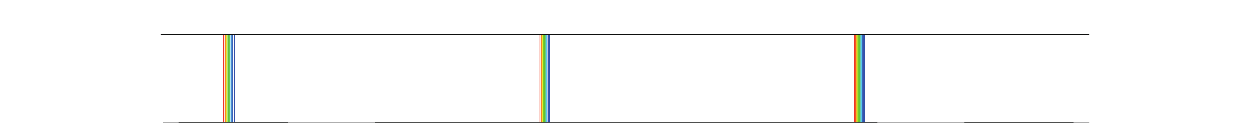}}
  \caption{Density contours for the odd-even grid perturbation problem.}
  \label{fig6-1}
\end{figure}

Quirk's odd-even decoupling problem \cite{quirk1994contribution} is a well-known test used to access the resistance of numerical schemes against shock instabilities. A plane normal shock wave moving from the left to the right at Mach 6 is computed on a regular grid system with a slight perturbation in one of the grid lines parallel to the direction of the flow. The computational domain is covered by an $800\times20$ structured grid involving the following grid perturbations:
\begin{equation}
y(i,j_{mid})=
\begin{cases}
y_{mid}+0.001 & \text{if i is even}\\
y_{mid}-0.001 & \text{if i is odd}
\end{cases}
\end{equation}
where $ y(i,j_{mid}) $ is the $ y $ coordinate of a vertex $ (i,11) $, $ y_{mid} $ is the $ y $ coordinate of the halfway line. The initial conditions are $\left( {\rho ,u,v,p} \right) = \left( {1.4,0,0,1} \right)$, the inlet boundary condition is set to post shock values, nonreflecting simple wave boundary conditions are imposed at the outlet. The upper and lower boundaries are considered as solid walls. This problem is solved by using different flux functions, along with the two-stage second-order Runge-Kutta method \cite{SHU1988439}. The CFL number is set to 0.5 for all the schemes to compute the solutions. Fig. \ref{fig6-1} shows the contour plots of density at the time levels $t=0$, $t=25$ and $t=50$, where there are $20$ contour levels varying from $1.5$ to $6.0$.

As shown in Fig. \ref{fig6-1}, the shocks were destroyed by all the original low diffusion approximate Riemann solvers. Compared with the Roe and HLLEM solvers, the Godunov's exact Riemann solver and HLLC scheme are less prone to the irregularities. Whereas, typical carbuncles were developed and could be observed clearly. With the entropy-control technique, all the schemes produce no odd-even decoupling along the shock. The shock waves were kept all the way through.

\subsection{Shock diffraction}
\label{sec6.2}
\begin{figure}
  \centering
  \subfigure[Godunov]{
		\begin{minipage}[b]{0.35\textwidth}
			\includegraphics[width=1.9in]{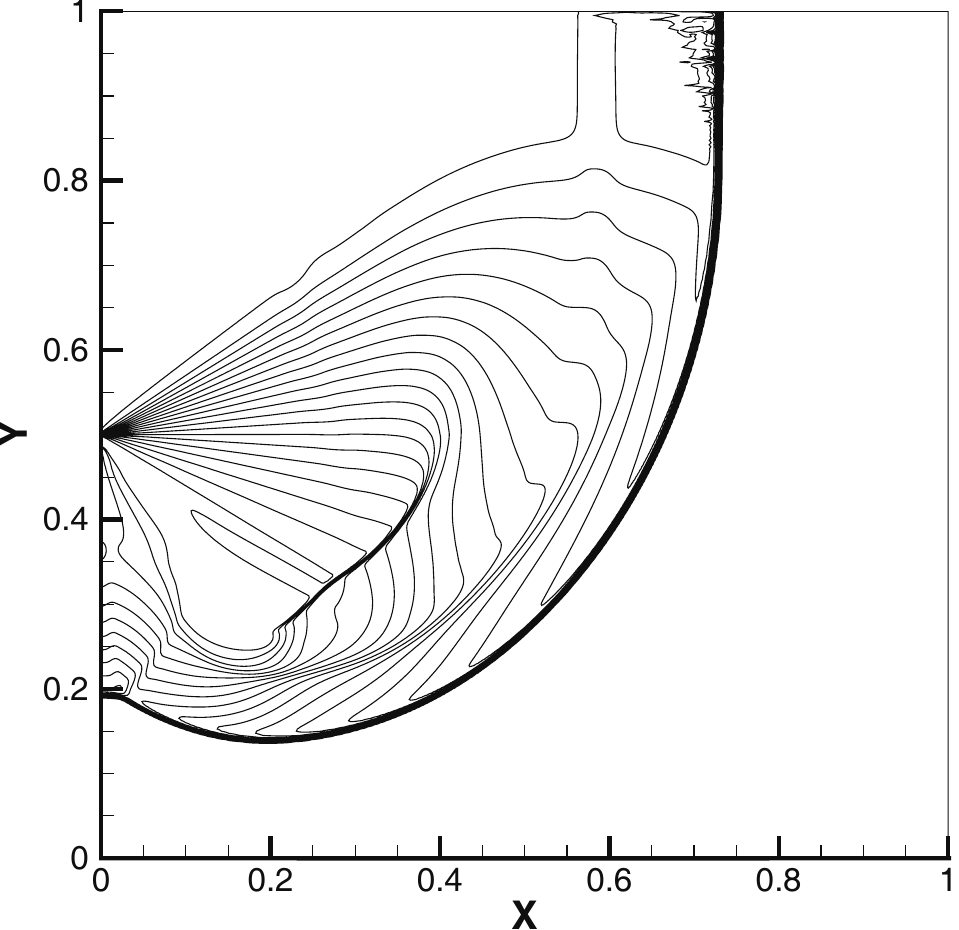}
		\end{minipage}
	}
	\subfigure[Godunov-EC]{
		\begin{minipage}[b]{0.35\textwidth}
			\includegraphics[width=1.9in]{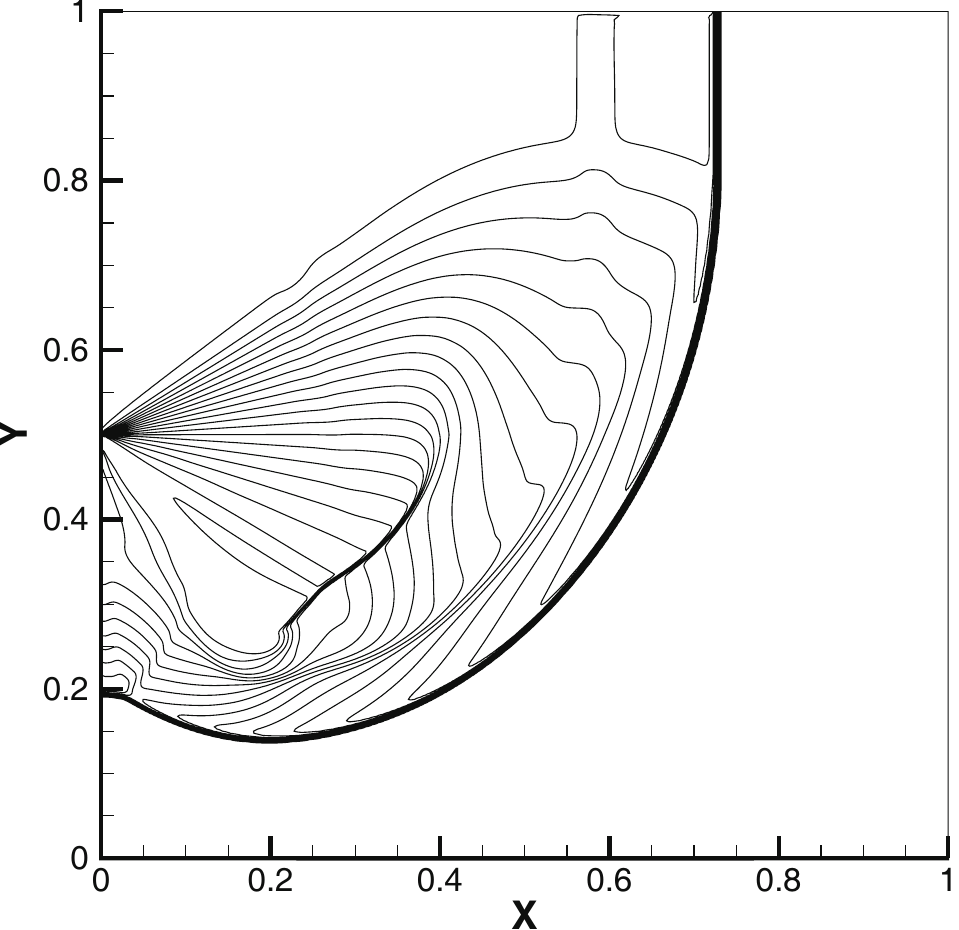}
		\end{minipage}
  }
	\subfigure[Roe]{
		\begin{minipage}[b]{0.35\textwidth}
			\includegraphics[width=1.9in]{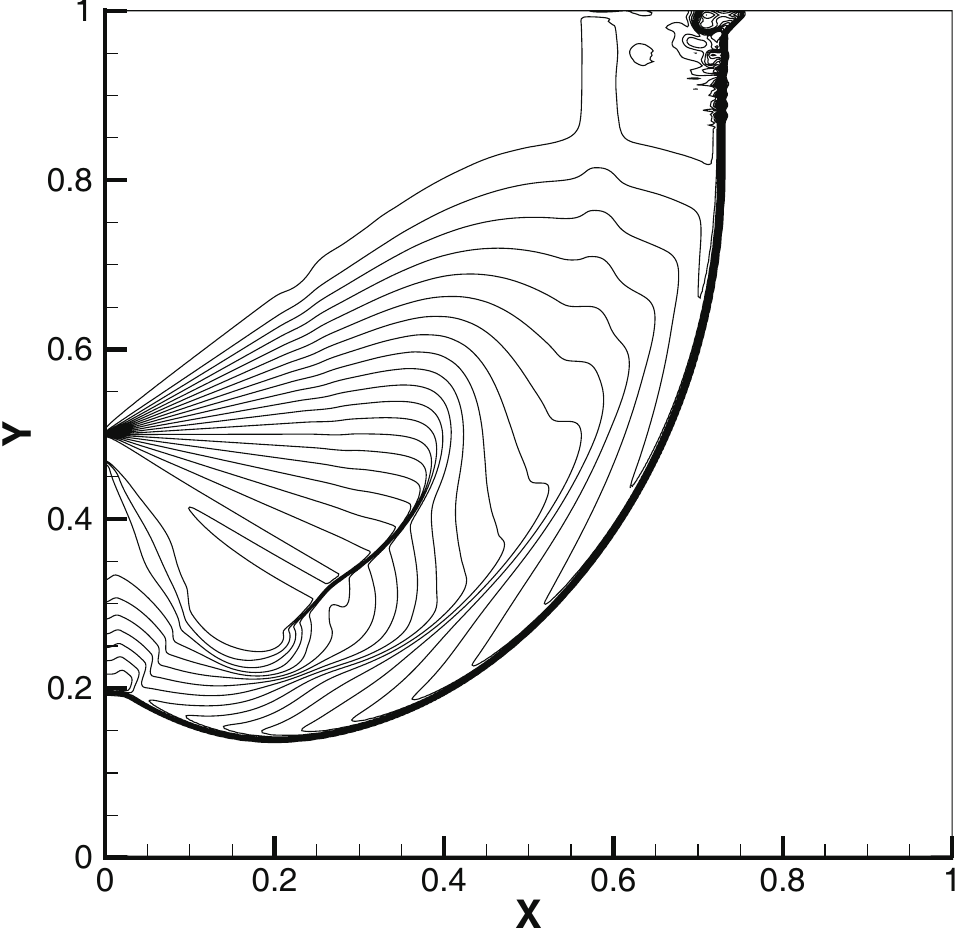}
		\end{minipage}
	}
	\subfigure[Roe-EC]{
		\begin{minipage}[b]{0.35\textwidth}
			\includegraphics[width=1.9in]{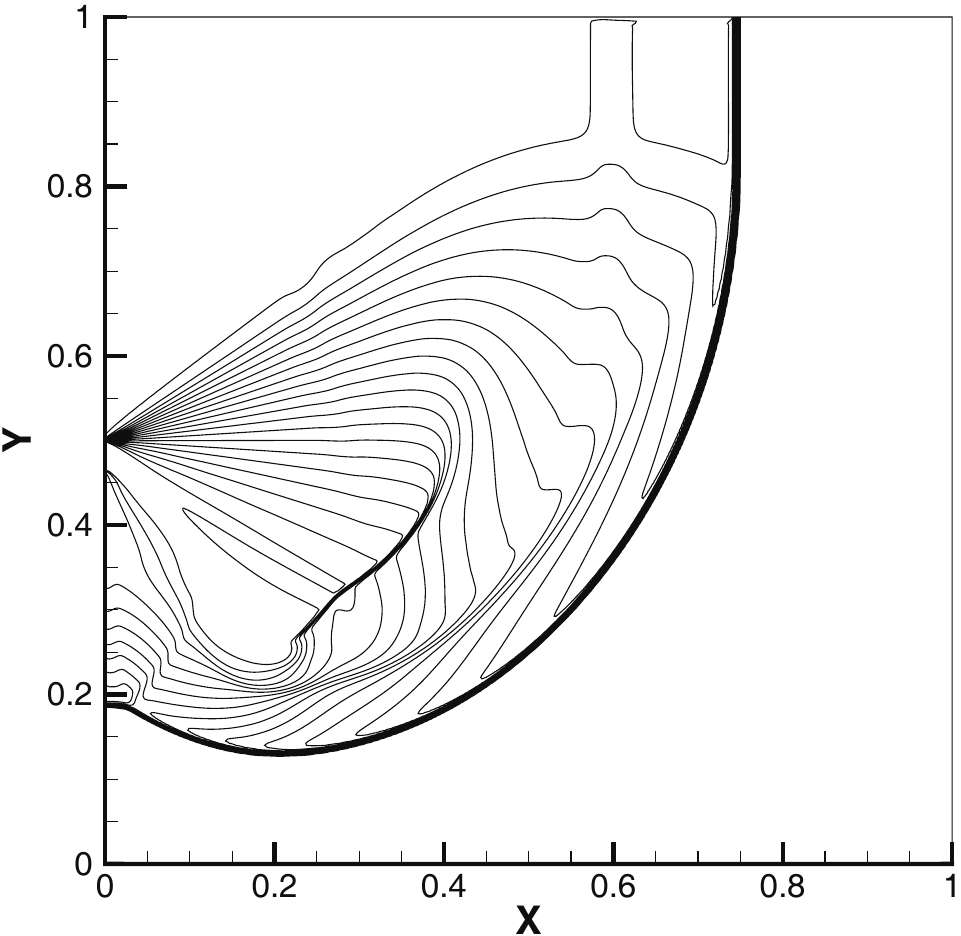}
		\end{minipage}
  }
	\subfigure[HLLEM]{
		\begin{minipage}[b]{0.35\textwidth}
			\includegraphics[width=1.9in]{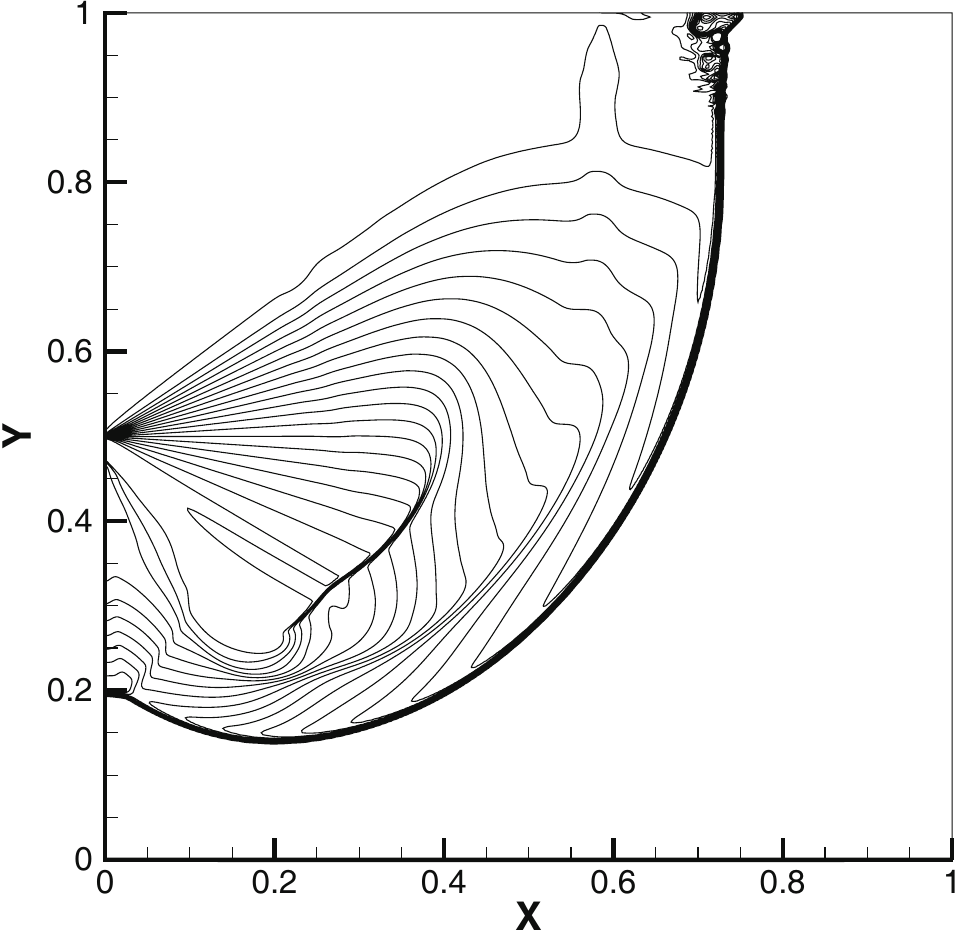}
		\end{minipage}
	}
	\subfigure[HLLEM-EC]{
		\begin{minipage}[b]{0.35\textwidth}
			\includegraphics[width=1.9in]{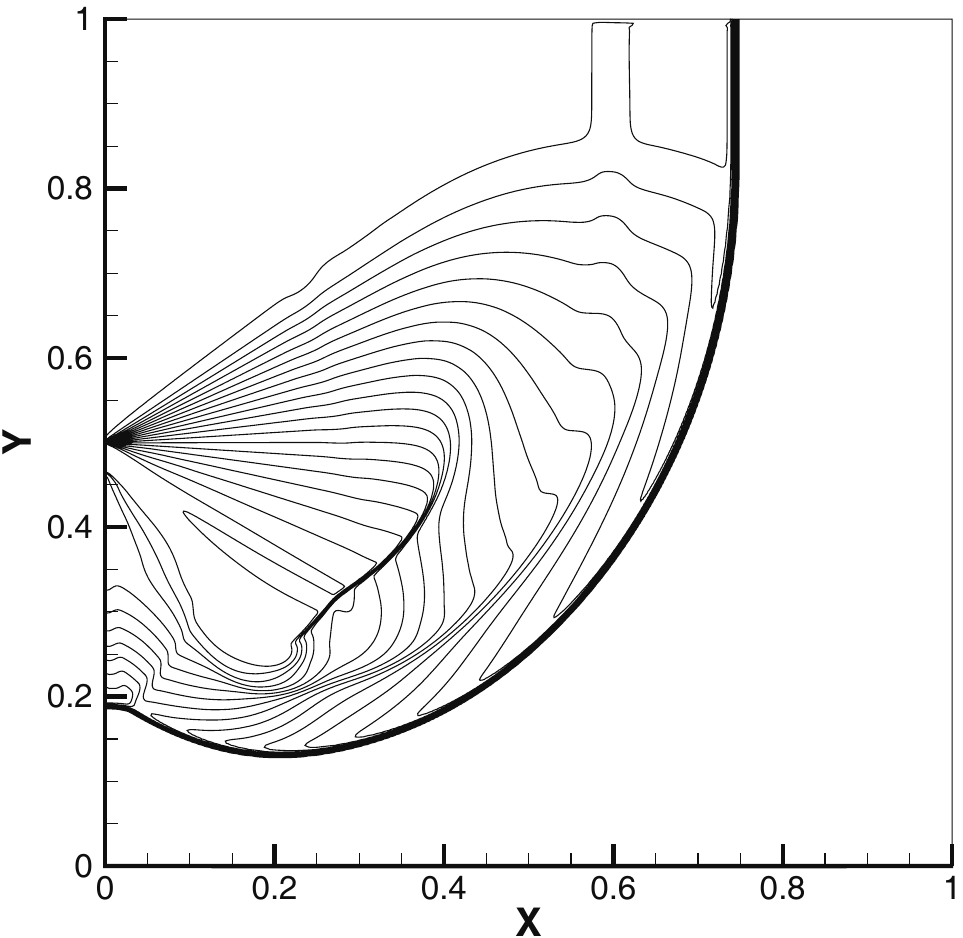}
		\end{minipage}
  }
	\subfigure[HLLC]{
		\begin{minipage}[b]{0.35\textwidth}
			\includegraphics[width=1.9in]{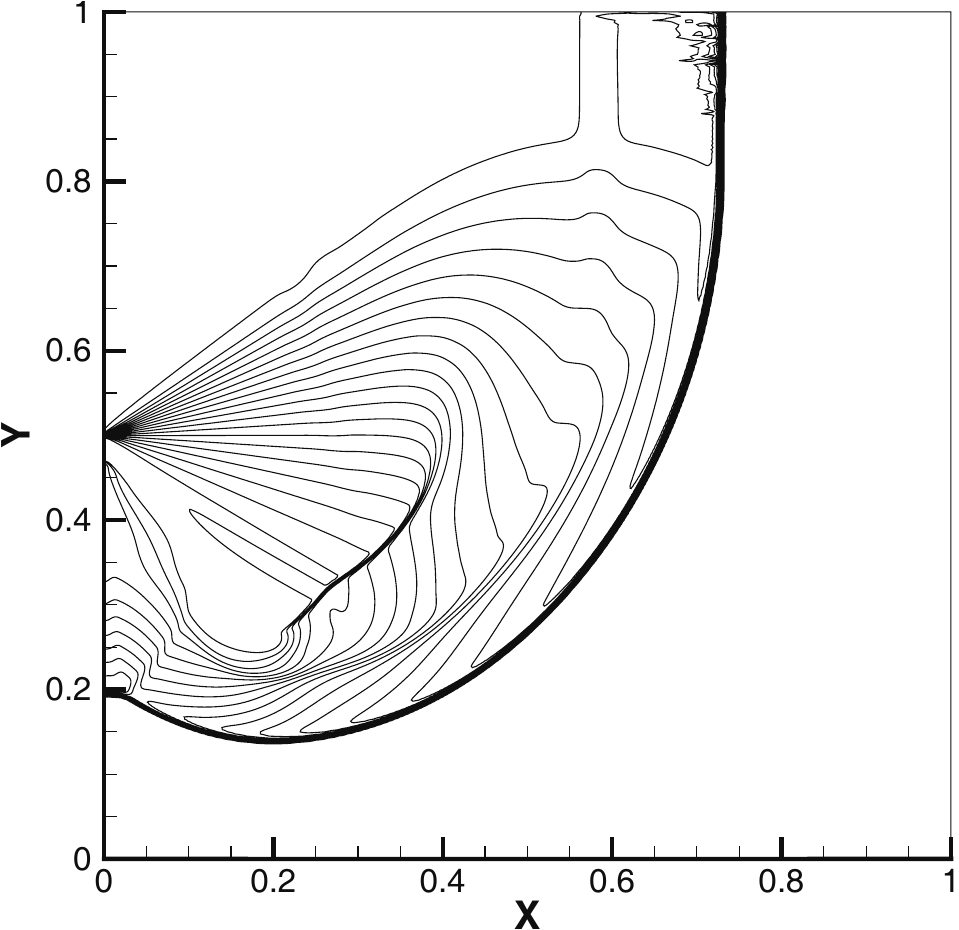}
		\end{minipage}
	}
	\subfigure[HLLC-EC]{
		\begin{minipage}[b]{0.35\textwidth}
			\includegraphics[width=1.9in]{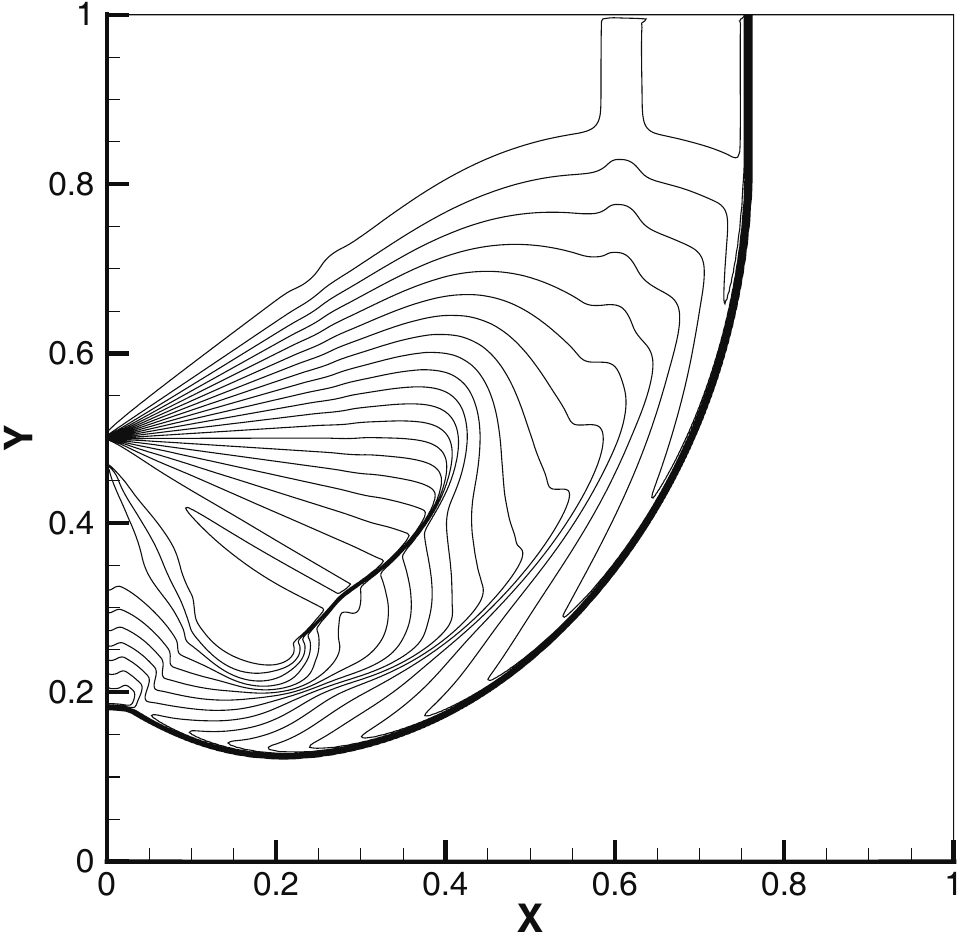}
		\end{minipage}
  }
	\caption{Shock diffraction: density contours at $t=0.15$.}
	\label{fig6-2}
\end{figure}

The problem consists of the diffraction of a plane normal shock wave moving at the Mach number 5.09 around a $90^{\circ}$ corner.  The computational domain is a unit square $\left[ {0,1} \right] \times \left[ {0,1} \right]$ that is discretized into $400\times400$  uniform cells. The corner is located at the midpoint of the left boundary: the lower half is treated as a wall; the top half is taken as an inflow. The shock is initially located at the midpoint of the left boundary. To the right of the shock, the domain is initialized to pre shock conditions of $\rho =1.4$, $p=1$, $u=0$ and $v=0$. The domain to the left of the shock is initialized to post shock conditions. The inflow boundary condition is imposed at the left boundary and the slip boundary condition is imposed at the corner and the top boundary. The right and bottom boundaries are taken as outflow. The simulations are conducted by 2nd order accurate schemes with the CFL value of 0.5. The plots shown in Fig. \ref{fig6-2} have been generated at time $t=0.15$. A total of 20 density contour levels varying from 0.5 to 5.0 have been illustrated. As shown, solutions computed by different original flux functions all exhibit obvious post-shock oscillations. Whereas, with the proposed entropy-control technique, these post-shock oscillations are completely eliminated and the instabilities are not observed. We remark that the first-order accurate numerical results show the similar phenomena.

\subsection{Double Mach reflection problem}
\label{sec6.3}
\begin{figure}[htbp]
  \centering  %图片全局居中
  \subfigure[Godunov]{
  \label{fig6.3.1}
  \includegraphics[width=0.48\textwidth]{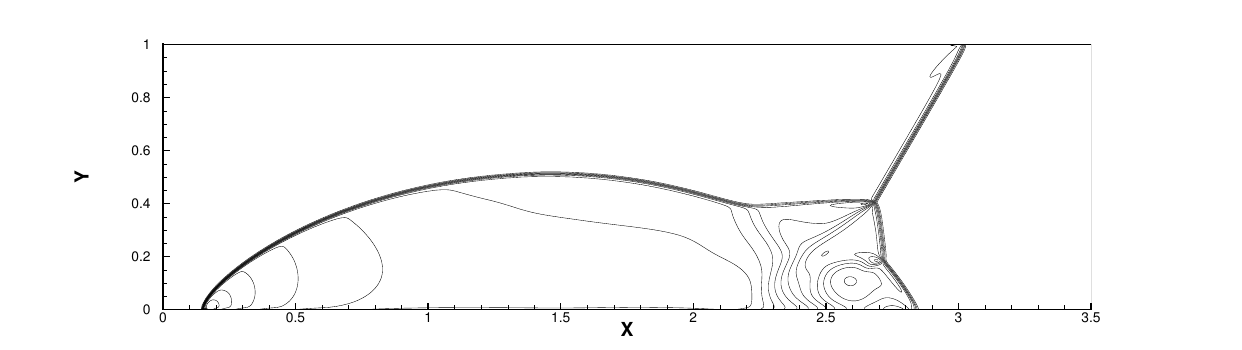}}
  \subfigure[Godunov-EC]{
  \label{fig6.3.2}
  \includegraphics[width=0.48\textwidth]{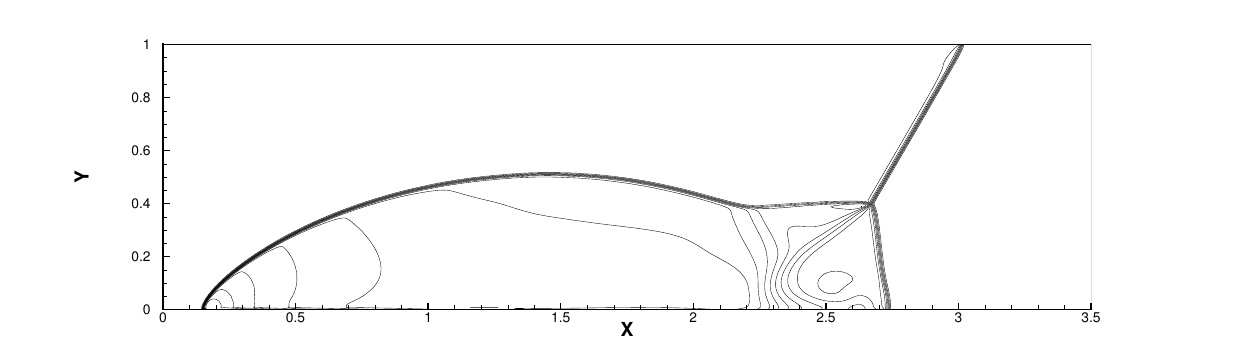}}
  \subfigure[Roe]{
  \label{fig6.3.3}
  \includegraphics[width=0.48\textwidth]{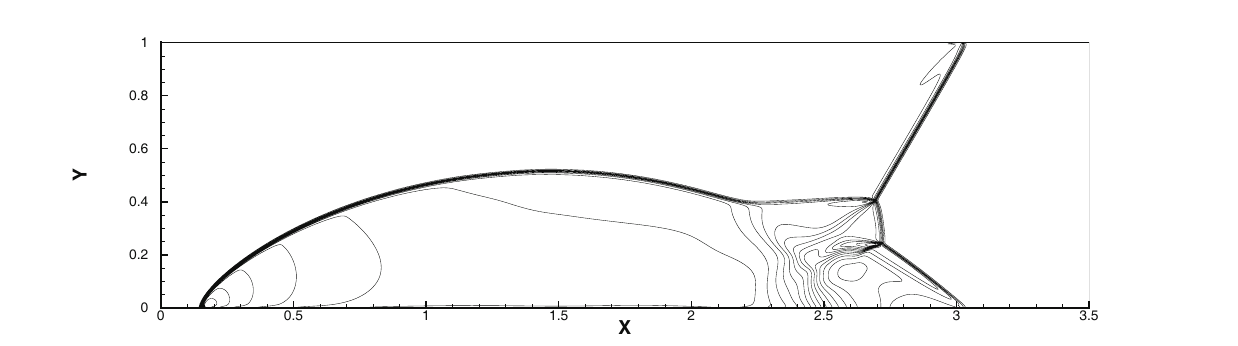}}
  \subfigure[Roe-EC]{
  \label{fig6.3.4}
  \includegraphics[width=0.48\textwidth]{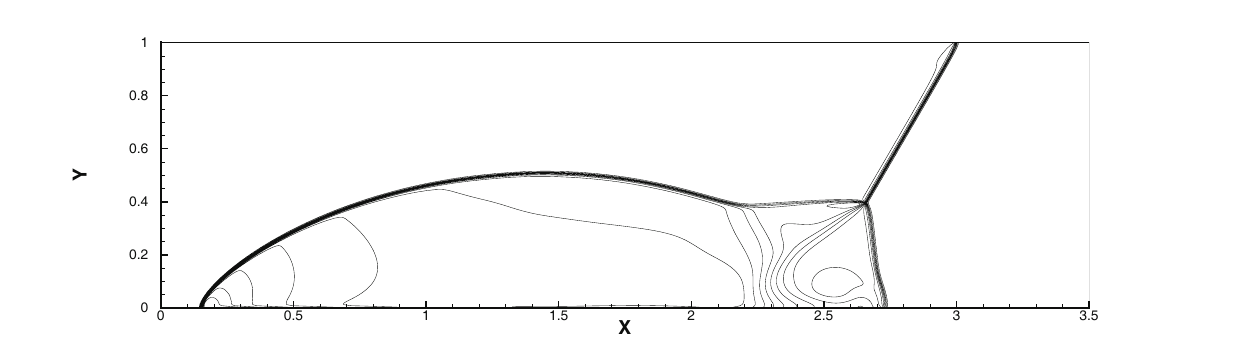}}
  \subfigure[HLLEM]{
  \label{fig6.3.5}
  \includegraphics[width=0.48\textwidth]{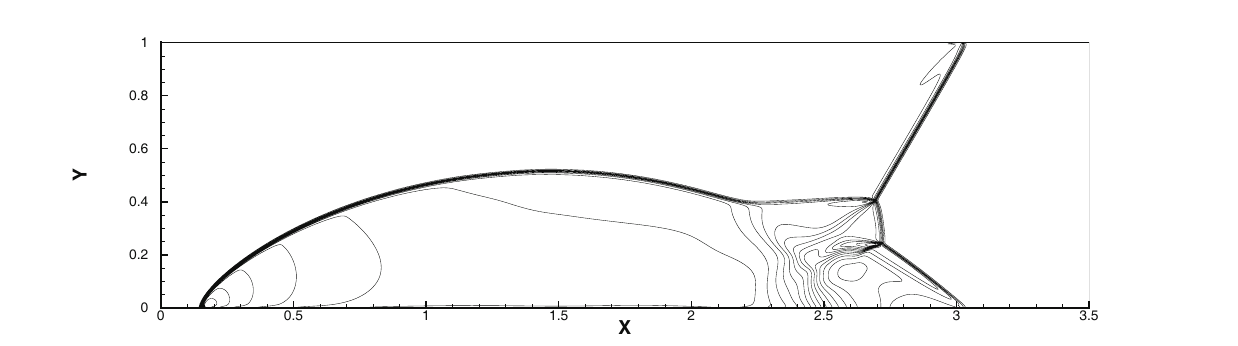}}
  \subfigure[HLLEM-EC]{
  \label{fig6.3.6}
  \includegraphics[width=0.48\textwidth]{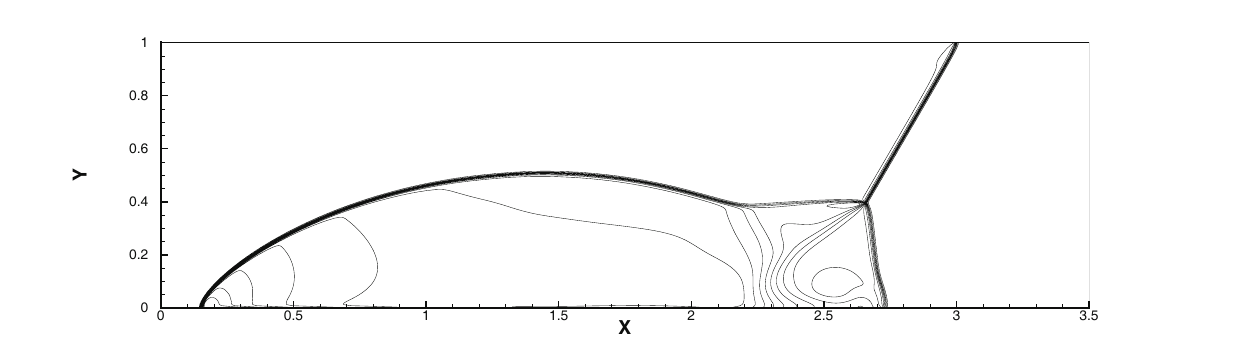}}
  \subfigure[HLLC]{
  \label{fig6.3.7}
  \includegraphics[width=0.48\textwidth]{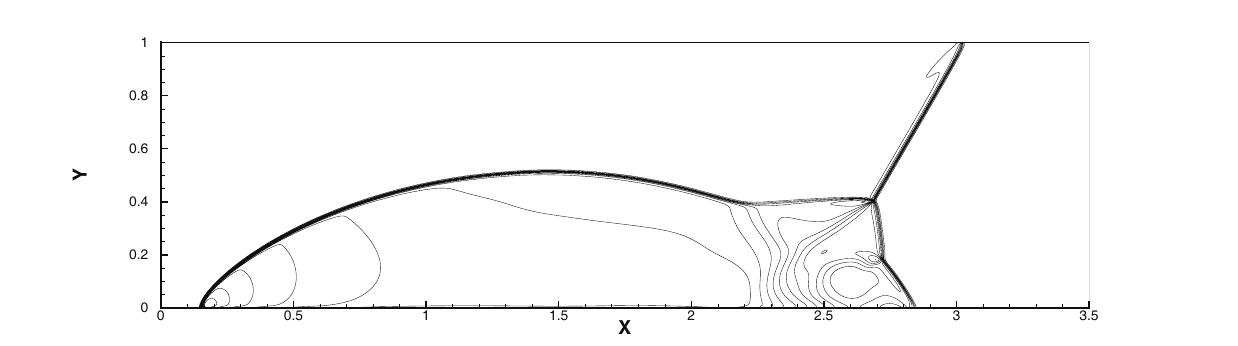}}
  \subfigure[HLLC-EC]{
  \label{fig6.3.8}
  \includegraphics[width=0.48\textwidth]{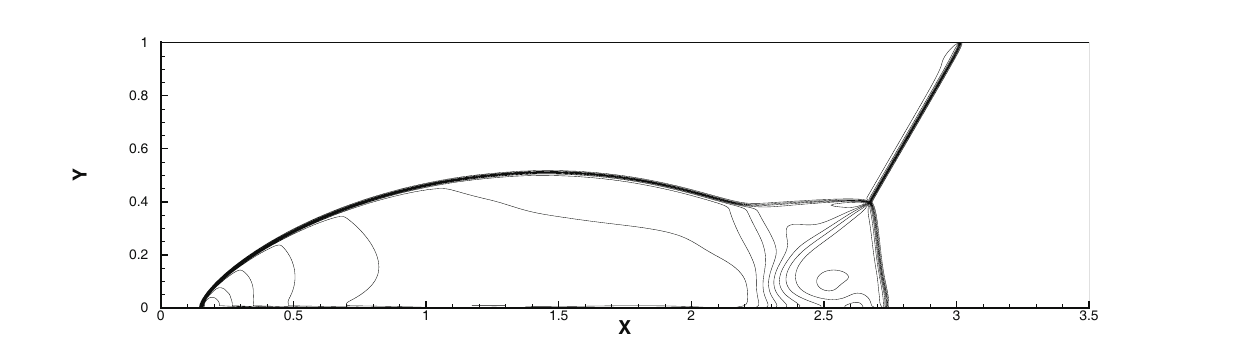}}
  \caption{Double Mach reflection: density contours at $t=0.2$.}
  \label{fig6-3}
\end{figure}

It is well-known that many low diffusion Riemann solvers produce kinked Mach stems in the double Mach reflection problem. This test case was first studied by Woodward and Colella \cite{Woodward1984The}, and later by other scholars to test the performance of numerical schemes. The computational domain is $\left[ {0,\;4} \right] \times \left[ {0,\;1} \right]$ which has been divided into 960 cells along the length and 240 cells along the width. The shock with a Mach number of 10 is initially set up to be inclined at an angle of 60 with the bottom reflecting wall. The domain in front of the shock is initialized with pre shock values given as $\rho=1.4$, $u = 0$, $v = 0$, $p = 1$. The domain behind the shock is initialized to post shock values. The computations are performed by first-order numerical schemes and the third-order TVD Runge-Kutta time discretization \cite{SHU1988439} with $\rm{CFL}=0.5$ up to $t=0.2$. The density contours computed by different schemes are shown in Fig. \ref{fig6-3}, where 20 contour levels varying from 2.0 to 20.0 are used.

As shown, all the original methods produce the unphysical triple points, i.e., the kinked Mach stems. With the proposed entropy-control technique, these modified schemes are all able to resolve shocks without any irregularities and the kinked Mach stems are barely noticeable.

\subsection{Hypersonic flow over a cylinder}
\label{sec6.4}
The simulation of hypersonic inviscid flow over a cylinder is a typical problem to assess the performance of a scheme with respect to the carbuncle. This test problem is computed on a quadrilateral grid with 60 cells in the radial direction, 280 cells in the circumferential direction. The freestream Mach number is 20. The computational domain has been initialized with values $\rho=1.4$, $p=1.4$, $u=20$ and $v=0$. The simulations are conducted by using different flux functions, along with the LU-SGS approach \cite{yoon1988lower}. Computations are conducted for 50,000 time steps with $\rm{CFL}=0.5$.

\begin{figure}
  \centering
  \subfigure[Godunov]{
		\begin{minipage}[b]{0.23\textwidth}
			\includegraphics[width=0.5\textwidth]{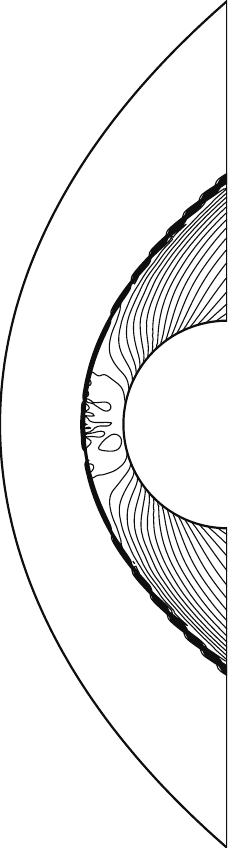}
		\end{minipage}
	}
	\subfigure[Godunov-EC]{
		\begin{minipage}[b]{0.23\textwidth}
			\includegraphics[width=0.5\textwidth]{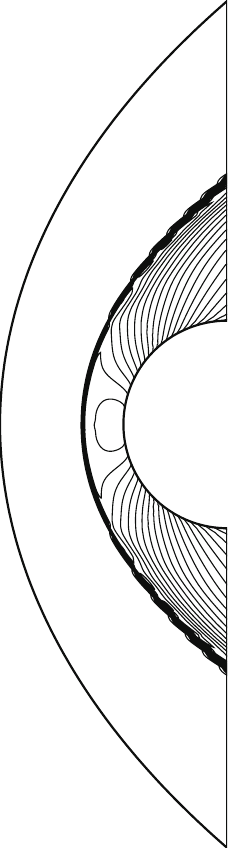}
		\end{minipage}
  }
	\subfigure[Roe]{
		\begin{minipage}[b]{0.23\textwidth}
			\includegraphics[width=0.5\textwidth]{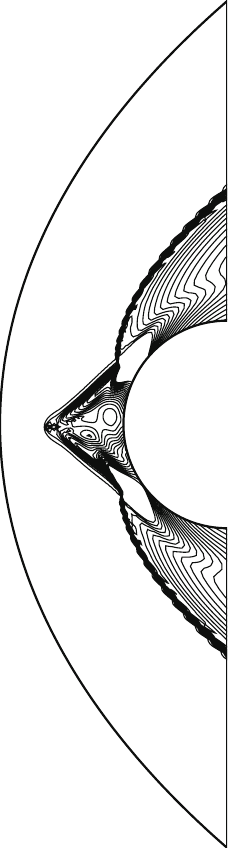}
		\end{minipage}
	}
	\subfigure[Roe-EC]{
		\begin{minipage}[b]{0.23\textwidth}
			\includegraphics[width=0.5\textwidth]{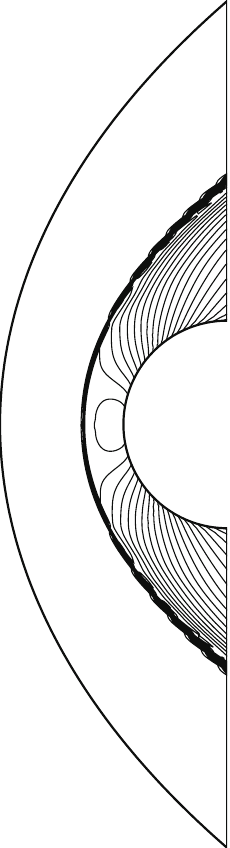}
		\end{minipage}
  }
	\subfigure[HLLEM]{
		\begin{minipage}[b]{0.23\textwidth}
			\includegraphics[width=0.5\textwidth]{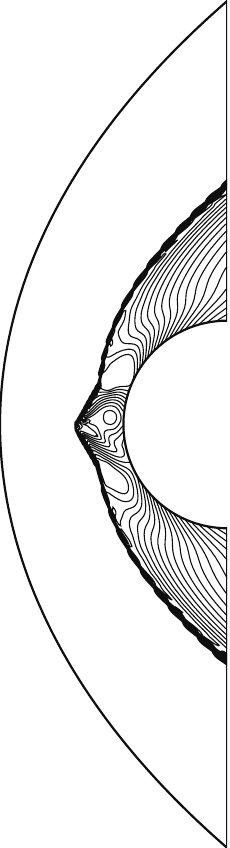}
		\end{minipage}
	}
	\subfigure[HLLEM-EC]{
		\begin{minipage}[b]{0.23\textwidth}
			\includegraphics[width=0.5\textwidth]{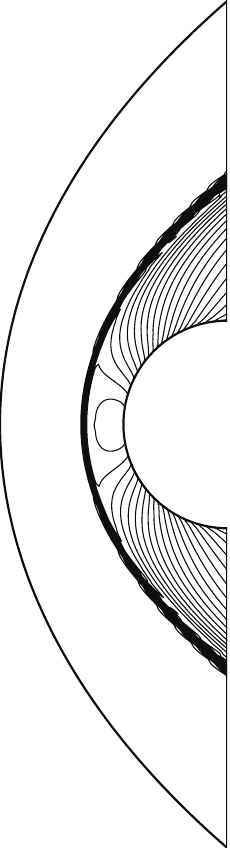}
		\end{minipage}
  }
	\subfigure[HLLC]{
		\begin{minipage}[b]{0.23\textwidth}
			\includegraphics[width=0.5\textwidth]{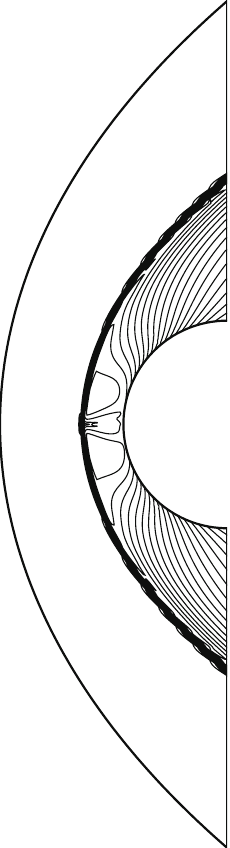}
		\end{minipage}
	}
	\subfigure[HLLC-EC]{
		\begin{minipage}[b]{0.23\textwidth}
			\includegraphics[width=0.5\textwidth]{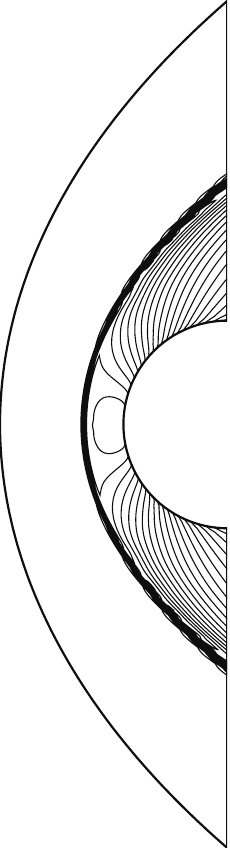}
		\end{minipage}
  }
	\caption{Density contours for hypersonic flow over a cylinder.}
	\label{fig6-4}
\end{figure}

\begin{figure}[htbp]
	\centering
	\includegraphics[width=3.5in]{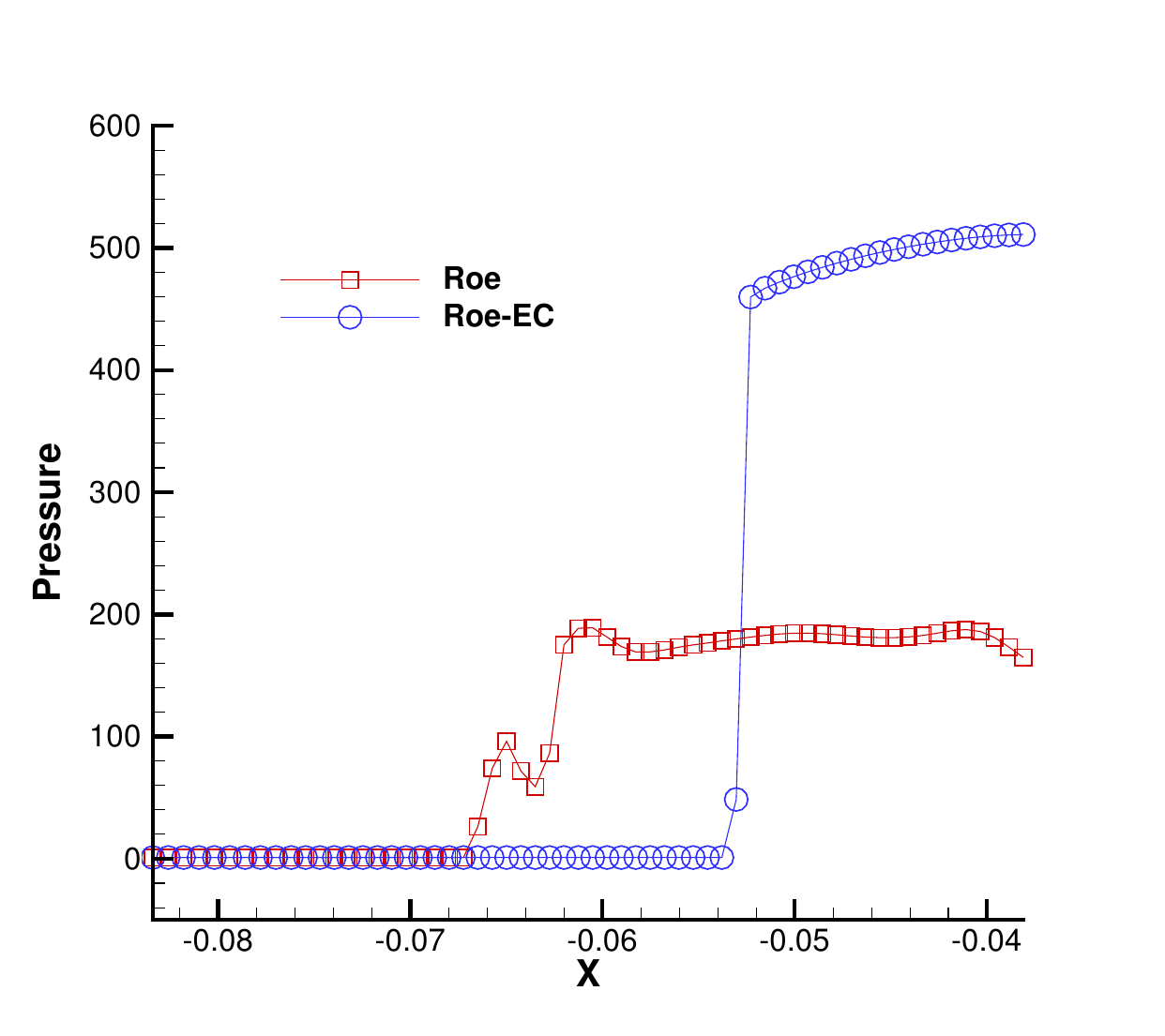}
    \caption{Pressure profiles along the stagnation line for different schemes.}
	\label{fig6-4(2)}
\end{figure}

In Fig. \ref{fig6-4}, the density contours by different approximate Riemann solvers are demonstrated, where 20 contour levels varying from 1.5 to 8.5 are used. As shown in Fig. \ref{fig6-4}, all the unmodified numerical schemes produce the carbuncle phenomena. The pressure profiles along the stagnation line for different schemes have been plotted in Fig. \ref{fig6-4(2)}. As shown, the Roe scheme does not produce a correct solution due to the carbuncle. Whereas, the Roe-EC solver is able to produce a correct pressure profile with only one interior point. For other approximate Riemann solvers, similar results are obtained but not shown for clarity.

\subsection{Flat plate boundary layer}
\label{sec6.5}
\begin{figure}[htbp]
	\centering
	\includegraphics[width=5.0in]{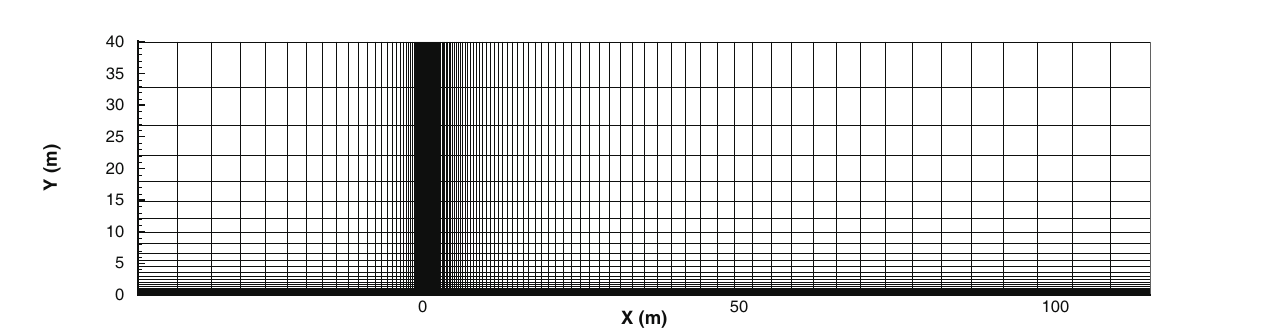}
	\caption{Mesh distribution for the laminar boundary layer computation (the minimal value of $\Delta y$ is 0.05).}
	\label{fig6-5-1}
\end{figure}

\begin{figure}[htbp]
	\centering
	\includegraphics[width=3.5in]{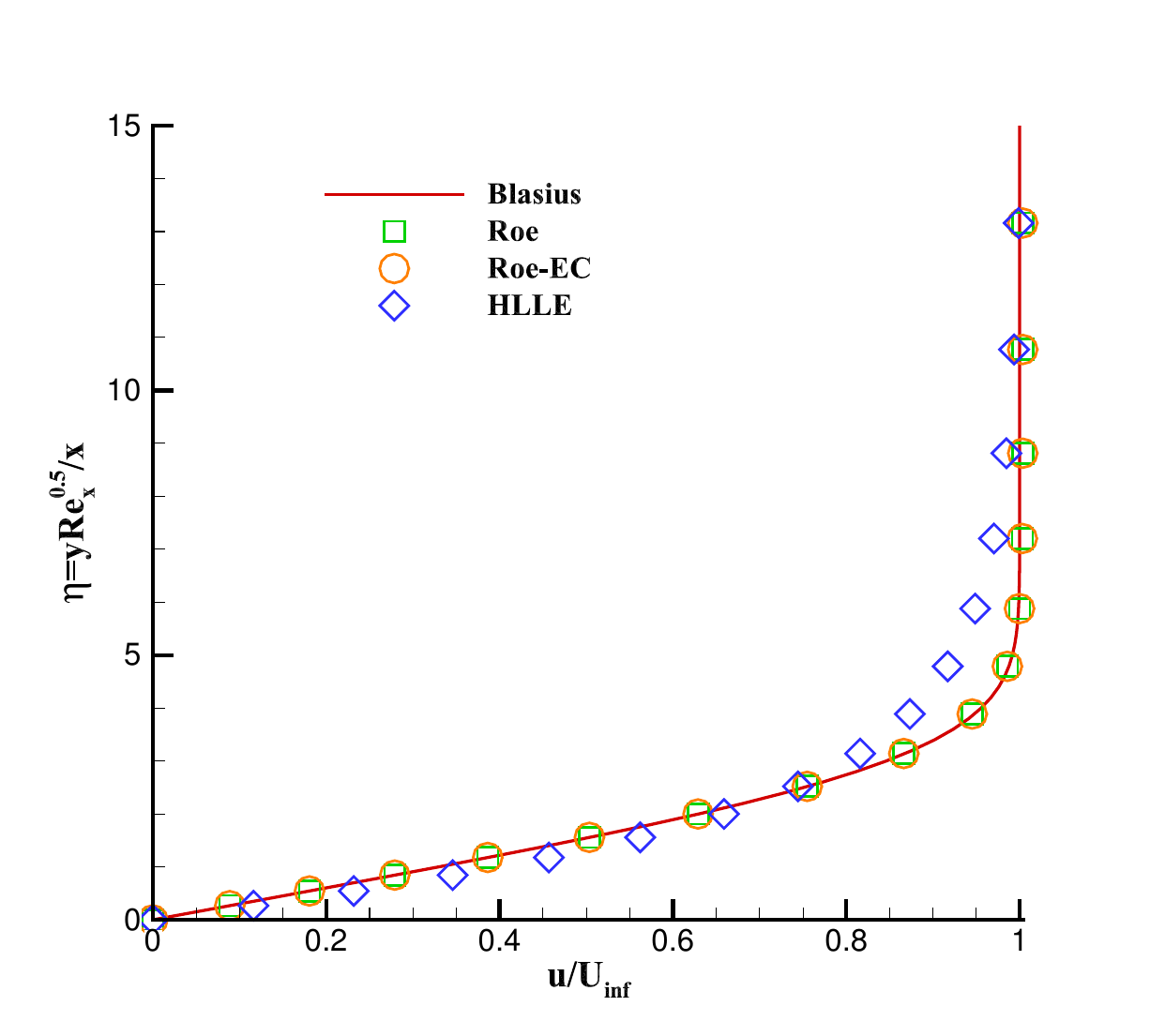}
	\caption{Velocity profiles for the flat plate boundary layer problem.}
	\label{fig6-5-2}
\end{figure}

A laminar boundary layer problem is used here to assess the ability of the proposed scheme to resolve shear layers. The Mach number of the freestream is $Ma = 0.3$  and the Reynolds number is ${\mathop{\rm Re}\nolimits}  = {10^5}$. The problem is computed by a second-order Navier-Stokes code. In Fig. \ref{fig6-5-1}, the computational domain is discretized into $120\times30$ non-uniform cells. The non-slip adiabatic boundary condition is imposed at the plate and the bottom boundary before the flat plate is treated as a symmetry plane. The non-reflecting boundary condition based on the Riemann invariants is adopted for the other boundaries. The computations were conducted for 50,000 steps with $\rm{CFL}=0.5$ , and all the computations achieved at least three orders of magnitude reductions of the density residuals. In Fig. \ref{fig6-5-2}, velocity profiles computed by different schemes are compared. As shown, the Roe scheme with an entropy-control technique gives an almost identical result with that of the original Roe flux. The solutions match with the exact Blasius solutions very well. The entropy-control term does not decrease the accuracy of numerical schemes for resolving shear layers. Other flux functions give the same results, but not shown for clarity. Results computed by the HLLE scheme are also included for comparison. It can be observed that it gives a  dissipative and inaccurate solution.

\subsection{Elling's physical carbuncle problem}
\label{sec6.6}
\begin{figure}[htbp]
	\centering
	\includegraphics[width=2.8in]{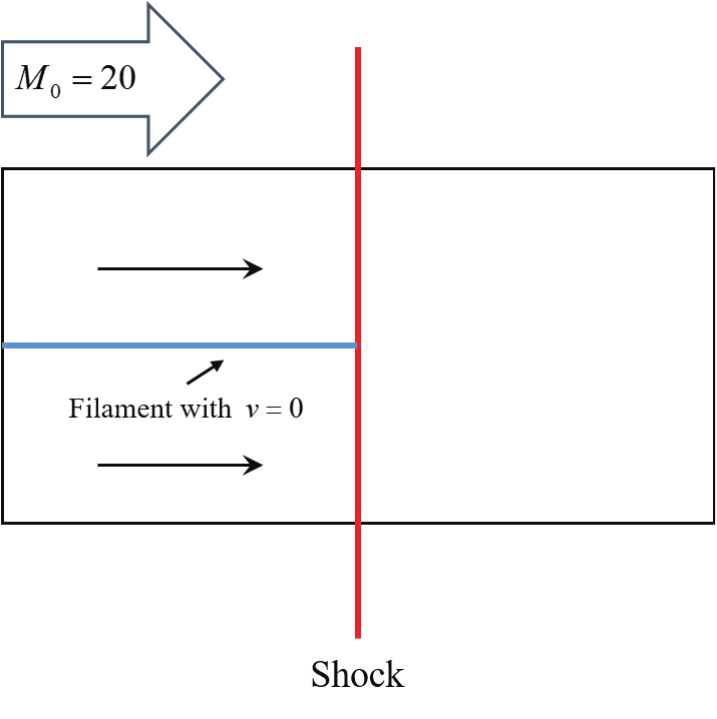}
	\caption{Numerical setup for Elling's test}
	\label{fig6-6-1}
\end{figure}

The last test case considered is a challenging problem. It is first proposed by Elling \cite{elling2009carbuncle} and later studied by other scholars \cite{kemm2018heuristical} to test numerical schemes to resolve physical carbuncle problem. In references \cite{elling2005nonuniqueness,elling2009carbuncle}, the author proposes that carbuncles may be a special class of entropy solutions which can be physically correct in some circumstances. The description of this problem is presented in Fig. \ref{fig6-6-1}. It is set up to model the interaction of a vortex filament with a strong shock. Numerical schemes that are able to avoid unphysical carbuncles must also allow this physical carbuncle-like structure. Thus, it is a suitable test case to access whether the improvement for carbuncles will reduce the accuracy of numerical methods on resolving shear layers.

\begin{figure}[htbp]
  \centering  %图片全局居中
  \subfigure[Godunov]{
  \label{fig6.6.7}
  \includegraphics[width=0.48\textwidth]{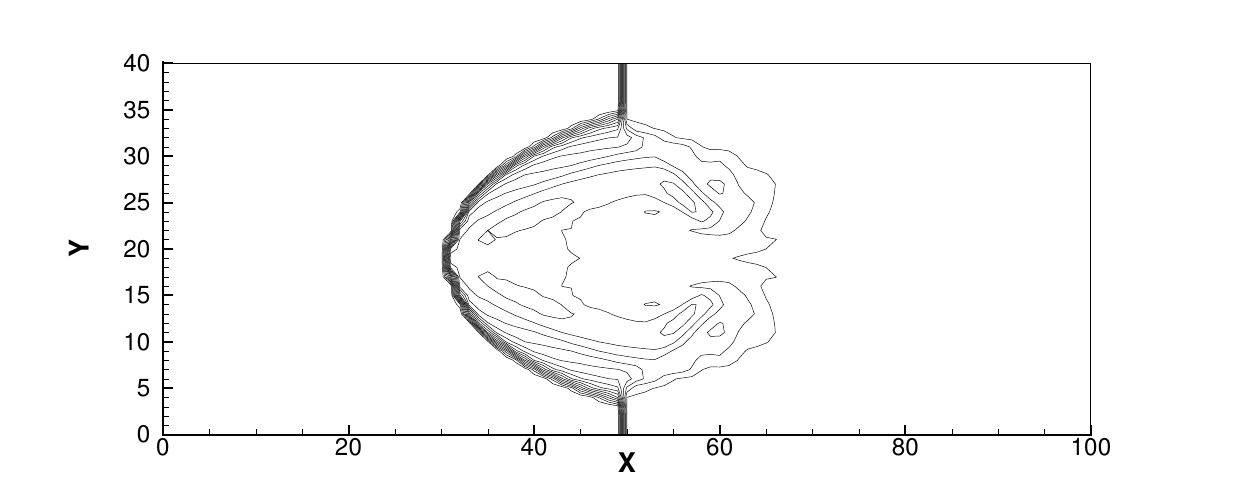}}
  \subfigure[Godunov-EC]{
  \label{fig6.6.8}
  \includegraphics[width=0.48\textwidth]{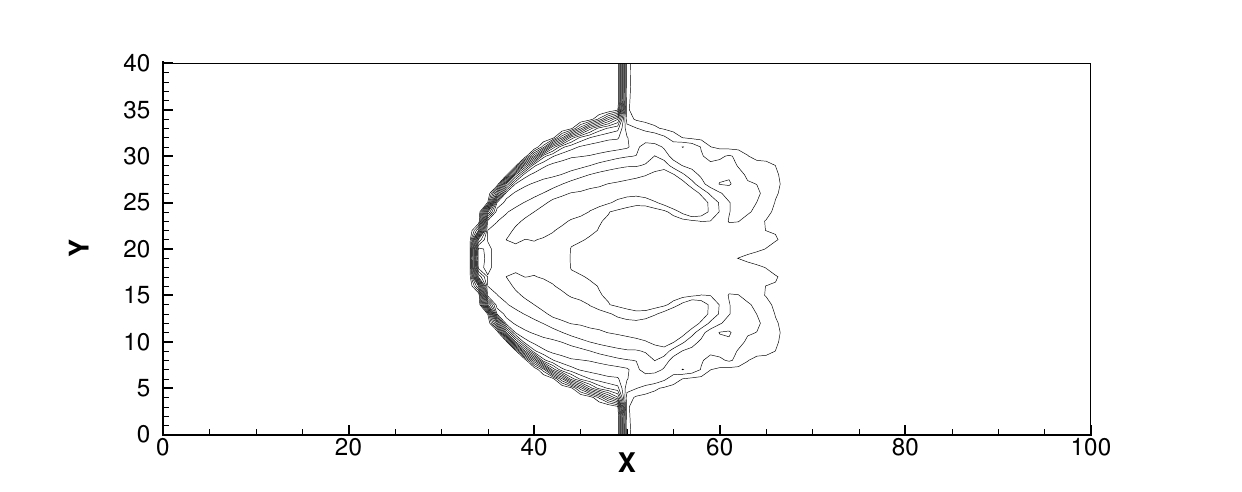}}
  \subfigure[Roe]{
  \label{fig6.6.1}
  \includegraphics[width=0.48\textwidth]{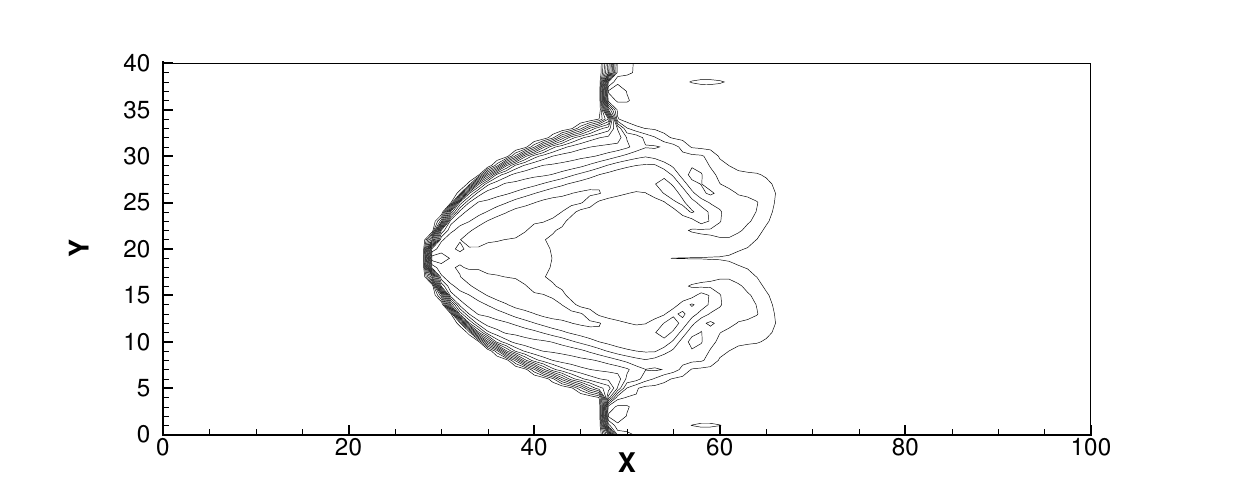}}
  \subfigure[Roe-EC]{
  \label{fig6.6.2}
  \includegraphics[width=0.48\textwidth]{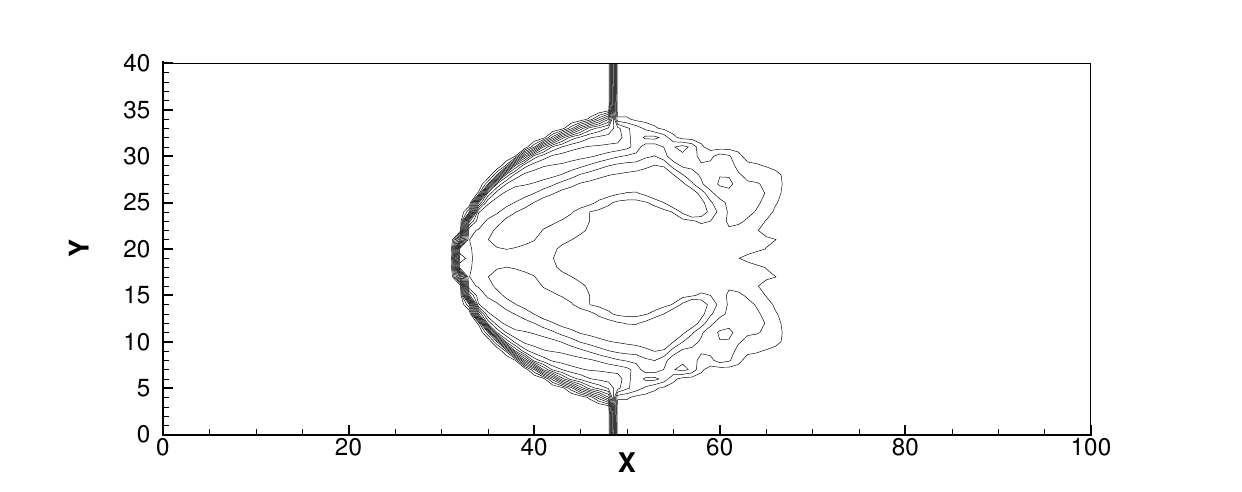}}
  \subfigure[HLLEM]{
  \label{fig6.6.3}
  \includegraphics[width=0.48\textwidth]{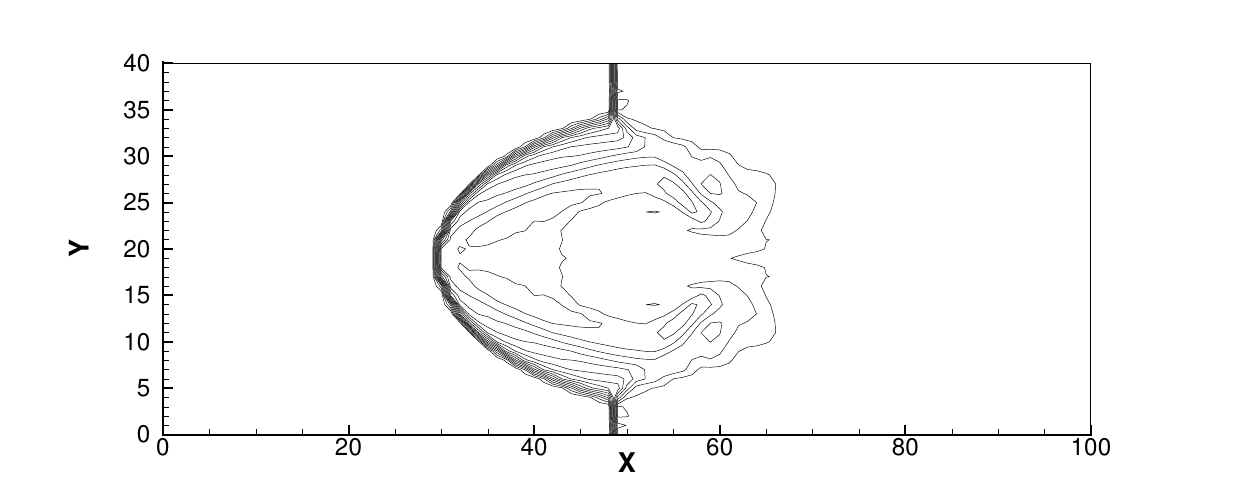}}
  \subfigure[HLLEM-EC]{
  \label{fig6.6.4}
  \includegraphics[width=0.48\textwidth]{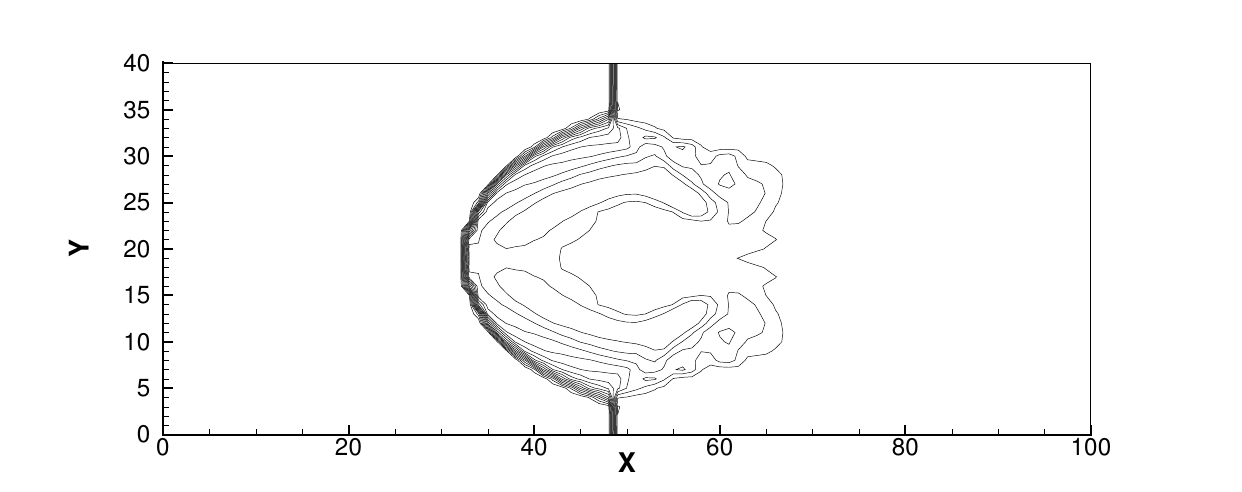}}
  \subfigure[HLLC]{
  \label{fig6.6.5}
  \includegraphics[width=0.48\textwidth]{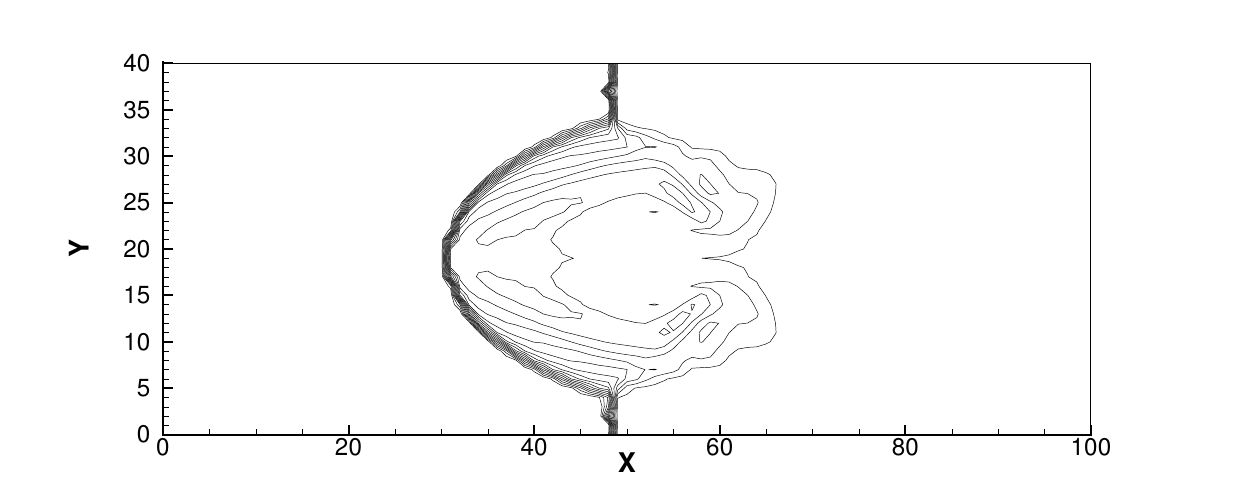}}
  \subfigure[HLLC-EC]{
  \label{fig6.6.6}
  \includegraphics[width=0.48\textwidth]{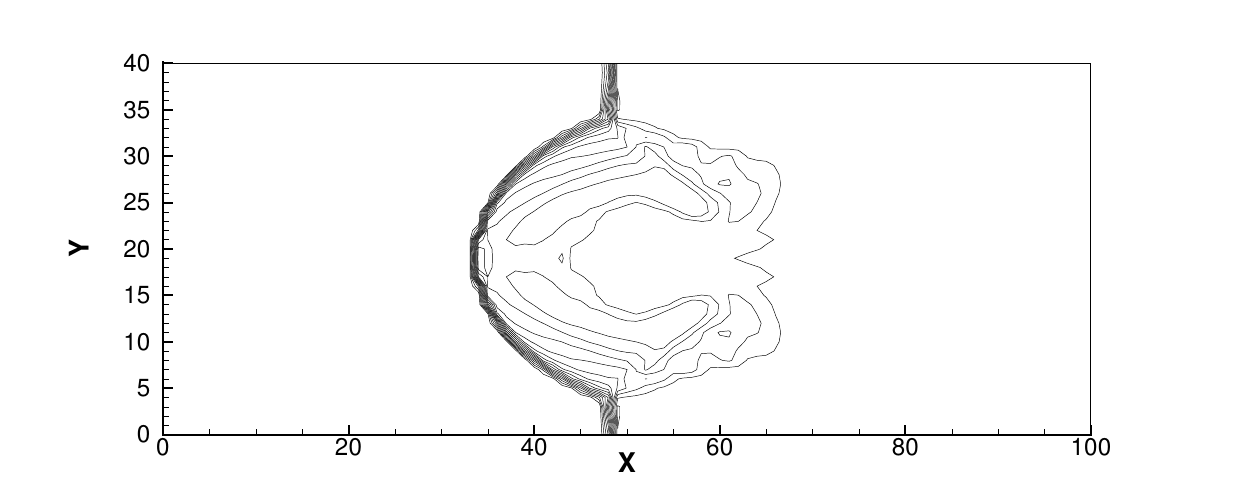}}
  \subfigure[HLLE]{
  \label{fig6.6.9}
  \includegraphics[width=0.48\textwidth]{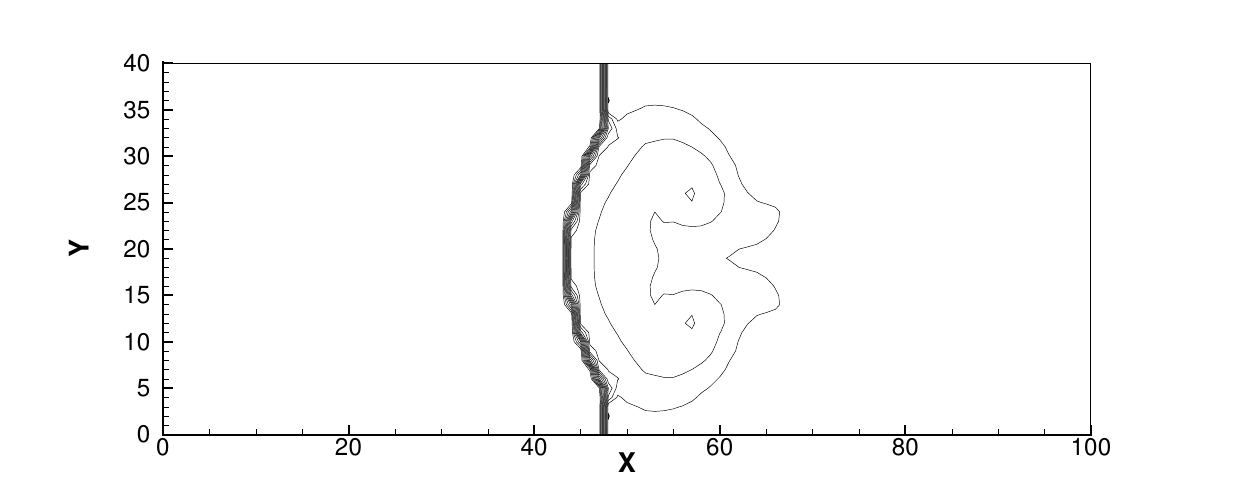}}
  \subfigure[Roe-HLLE]{
  \label{fig6.6.10}
  \includegraphics[width=0.48\textwidth]{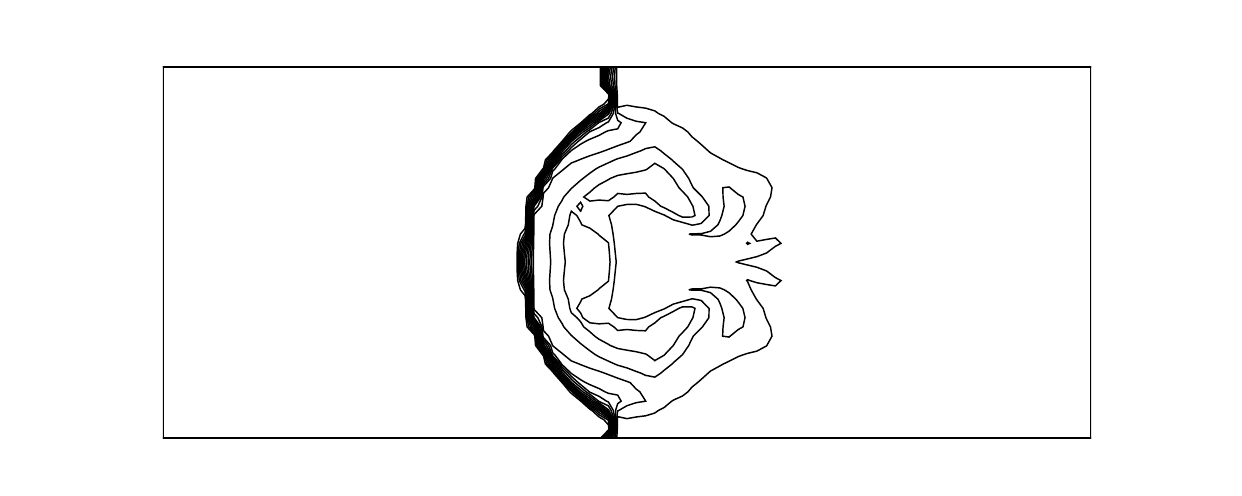}}
  \caption{Entropy contours computed by different solvers at $t=100$.}
  \label{fig6-6-2}
\end{figure}

Following Kemm \cite{kemm2018heuristical}, this case is tested in the computational domain $[0,100]\times[0,4]$  which has been divided into 100 cells along the length and 40 cells along the width. The shock of a Mach number $M_0=20$ is located at a cell interface. The domain pre to the shock is initialized with $\rho=1$, $u=1$ , $v=0$, $p=\frac{1}{\gamma M_0^2}$. The post shock condition is calculated by the Rankine-Hugoniot relations. In the supersonic inflow region, only the middle slice of the computational grid is changed and the velocity is artificially set to zero. Computations are performed by second-order accurate numerical schemes using the third-order TVD Runge-Kutta time discretization \cite{SHU1988439} with $\rm{CFL}=0.2$ up to $t=100$. In Fig. \ref{fig6-6-2}, the entropy contours $s=\log(p/{\rho}^{\gamma})$ computed by different approximate Riemann solvers are demonstrated, where 15 contour levels varying from -6.0 to -2.5 are used. As shown, the numerical results computed by Godunov's exact Riemann solver, Roe, HLLEM and HLLC are rather similar. With their modified versions, the physical carbuncles are properly reproduced. This indicates that the proposed entropy-control technique involves minimal diffusion on shear layers. Whereas, the HLLE scheme smears the physical carbuncle severely, demonstrating its excessive amount of shear viscosity. The last picture presents the result computed by the hybrid solver Roe-HLLE suggested in \cite{quirk1994contribution}. This type of numerical methods usually apply a switching function to combine a complete Riemann solver (e.g. Roe scheme, HLLC scheme, etc.) with an incomplete Riemann solver (e.g. HLLE scheme, van Leer scheme, etc.) to eliminate shock instabilities while maintaining high resolution in smooth regions. As shown, the physical carbuncle is also destroyed. This indicates that hybrid approximate Riemann solvers may lose their high resolution in the vicinity of shock-vortex interactions.

\section{Conclusion}
\label{sec7}
In the current study, we devote our efforts to developing a general approach to improve the robustness of Godunov-type schemes at strong shocks without compromising their accuracy on discontinuities. To this end, a shock instability analysis has been carried out by combining two interlinked techniques: dissipation analysis of Godunov-type schemes and numerical experiments. Results of the stability analysis demonstrate that the numerical shock instability of Godunov-type schemes is due to inappropriate entropy production inside the numerical shock structure. Moreover, if enough entropy production is guaranteed inside shocks, then the instability problem can be successfully eliminated. Such results allow to propose a general approach that can be applied to suppressing numerical shock instabilities of various Godunov-type schemes. The essential component of this approach is an entropy-control term that plays a role in increasing the entropy production in the numerical shock structure. A further linearized analysis illustrates that this entropy-control term is equivalent to reducing mass flux perturbations behind shocks, which is found to trigger the shock instability in the view of perturbations. Several stringent shock wave problems are conducted to indicate that the proposed entropy-control approach can effectively improve the robustness of low diffusion Godunov-type schemes at strong shocks. Meanwhile, comparisons of computed solutions from a boundary layer problem and a shock-vortex interaction problem indicate that these modified methods preserve the accuracy of the original solvers on resolving shear layers.

The proposed entropy-control technique is expected to also applied to approximations to other hyperbolic systems of conservation laws, such as the shallow water equations and the MHD equations. Extensions to cases of high-order schemes will also be considered in further investigations.

\section*{Acknowledgements}	
This work was supported by the National Natural Science Foundation of China (11472004), the Foundation of Innovation of NUDT (B150106).

\section*{Appendix A}
Here the modified equation for the HLLEM-$\rho$ scheme is presented. As illustrated in section {\ref{sec3.3.1}}, the modified equation of the HLLEM-${\rho}$ scheme can be derived if its numerical viscosity coefficient matrix is determined. The flux function of the HLLEM-${\rho}$ can be expressed as
\begin{equation}\label{A.1}
  {\bf{F}}_{{\rm{HLLEM}}-{\rho}} = \frac{1}{2}\left( {{{\bf{F}}_L} + {{\bf{F}}_R}} \right) - \frac{1}{2}{ {\bf{Q}}_{i + 1/2}}\left( {{{\bf{U}}_R} - {{\bf{U}}_L}} \right)
\end{equation}
where the numerical viscosity coefficient matrix ${\bf{Q}}_{i + 1/2}$ is defined by
\begin{equation}\label{A.2}
  {\bf{Q}}_{i + 1/2}=\frac{S_L+S_R}{S_R-S_L}\widehat{{\bf{A}}}-\frac{2{S_L}{S_R}}{S_R-S_L}{\bf{I}}+\frac{2{S_L}{S_R}}{S_R-S_L}{\hat{\delta}}_2\widehat{\bf{R}}{\bf{T}}\widehat{\bf{R}}^{-1}-\frac{2{S_L}{S_R}}{S_R-S_L}{\hat{\delta}}_2\widehat{\bf{R}}{\bf{T}}\widehat{\bf{R}}^{*,-1}
\end{equation}
with the following relation
\begin{equation}\label{A.3}
  \widehat{\alpha} _2^*{{\widehat{\bf{R}}}_2} = \widehat{\bf{R}}{\bf{T}}\widehat{\bf{R}}^{*,-1} \cdot \left( {{{\bf{U}}_{R}} - {{\bf{U}}_{L}}} \right).
\end{equation}
In Eq. (\ref{A.3}), $\widehat{\alpha}_2^*$ denotes the modified wave strength corresponding to entropy waves and is defined in Eq. (\ref{eq3.3.2.2}), $\widehat{\bf{R}}^{*,-1}$ is determined by Eq. (\ref{A.3}).

The modified equation corresponding to the HLLEM-$\rho$ scheme can be written as
\begin{equation}\label{A.4}
  {\bf{U}}_t + {\bf{F}\left({{\bf{U}}}\right)}_x = \frac{1}{2} { {\bf{Q}}_{{\rm{HLLEM}}-{\rho}}\left({{\bf{U}}}\right) }  {{\bf{U}}_{xx}} \Delta x
\end{equation}
where the viscosity matrix can be obtained as
\begin{equation}\label{A.5}
  {\bf{Q}}_{{\rm{HLLEM}}-{\rho}} \left( {\bf{U}} \right) = \frac{{{S_R} + {S_L}}}{{{S_R} - {S_L}}} {\bf{A}} \left( {\bf{U}} \right) - \frac{{2{S_L}{S_R}}}{{{S_R} - {S_L}}}{\bf{I}} + \frac{{2{S_R}{S_L}}}{{{S_R} - {S_L}}} {\delta _2} {\bf{R}}\left( {\bf{U}} \right){\bf{T}}{\bf{R}}^{-1}\left( {\bf{U}} \right)-\frac{2{S_L}{S_R}}{S_R-S_L}{\delta}_2{\bf{R}} \left( {\bf{U}} \right) {\bf{T}}{\bf{R}}^{*,-1} \left( {\bf{U}} \right).
\end{equation}
The modified equation in characteristic variables can be obtained following the procedure presented in Eq. (\ref{eq3.3.1.12}), it can be written as
\begin{equation}\label{A.6}
  {{\bf{W}}_t} + \Lambda {{\bf{W}}_x} = \frac{1}{2}{\bf{V}}{{\bf{W}}_{xx}}\Delta x + (f_{\rho} - 1) \frac{{{S_{\rm{L}}}{S_R}}}{{{S_{\rm{R}}} - {S_{\rm{L}}}}}{\delta _2}{\bf{T}} {\bf{W}}_{xx}^1 \Delta x ,
\end{equation}
where the viscosity coefficient matrix $\bf{V}$ is defined in (\ref{eq3.3.1.15}), the characteristic variables $\rm{d}{\bf{W}}^1$ and $\rm{d}{\bf{W}}^2$ are defined in Eq. (\ref{eq3.3.2.11a}) and Eq. (\ref{eq3.3.2.11b}).

To access the numerical viscosity corresponding to the entropy wave, we need to consider the following modified equation for the entropy wave ${\rm{d}}w_2$,
\begin{equation}\label{A.7}
  \frac{\partial w_2 }{\partial t} + {\lambda}_2 \frac{\partial w_2 }{\partial x} = \frac{1}{2} \mu_2 \frac{\partial^2}{\partial x^2} w_2 \Delta x+(f_{\rho}-1) \frac{{{S_L}{S_R}}}{{{S_R} - {S_L}}}{\delta _2} \frac{\partial^2}{\partial x^2} w_2^1 \Delta x,
\end{equation}
where the numerical viscosity coefficient  $\mu_2$ is defined in Eq. (\ref{eq3.3.1.17}). Following Eq. (\ref{eq3.3.2.11a}) and (\ref{eq3.3.2.11b}), the entropy wave ${\rm{d}}w_2$ can be splitted into the following form,
\begin{equation}\label{A.8}
  {\rm{d}}w_2={\rm{d}}w_2^1-{\rm{d}}w_2^2
\end{equation}
with
\begin{equation}\label{A.9}
{\rm{d}}w_2^1 = {{\rm{d}}\rho }, \quad
{\rm{d}}w_2^2 = {\frac{{{\rm{d}}p}}{{{a^2}}}} .
\end{equation}
Considering (\ref{A.8}) and (\ref{A.9}), Eq. (\ref{A.7}) can be splitted into the following subsystems,
\begin{align}
  \frac{{\partial w_2^1}}{{\partial t}} + {\lambda _2}\frac{{\partial w_2^1}}{{\partial x}} & = \frac{1}{2}{\mu _2}\frac{{{\partial ^2}}}{{\partial {x^2}}}w_2^1\Delta x + \left( {f_{\rho}} - 1 \right)\frac{{{S_L}{S_R}}}{{{S_R} - {S_L}}}{\delta _2}\frac{{{\partial ^2}}}{{\partial {x^2}}}w_2^1\Delta x, \label{A.10} \\
  \frac{{\partial w_2^2}}{{\partial t}} + {\lambda _2}\frac{{\partial w_2^2}}{{\partial x}} &= \frac{1}{2}{\mu _2}\frac{{{\partial ^2}}}{{\partial {x^2}}}w_2^2\Delta x . \label{A.11}
\end{align}
Due to the nonpositivity of the last term at the right side of Eq. (\ref{A.10}), a larger $f_{\rho}$ will lead to less diffusion on ${\rm{d}}w_2^1$. Combing with the relation (\ref{A.8}), an increased entropy will be obtained.

\section*{Appendix B}
Here we give a detailed description of how to obtain the formulas presented in Eq. (\ref{eq4.2.7}) and Eq. (\ref{eq4.2.8}). At time $t^n$, it is assumed that the states along the $y$ direction inside the shock structure are distributed in a saw-toothed manner. These solutions are given by
\begin{equation}\label{B.1}
\rho_{j}^n=\rho_{i,j}^{*,n}-\delta {\rho}^n, \quad \left( {\rho u} \right)_{i,j}^n = {\left( {\rho u} \right)_{i,j}^{*,n}}-\delta ({\rho u})^n, \quad \left( {\rho v} \right)_{i,j}^n = {\left( {\rho v} \right)_{i,j}^{*,n}} - \delta ({\rho v})^n, \quad p_{i,j}^n = {p_{i,j}^{*,n}} - \delta p^n,
\end{equation}
and
\begin{equation}\label{B.2}
  \rho _{i,j \pm 1}^n = \rho_{i,j \pm 1}^{*,n} + \delta {\rho}^n, \quad \left( {\rho u} \right)_{i,j \pm 1}^n = {\left( {\rho u} \right)_{i,j \pm 1}^{*,n}}+\delta ({\rho u})^n, \quad \left( {\rho v} \right)_{i,j \pm 1}^n = {\left( {\rho v} \right)_{i,j \pm 1}^{*,n}} + \delta ({\rho v})^n, \quad p_{i,j \pm 1}^n = {p_{i,j \pm 1}^{*,n}} + \delta {p^n},
\end{equation}
where ${()}^{*}$ represent the stable steady solutions. In what follows, we omit the subscript $i$ for clarity. In the two-dimensional case, we need to clarify how the perturbations will promote the perturbed mass flux in the transverse direction. Hence, the following momentum equation is considered,
\begin{equation}\label{B.3}
  \left( {\rho v} \right)_j^{n + 1} = \left( {\rho v} \right)_j^n - \frac{{\Delta t}}{{\Delta y}}\left[ {\left( {\rho {v^2} + p} \right)_{j + 1/2}^n - \left( {\rho {v^2} + p} \right)_{j - 1/2}^n} \right].
\end{equation}
For the flux function HLLEM-${\rho}$, the numerical fluxes at the interfaces in (\ref{B.3}) can be written as
\begin{flalign}\label{B.4}
\begin{split}
\left( {\rho {v^2} + p} \right)_{j + 1/2}^n &= \frac{S_{R,j+1/2}}{{S_{R,j+1/2}}-{S_{L,j+1/2}}} \left( {\rho {v^2} + p} \right)_j^n - \frac{S_{L,j+1/2}}{{S_{R,j+1/2}}-{S_{L,j+1/2}}}  \left( {\rho {v^2} + p} \right)_{j + 1}^n\\
&\quad + \frac{{S_{L,j+1/2}}{S_{R,j+1/2}}}{{S_{R,j+1/2}} - {S_{L,j+1/2}}}\left[ {\left( {\rho v} \right)_{j + 1}^n - \left( {\rho v} \right)_j^n - {{\widehat \delta }_{2,j + 1/2}}\left( {{\rho _{j + 1}} - {\rho _j} - {f_p}\frac{{{p_{j + 1}} - {p_j}}}{{\widehat a_{j + 1/2}^2}}} \right){{\widehat v}_{j + 1/2}}} \right],\\
\left( {\rho {v^2} + p} \right)_{j - 1/2}^n &= \frac{S_{R,j-1/2}}{{S_{R,j-1/2}}-{S_{L,j-1/2}}} \left( {\rho {v^2} + p} \right)_{j - 1}^n - \frac{S_{L,j-1/2}}{{S_{R,j-1/2}}-{S_{L,j-1/2}}} \left( {\rho {v^2} + p} \right)_j^n\\
&\quad + \frac{{S_{L,j-1/2}}{S_{R,j-1/2}}}{{S_{R,j-1/2}} - {S_{L,j-1/2}}}  \left[ {\left( {\rho v} \right)_j^n - \left( {\rho v} \right)_{j - 1}^n - {{\widehat \delta }_{2,j - 1/2}}\left( {\rho _j^n - \rho _{j - 1}^n - {f_p}\frac{{p_j^n - p_{j - 1}^n}}{{{{\widehat a}^2}}}} \right){{\widehat v}_{j - 1/2}}} \right],
\end{split}
\end{flalign}
where ${\widehat {\left(  \cdot  \right)}_{j + 1/2}}$ denote Roe's averaged variables between states in cells $(i,j)$ and $(i,j+1)$. The Roe's averaged varibles used can be approximated as
\begin{equation}\label{B.5}
r = \sqrt {\frac{{\rho _{j + 1}^n}}{{\rho _j^n}}}  = \sqrt {\frac{{{\rho ^{*,n}} + \delta {\rho ^n}}}{{{\rho ^{*,n}} - \delta {\rho ^n}}}}  = 1 + \frac{1}{{{\rho ^{*,n}}}} \cdot \delta {\rho ^n}
\end{equation}
and
\begin{equation}\label{B.6}
{\widehat v_{j + 1/2}^n} = \frac{{v_j^n + rv_{j + 1}^n}}{{1 + r}} = \frac{{v_j^{*,n} - \delta {v^n} + r\left( {v_{j + 1}^{*,n} + \delta {v^n}} \right)}}{{1 + r}} = {v^{*,n}} + \frac{{\delta {\rho ^n}}}{{2{\rho ^{*,n}} + \delta {\rho ^n}}}\delta {v^n} \approx {v^{*,n}}
\end{equation}
where the second order and even higher order perturbation terms are suppressed in the context of linearized analysis. Similarly, the Roe's averaged sound speed can be written by
\begin{equation}\label{B.7}
  {\widehat a_{j + 1/2}^n} \approx {a^{*,n}} .
\end{equation}
Inserting Eqs. (\ref{B.1}), (\ref{B.2}) and (\ref{B.5}~\ref{B.7}) into Eqs. (\ref{B.4}), one can obtain,
\begin{flalign}\label{B.8}
  \begin{split}
  \left( {\rho {v^2} + p} \right)_{j + 1/2}^n &=   {\left( {\rho v} \right)^{*,n}} \cdot {v^{*,n}} + {p^{*,n}} - \frac{{{v^{*,n}}}}{{{a^{*,n}}}}\left[ {{{\left( {\rho v} \right)}^{*,n}} \cdot \delta {v^n} + {v^{*,n}} \cdot \delta {{\left( {\rho v} \right)}^n} + \delta {p^n}} \right]\\
  &\quad + \frac{{\left( {{v^{*,n}} - {a^{*,n}}} \right)\left( {{v^{*,n}} + {a^{*,n}}} \right)}}{{{a^{*,n}}}} \cdot \delta {\left( {\rho v} \right)^n} - \left( {{v^{*,n}} - {a^{*,n}}} \right){v^{*,n}} \cdot \delta {\rho ^n} + \left( {{v^{*,n}} - {a^{*,n}}} \right){f_p}\frac{{\delta {p^n}}}{{{{\left( {{a^{*,n}}} \right)}^2}}}{v^{*,n}},\\
  \left( {\rho {v^2} + p} \right)_{j - 1/2}^n &= {\left( {\rho v} \right)^{*,n}} \cdot {v^{*,n}} + {p^{*,n}} + \frac{{{v^{*,n}}}}{{{a^{*,n}}}}\left[ {{{\left( {\rho v} \right)}^{*,n}} \cdot \delta {v^n} + {v^{*,n}} \cdot \delta {{\left( {\rho v} \right)}^n} + \delta {p^n}} \right]\\
  &\quad - \frac{{\left( {{v^{*,n}} - {a^{*,n}}} \right)\left( {{v^{*,n}} + {a^{*,n}}} \right)}}{{{a^{*,n}}}} \cdot \delta {\left( {\rho v} \right)^n} + \left( {{v^{*,n}} - {a^{*,n}}} \right){v^{*,n}} \cdot \delta {\rho ^n} - \left( {{v^{*,n}} - {a^{*,n}}} \right){f_p}\frac{{\delta {p^n}}}{{{{\left( {{a^{*,n}}} \right)}^2}}}{v^{*,n}}.
  \end{split}
\end{flalign}
Inserting Eqs. (\ref{B.8}) into Eq. (\ref{B.3}), it can be obtained that
\begin{flalign}\label{B.9}
  \begin{split}
(\rho v)_j^{n+1}-(\rho v)_j^{n} & =(\rho v)_j^{*,n+1}-(\rho v)_j^{*,n}+\delta (\rho v)_j^{n+1}-\delta (\rho v)_j^{n} \\
& \quad + 2\frac{{\Delta t}}{{\Delta y}}\left\{ {{{\left( {{v^{*,n}}} \right)}^2} \cdot \delta {\rho ^n} + {\rho ^{*,n}}\frac{{{{\left( {{v^{*,n}}} \right)}^2} + {{\left( {{a^{*,n}}} \right)}^2}}}{{{a^{*,n}}}} \cdot \delta {v^n} + \left[ {\frac{{{v^{*,n}}}}{{{a^{*,n}}}} + \frac{{{v^{*,n}}\left( {{a^{*,n}} - {v^{*,n}}} \right)}}{{{a^{*,n}}}}{f_p}} \right] \cdot \delta {p^n}} \right\}
\end{split}
\end{flalign}
It should be noted that
\begin{equation}\label{B.10}
  (\rho v)_j^{*,n+1}-(\rho v)_j^{*,n}=0 .
\end{equation}
Subtracting Eq. (\ref{B.10}) from Eq. (\ref{B.9}), it can be obtained that
\begin{equation}\label{B.11}
\delta (\rho v)_j^{n+1}-\delta (\rho v)_j^{n} = 2\frac{{\Delta t}}{{\Delta y}}\left\{ {{{\left( {{v^{*,n}}} \right)}^2} \cdot \delta {\rho ^n} + {\rho ^{*,n}}\frac{{{{\left( {{v^{*,n}}} \right)}^2} + {{\left( {{a^{*,n}}} \right)}^2}}}{{{a^{*,n}}}} \cdot \delta {v^n} + \left[ {\frac{{{v^{*,n}}}}{{{a^{*,n}}}} + \frac{{{v^{*,n}}\left( {{a^{*,n}} - {v^{*,n}}} \right)}}{{{a^{*,n}}}}{f_p}} \right] \cdot \delta {p^n}} \right\}
\end{equation}
where the term $\frac{\Delta t}{\Delta y}$ can be approximated as
\begin{equation}\label{B.12}
\frac{{\Delta t}}{{\Delta y}} = \frac{{\nu \Delta y}}{{{{\left( {\left| v \right| + a} \right)}_{\max }}\Delta y}} \approx \frac{\nu }{{{v^{*,n}} + {a^{*,n}}}}
\end{equation}
In Eq. (\ref{B.12}), $\nu$ denotes the Courant number.

Inserting (\ref{B.12}) into (\ref{B.11}), we can obtain the following relationship
\begin{equation}\label{B.13}
    \delta \left( {\rho v} \right)_j^{n + 1} - \delta \left( {\rho v} \right)_j^{n} = {\theta _\rho } \cdot \delta \rho ^n{\rm{ + }}{\theta _v} \cdot \delta v^n + {\theta _p} \cdot \delta p^n
\end{equation}
with
\begin{equation} \label{B.14}
\begin{aligned}
{\theta}_{\rho} & = 2\frac{\upsilon }{{{v^{*,n}} + {a^{*,n}}}} {{\left( {{v^{*,n}}} \right)}^2}  \\
{\theta}_v & = 2\frac{\upsilon }{{{v^{*,n}} + {a^{*,n}}}} {\rho ^{*,n}}\frac{{{{\left( {{v^{*,n}}} \right)}^2} + {{\left( {{a^{*,n}}} \right)}^2}}}{{{a^{*,n}}}}       \\
{\theta}_p & = 2\frac{\upsilon }{{{v^{*,n}} + {a^{*,n}}}} \left[ {\frac{{{v^{*,n}}}}{{{a^{*,n}}}} + {f_p}\frac{{{v^{*,n}}\left( {{a^{*,n}} - {v^{*,n}}} \right)}}{{{a^{*,n}}}}} \right].
\end{aligned}
\end{equation}

\section*{References}
	
	\bibliography{manuscript}

\begin{thebibliography}{10}
\expandafter\ifx\csname url\endcsname\relax
  \def\url#1{\texttt{#1}}\fi
\expandafter\ifx\csname urlprefix\endcsname\relax\def\urlprefix{URL }\fi
\expandafter\ifx\csname href\endcsname\relax
  \def\href#1#2{#2} \def\path#1{#1}\fi

\bibitem{bonfiglioli2016moretti}
A.~Bonfiglioli, R.~Paciorri, F.~Nasuti, M.~Onofri, {Moretti's Shock-Fitting
  Methods on Structured and Unstructured Meshes}, in: Handbook of Numerical
  Analysis, Vol.~17, Elsevier, 2016, pp. 403--439.

\bibitem{PEERY1988}
K.~Peery, S.~Imlay, {Blunt-body flow simulations}, in: 24th Joint Propulsion
  Conference, Joint Propulsion Conferences, American Institute of Aeronautics
  and Astronautics, 1988, p. 2904.
\newblock \href {https://doi.org/doi:10.2514/6.1988-2904}
  {\path{doi:doi:10.2514/6.1988-2904}}.

\bibitem{ismail2006toward}
F.~Ismail, {Toward a reliable prediction of shocks in hypersonic flow:
  resolving carbuncles with entropy and vorticity control}, Ph.d. thesis,
  University of Michigan (2006).

\bibitem{shen2014}
Z.~Shen, W.~Yan, G.~Yuan, {A stability analysis of hybrid schemes to cure shock
  instability}, Communications in Computational Physics 15~(5) (2014)
  1320--1342.
\newblock \href {https://doi.org/10.4208/cicp.210513.091013a}
  {\path{doi:10.4208/cicp.210513.091013a}}.

\bibitem{Rodionov2017}
A.~V. Rodionov, {Artificial viscosity in Godunov-type schemes to cure the
  carbuncle phenomenon}, Journal of Computational Physics 345 (2017) 308--329.
\newblock \href {https://doi.org/10.1016/j.jcp.2017.05.024}
  {\path{doi:10.1016/j.jcp.2017.05.024}}.

\bibitem{Xie2017}
W.~Xie, W.~Li, H.~Li, Z.~Tian, S.~Pan, {On numerical instabilities of
  Godunov-type schemes for strong shocks}, Journal of Computational Physics 350
  (2017) 607--637.
\newblock \href {https://doi.org/10.1016/j.jcp.2017.08.063}
  {\path{doi:10.1016/j.jcp.2017.08.063}}.

\bibitem{quirk1994contribution}
J.~J. Quirk, {A contribution to the great Riemann solver debate}, International
  Journal for Numerical Methods in Fluids 18~(6) (1994) 555--574.

\bibitem{bader2014carbuncle}
G.~Bader, F.~Kemm, The carbuncle phenomenon in shallow water simulations, in:
  The 2nd International Conference on Computational Science and Engineering
  (ICCSE-2014)(Ho Chi Minh City, Vietnam), 2014.

\bibitem{kemm2014note}
F.~Kemm, {A note on the carbuncle phenomenon in shallow water simulations},
  ZAMM-Journal of Applied Mathematics and Mechanics/Zeitschrift f{\"{u}}r
  Angewandte Mathematik und Mechanik 94~(6) (2014) 516--521.

\bibitem{navas2018improved}
A.~Navas-Montilla, J.~Murillo, Improved riemann solvers for an accurate
  resolution of 1d and 2d shock profiles with application to hydraulic jumps,
  Journal of Computational Physics 378 (2019) 445 -- 476.
\newblock \href {https://doi.org/https://doi.org/10.1016/j.jcp.2018.11.023}
  {\path{doi:https://doi.org/10.1016/j.jcp.2018.11.023}}.

\bibitem{hanawa2008}
T.~Hanawa, H.~Mikami, T.~Matsumoto, Improving shock irregularities based on the
  characteristics of the mhd equations, Journal of Computational Physics
  227~(16) (2008) 7952--7976.

\bibitem{Lax2008}
P.~D. Lax, {Mathematics and physics}, Bulletin of the American Mathematical
  Society 45~(1) (2008) 135--152.

\bibitem{Fjordholm2015}
U.~S. Fjordholm, R.~Kaeppeli, S.~Mishra, E.~Tadmor, {Construction of
  Approximate Entropy Measure-Valued Solutions for Hyperbolic Systems of
  Conservation Laws}, Foundations of Computational Mathematics 17~(3) (2017)
  763--827.
\newblock \href {https://doi.org/10.1007/s10208-015-9299-z}
  {\path{doi:10.1007/s10208-015-9299-z}}.

\bibitem{godunov1959finite}
S.~Godunov, A finite difference method for the computation of discontinuous
  solutions of the equations of fluid dynamics., Sbornik: Mathematics 47~(8-9)
  (1959) 357--393.

\bibitem{harten1997high}
A.~Harten, {High resolution schemes for hyperbolic conservation laws}, Journal
  of Computational Physics 135~(2) (1997) 260--278.

\bibitem{Einfeldt1988}
B.~Einfeldt, {On Godunov-Type Methods for Gas Dynamics}, SIAM Journal on
  Numerical Analysis 25~(2) (1988) 294--318.
\newblock \href {https://doi.org/10.1137/0725021} {\path{doi:10.1137/0725021}}.

\bibitem{Einfeldt1991}
B.~Einfeldt, C.~D. Munz, P.~L. Roe, B.~Sj{\"{o}}green, {On Godunov-type methods
  near low densities}, Journal of Computational Physics 92~(2) (1991) 273--295.
\newblock \href {https://doi.org/10.1016/0021-9991(91)90211-3}
  {\path{doi:10.1016/0021-9991(91)90211-3}}.

\bibitem{Osher1982}
S.~Osher, F.~Solomon, Upwind difference schemes for hyperbolic systems of
  conservation laws, Mathematics of computation 38~(158) (1982) 339--374.

\bibitem{tadmor2003entropy}
E.~Tadmor, {Entropy stability theory for difference approximations of nonlinear
  conservation laws and related time-dependent problems}, Acta Numerica 12
  (2003) 451--512.

\bibitem{tadmor2016entropy}
E.~Tadmor, {Entropy stable schemes}, Handbook of Numerical Analysis 17 (2016)
  467--493.

\bibitem{Bultelle1998}
M.~Bultelle, M.~Grassin, D.~Serre, {Unstable Godunov discrete profiles for
  steady shock waves}, SIAM Journal on Numerical Analysis 35~(6) (1998)
  2272--2297.
\newblock \href {https://doi.org/10.1137/S0036142996312288}
  {\path{doi:10.1137/S0036142996312288}}.

\bibitem{Barth1989}
T.~J. Barth, Some notes on shock resolving flux functions. part 1: Stationary
  characteristics, Tech. Rep. NASA-TM-101087, NASA Ames Research Center (1989).

\bibitem{de2009euler}
C.~{De Lellis}, L.~{Sz{\'{e}}kelyhidi Jr}, {The Euler equations as a
  differential inclusion}, Annals of mathematics (2009) 1417--1436.

\bibitem{chiodaroli2015global}
E.~Chiodaroli, C.~{De Lellis}, O.~Kreml, {Global Ill-Posedness of the
  Isentropic System of Gas Dynamics}, Communications on Pure and Applied
  Mathematics 68~(7) (2015) 1157--1190.

\bibitem{baba2018non}
H.~A. Baba, C.~Klingenberg, O.~Kreml, V.~Macha, S.~Markfelder, {Non-uniqueness
  of admissible weak solution to the Riemann problem for the full Euler system
  in 2D}, arXiv preprint arXiv:1805.11354 (2018).

\bibitem{elling2005nonuniqueness}
V.~Elling, {Nonuniqueness of entropy solutions and the carbuncle phenomenon},
  in: Proceedings of the 10th Conference on Hyperbolic Problems (HYP2004),
  Vol.~1, 2005, pp. 375--382.

\bibitem{elling2009carbuncle}
V.~Elling, {The carbuncle phenomenon is incurable}, Acta Mathematica Scientia
  29~(6) (2009) 1647--1656.

\bibitem{Robinet2000Shock}
J.~C. Robinet, J.~Gressier, G.~Casalis, J.~M. Moschetta, {Shock wave
  instability and the carbuncle phenomenon: same intrinsic origin?}, Journal of
  Fluid Mechanics 417~(417) (2000) 237--263.

\bibitem{ismail2009affordable}
F.~Ismail, P.~L. Roe, {Affordable, entropy-consistent Euler flux functions II:
  Entropy production at shocks}, Journal of Computational Physics 228~(15)
  (2009) 5410--5436.

\bibitem{roe2006affordable}
P.~L. Roe, {Affordable, entropy consistent flux functions}, in: Eleventh
  International Conference on Hyperbolic Problems: Theory, Numerics and
  Applications, Lyon, 2006.

\bibitem{Wada1997}
Y.~Wada, M.-S. Liou, An accurate and robust flux splitting scheme for shock and
  contact discontinuities, SIAM J. Sci. Comput. 18~(3) (1997) 633--657.
\newblock \href {https://doi.org/10.1137/S1064827595287626}
  {\path{doi:10.1137/S1064827595287626}}.

\bibitem{xu2001dissipative}
K.~Xu, Z.~Li, {Dissipative mechanism in Godunov-type schemes}, International
  journal for numerical methods in fluids 37~(1) (2001) 1--22.

\bibitem{Dumbser2004}
M.~Dumbser, J.~M. Moschetta, J.~Gressier, {A matrix stability analysis of the
  carbuncle phenomenon}, Journal of Computational Physics 197~(2) (2004)
  647--670.
\newblock \href {https://doi.org/10.1016/j.jcp.2003.12.013}
  {\path{doi:10.1016/j.jcp.2003.12.013}}.

\bibitem{chauvat2005shock}
Y.~Chauvat, J.-M. Moschetta, J.~Gressier, {Shock wave numerical structure and
  the carbuncle phenomenon}, International Journal for Numerical Methods in
  Fluids 47~(8-9) (2005) 903--909.

\bibitem{zaide2012numerical}
D.~W.-M. Zaide, {Numerical shockwave anomalies}, Ph.D. thesis, University of
  Michigan (2012).

\bibitem{coquel1995hybrid}
F.~Coquel, M.-S. Liou, Hybrid upwind splitting (hus) by a field-by-field
  decomposition, NASA STI/Recon Technical Report N 95 (1995).

\bibitem{Kim2009}
S.~D. Kim, B.~J. Lee, H.~J. Lee, I.~S. Jeung, {Robust HLLC Riemann solver with
  weighted average flux scheme for strong shock}, Journal of Computational
  Physics 228~(20) (2009) 7634--7642.
\newblock \href {https://doi.org/10.1016/j.jcp.2009.07.006}
  {\path{doi:10.1016/j.jcp.2009.07.006}}.

\bibitem{Kim2010Realization}
S.~D. Kim, B.~J. Lee, H.~J. Lee, I.~S. Jeung, J.~Y. Choi, {Realization of
  contact resolving approximate Riemann solvers for strong shock and expansion
  flows}, International Journal for Numerical Methods in Fluids 62~(10) (2010)
  1107--1133.

\bibitem{zhang2017robust}
F.~Zhang, J.~Liu, B.~Chen, W.~Zhong, A robust low-dissipation ausm-family
  scheme for numerical shock stability on unstructured grids, International
  Journal for Numerical Methods in Fluids 84~(3) (2017) 135--151.

\bibitem{REN20031379}
Y.-X. Ren, A robust shock-capturing scheme based on rotated riemann solvers,
  Computers \& Fluids 32~(10) (2003) 1379 -- 1403.
\newblock \href {https://doi.org/https://doi.org/10.1016/S0045-7930(02)00114-7}
  {\path{doi:https://doi.org/10.1016/S0045-7930(02)00114-7}}.

\bibitem{nishikawa2008very}
H.~Nishikawa, K.~Kitamura, {Very simple, carbuncle-free,
  boundary-layer-resolving, rotated-hybrid Riemann solvers}, Journal of
  Computational Physics 227~(4) (2008) 2560--2581.

\bibitem{zhang2016evaluation}
F.~Zhang, J.~Liu, B.~Chen, W.~Zhong, Evaluation of rotated upwind schemes for
  contact discontinuity and strong shock, Computers \& Fluids 134 (2016)
  11--22.

\bibitem{SIMON2018144}
S.~Simon, J.~C. Mandal, {A cure for numerical shock instability in HLLC Riemann
  solver using antidiffusion control}, Computers {\&} Fluids 174 (2018)
  144--166.

\bibitem{powers2015physical}
J.~M. Powers, J.~D. Bruns, A.~Jemcov, Physical diffusion cures the carbuncle
  phenomenon, in: 53rd AIAA Aerospace Sciences Meeting, 2015, p. 0579.

\bibitem{rodionov2018artificial}
A.~V. Rodionov, Artificial viscosity to cure the carbuncle phenomenon: The
  three-dimensional case, Journal of Computational Physics 361 (2018) 50--55.

\bibitem{kitamura2013towards}
K.~Kitamura, E.~Shima, Towards shock-stable and accurate hypersonic heating
  computations: A new pressure flux for ausm-family schemes, Journal of
  Computational Physics 245 (2013) 62--83.

\bibitem{SIMON2019477}
S.~Simon, J.~Mandal, A simple cure for numerical shock instability in the hllc
  riemann solver, Journal of Computational Physics 378 (2019) 477 -- 496.
\newblock \href {https://doi.org/10.1016/j.jcp.2018.11.022}
  {\path{doi:10.1016/j.jcp.2018.11.022}}.

\bibitem{Toro1994}
E.~F. Toro, M.~Spruce, W.~Speares, {Restoration of the contact surface in the
  HLL-Riemann solver}, Shock Waves 4~(1) (1994) 25--34.
\newblock \href {https://doi.org/10.1007/BF01414629}
  {\path{doi:10.1007/BF01414629}}.

\bibitem{zaide2011shock}
D.~Zaide, P.~Roe, Shock capturing anomalies and the jump conditions in one
  dimension, in: 20th AIAA Computational Fluid Dynamics Conference, 2011, p.
  3686.

\bibitem{zaide2012flux}
D.~W. Zaide, P.~L. Roe, Flux functions for reducing numerical shockwave
  anomalies, ICCFD7, Big Island, Hawaii (2012) 9--13.

\bibitem{Xie2019}
W.~Xie, Y.~Zhang, Q.~Chang, H.~Li, {Towards an Accurate and Robust Roe-Type
  Scheme for All Mach Number Flows}, Advances in Applied Mathematics and
  Mechanics 11~(1) (2019) 1--36.
\newblock \href {https://doi.org/10.4208/aamm.OA-2018-0141}
  {\path{doi:10.4208/aamm.OA-2018-0141}}.

\bibitem{roe1997approximate}
P.~L. Roe, {Approximate Riemann solvers, parameter vectors, and difference
  schemes}, Journal of computational Physics 135~(2) (1997) 250--258.

\bibitem{harten1983upstream}
A.~Harten, P.~D. Lax, B.~{Van Leer}, {On upstream differencing and Godunov-type
  scheme for hyperbolic conservation laws}, SIAM review 25~(1) (1983) 35--61.

\bibitem{Batten19972}
P.~Batten, N.~Clarke, C.~Lambert, D.~M. Causon, On the choice of wavespeeds for
  the hllc riemann solver, SIAM Journal on Scientific Computing 18~(6) (1997)
  1553--1570.

\bibitem{Davis1988}
S.~F. Davis, {Simplified Second-Order Godunov-Type Methods}, SIAM Journal on
  Scientific and Statistical Computing 9~(3) (1988) 445--473.
\newblock \href {https://doi.org/10.1137/0909030} {\path{doi:10.1137/0909030}}.

\bibitem{park2003dissipation}
S.~H. Park, J.~H. Kwon, {On the dissipation mechanism of Godunov-type schemes},
  Journal of Computational Physics 188~(2) (2003) 524--542.

\bibitem{Kitamura2009}
K.~Kitamura, P.~Roe, F.~Ismail, {Evaluation of Euler Fluxes for Hypersonic Flow
  Computations}, AIAA Journal 47~(1) (2009) 44--53.
\newblock \href {https://doi.org/10.2514/1.33735} {\path{doi:10.2514/1.33735}}.

\bibitem{wahi2013numerical}
N.~Wahi, F.~Ismail, {Numerical shock instability on 1D Euler equations}, in:
  AIP Conference Proceedings, Vol. 1522, AIP, 2013, pp. 376--383.
\newblock \href {https://doi.org/10.1063/1.4801149}
  {\path{doi:10.1063/1.4801149}}.

\bibitem{van1982flux}
B.~van Leer, {Flux-vector splitting for the Euler equations}, in: Eighth
  International Conference on Numerical Methods in Fluid Dynamics, Springer,
  1982, pp. 507--512.

\bibitem{erpenbeck1962stability}
J.~J. Erpenbeck, Stability of step shocks, The Physics of Fluids 5~(10) (1962)
  1181--1187.

\bibitem{majda2012compressible}
A.~Majda, Compressible fluid flow and systems of conservation laws in several
  space variables, Vol.~53, Springer Science \& Business Media, 2012.

\bibitem{Nishikawa}
H.~Nishikawa, Free cfd codes,
  \url{http://www.ossanworld.com/cfdbooks/cfdcodes.html/}.

\bibitem{humpherys2009spectral}
J.~Humpherys, G.~Lyng, K.~Zumbrun, Spectral stability of ideal-gas shock
  layers, Archive for rational mechanics and analysis 194~(3) (2009)
  1029--1079.

\bibitem{humpherys2010}
J.~Humpherys, O.~Lafitte, K.~Zumbrun, Stability of viscous shock profiles in
  the high mach number limit, Communications in Mathematical Physics 293~(1)
  (2010) 1--36.

\bibitem{Humpherys2017}
J.~Humpherys, G.~Lyng, K.~Zumbrun, Multidimensional stability of
  large-amplitude navier--stokes shocks, Archive for Rational Mechanics and
  Analysis 226~(3) (2017) 923--973.
\newblock \href {https://doi.org/10.1007/s00205-017-1147-7}
  {\path{doi:10.1007/s00205-017-1147-7}}.

\bibitem{wang2014numerical}
Y.~Wang, J.~Li, Numerical defects of the hll scheme and dissipation matrices
  for the euler equations, SIAM Journal on Numerical Analysis 52~(1) (2014)
  207--219.

\bibitem{van1992design}
B.~Van~Leer, W.-T. Lee, P.~L. Roe, K.~G. Powell, C.-H. Tai, Design of optimally
  smoothing multistage schemes for the euler equations, Communications in
  applied numerical methods 8~(10) (1992) 761--769.

\bibitem{Kim2003}
S.~S. Kim, C.~Kim, O.~H. Rho, S.~K. Hong, {Cures for the shock instability:
  Development of a shock-stable Roe scheme}, Journal of Computational Physics
  185~(2) (2003) 342--374.
\newblock \href {https://doi.org/10.1016/S0021-9991(02)00037-2}
  {\path{doi:10.1016/S0021-9991(02)00037-2}}.

\bibitem{henderson2007grid}
S.~Henderson, J.~Menart, {Grid study on blunt bodies with the Carbuncle
  phenomenon}, in: 39th AIAA Thermophysics Conference, 2007, p. 3904.

\bibitem{leveque2002finite}
R.~J. LeVeque, Finite volume methods for hyperbolic problems, Vol.~31,
  Cambridge university press, 2002.

\bibitem{SHU1988439}
C.-W. Shu, S.~Osher, {Efficient implementation of essentially non-oscillatory
  shock-capturing schemes}, Journal of Computational Physics 77~(2) (1988)
  439--471.

\bibitem{Woodward1984The}
P.~Woodward, P.~Colella, {The numerical simulation of two-dimensional fluid
  flow with strong shocks}, Journal of Computational Physics 54~(1) (1984)
  115--173.

\bibitem{yoon1988lower}
S.~Yoon, A.~Jameson, Lower-upper symmetric-gauss-seidel method for the euler
  and navier-stokes equations, AIAA journal 26~(9) (1988) 1025--1026.

\bibitem{kemm2018heuristical}
F.~Kemm, {Heuristical and numerical considerations for the carbuncle
  phenomenon}, Applied Mathematics and Computation 320 (2018) 596--613.

\end{thebibliography}

\end{document}